\documentclass{aa}
\let\linenumbers\relax

\usepackage{hyperref}
\usepackage{subfigure}
\usepackage{caption}
\usepackage{graphicx}
\usepackage{txfonts}
\usepackage{lipsum}
\usepackage{lscape}
\usepackage{placeins}
\usepackage{multirow}
\usepackage{enumitem}
\usepackage{rotating}
\usepackage{xcolor}

\newcommand{\nwater}{N(\mathrm{solid\, H}_2\mathrm{O})}
\newcommand{\nmethanol}{N(\mathrm{solid\, CH}_3\mathrm{OH})}
\newcommand{\akstar}{A_K^{\mathrm{*}}}
\newcommand{\akmid}{A_K^{\mathrm{MIR}}}
\newcommand{\akred}{A_{K,0}^{\mathrm{*}}}

\begin{document}

\linenumbers

\title{Backlighting young stellar objects in the Central Molecular Zone}
\subtitle{An ensemble-averaged abundance structure of methanol ices}

\author{Yewon Kang\inst{1}
\and Deokkeun An\inst{1}\thanks{\email{deokkeun@ewha.ac.kr}}
\and Jiwon Han\inst{1,2}
\and Sang-Il Han\inst{1}
\and Dayoung Pyo\inst{1}
\and A.\ C.\ Adwin Boogert\inst{3}
\and Kee-Tae Kim\inst{4,5}
\and Do-Young Byun\inst{4,5}
}

\institute{Department of Science Education, Ewha Womans University, 52 Ewhayeodae-gil, Seodaemun-gu, Seoul 03760, Republic of Korea
\and
Department of Artificial Intelligence and Software, Ewha Womans University, 52 Ewhayeodae-gil, Seodaemun-gu, Seoul 03760, Republic of Korea
\and
Institute for Astronomy, University of Hawaii, 2680 Woodlawn Drive, Honolulu, HI 98622, USA
\and
Korea Astronomy and Space Science Institute, 776 Daedeok-daero, Yuseong-gu, Daejeon 34055, Republic of Korea
\and
University of Science and Technology, Korea (UST), 217 Gajeong-ro, Yuseong-gu, Daejeon 34113, Republic of Korea
}

\date{}

\abstract
{
The Central Molecular Zone (CMZ) of the Milky Way contains a substantial reservoir of dense molecular gas, where numerous young stellar objects (YSOs) and dense cloud cores have been identified, offering valuable opportunities to investigate star formation in the nuclear regions of spiral galaxies. However, the large distance and severe foreground extinction complicate detailed interpretation, particularly of infrared absorption features from various ice species that trace the chemical composition and evolutionary state of these embedded objects.
}{
To better characterise YSOs and dense cores in this region, we combined spectra from multiple YSOs, each likely backlit by a giant star, allowing us to probe their outer layers and derive an ensemble-averaged ice abundance profile.
}{
We obtained $L$-band spectra of 15 point-like sources with extremely red colours using GNIRS on Gemini North, enabling measurements of the CH$_3$OH absorption feature at $3.535\, \mu$m. A subset of these sources was also observed in the $K$ band. To better constrain the foreground extinction and H$_2$O ice column densities, we combined these data with $K$-band and mid-infrared spectra from our previous observations using NASA/IRTF and \textit{Spitzer}/IRS.
}{
We found that the CH$_3$OH abundance in the CH$_3$OH--CO$_2$ ice mixture (traced by the $15.4\ \mu$m shoulder) lies between $2$--$5\%$, reaffirming previous findings and confirming that it is systematically lower than the 5--15\% typically observed in the Galactic disk. Furthermore, by using the local excess of foreground extinction as a proxy for the projected distance between a backlit source and the centre of a YSO, we found that the CH$_3$OH abundance relative to solid CO$_2$ remains at the $\sim10\%$ level in the inner regions of the envelope, but increases sharply to $\sim30\%$ in the outer regions.
}{
The relatively low methanol ice abundance may reflect the unique chemical environment of the CMZ, including variations in elemental abundances and surface reaction pathways on dust grains. However, our results offer an alternative interpretation: since our sample is biased towards massive and luminous YSOs, intense heating from the central protostar may have caused substantial sublimation of methanol ice in the inner regions of their envelopes, thereby systematically lowering the observed CH$_3$OH/H$_2$O ice ratios.
}


\maketitle

\section{Introduction}\label{sec:intro}

As the nearest benchmark for studying nuclear molecular zones in spiral galaxies, the Central Molecular Zone (CMZ) of the Milky Way contains a substantial reservoir of dense molecular gas, accumulated through bar-driven inflows. While strong shear and turbulence are thought to inhibit star formation in this region \citep[e.g.][]{yusefzadeh:09, an:11}, a growing body of evidence indicates that the CMZ continues to host significant star-forming activity. This is supported by the presence of compact \ion{H}{II} regions \citep{gaume:95, meng:19, meng:22}, abundant H$_2$O and methanol masers \citep{caswell:10, longmore:17}, and a population of young stellar objects (YSOs) \citep{an:09, an:11, an:17, jang:22}, along with young massive star clusters containing O- and B-type stars \citep{mauerhan:10, dong:12}. More recently, ALMA observations have revealed a large number of deeply embedded protostars and associated bipolar SiO (5--4) outflows \citep{ginsburg:18, walker:21}, while JWST imaging has uncovered shocked molecular hydrogen structures and protostars, further confirming ongoing star formation in the region \citep{crowe:25}.

Among the various tracers of star formation, young stellar objects (YSOs) and compact cloud cores \citep[e.g.][]{hatchell:08, heyer:16} not only mark ongoing star-forming activity but also provide insights into the chemical composition of the natal molecular clouds that will give rise to the next generation of stars. In particular, infrared (IR) spectroscopy reveals a rich set of absorption features arising from the vibrational and rotational transitions of ice species such as H$_2$O, CO, CO$_2$, and CH$_3$OH \citep[e.g.][]{chiar:95, gerakines:99, whittet:98, pontoppidan:08, boogert:11, oberg:11, boogert:15, boogert:22}. These molecules form on icy dust grains under cold, dense conditions, beginning with the gravitational collapse of interstellar clouds, and are subsequently processed as the central protostar evolves. This processing includes thermal annealing, crystallization, molecular segregation, and eventual sublimation, each reflecting changes in the thermal and chemical environment \citep{boogert:15}. Therefore, observations of IR ice absorption bands provide a powerful diagnostic tool for probing the structural and evolutionary properties of YSOs and cloud cores.

In this context, both direct and indirect detections of methanol (CH$_3$OH) ice in point-like sources with extremely red colours in the CMZ \citep{an:09, an:11, an:17, jang:22} suggest that these objects are in an early stage of YSO evolution. The 15\ $\mu$m CO$_2$ absorption band is commonly observed towards quiescent dark clouds, where it typically exhibits a relatively narrow profile, comprising contributions from pure CO$_2$ ice and mixtures with H$_2$O (peaked at 15.3\ $\mu$m) and/or CO (15.1\ $\mu$m). In contrast, significantly broader 15\ $\mu$m features were found in the Spitzer/IRS mid-IR spectra of CMZ sources, due to the presence of a 15.4\ $\mu$m shoulder component arising from CO$_2$ ice mixed with species like CH$_3$OH. It is attributed to Lewis acid--base interactions that shift the CO$_2$ asymmetric stretching mode to longer wavelengths \citep{ehrenfreund:99, dartois:99}. Since methanol ice forms under low temperatures \citep[$T < 15$~K;][]{cuppen:09} and relatively high extinctions \citep[$A_V \ga 9$;][]{boogert:11, chiar:11, whittet:11}, it serves as a sensitive tracer of cold, dense environments characteristic of the earliest stages of star formation. Accordingly, \citet{an:11} interpreted the observed features as indirect evidence for the presence of CH$_3$OH ice and identified 16 objects as YSOs and 19 as candidate YSOs. This interpretation was later confirmed by follow-up near-IR spectroscopy \citep{an:17, jang:22}, which directly detected the 3.535\ $\mu$m absorption band of solid CH$_3$OH in many of the same sources. These results demonstrated that methanol ice can comprise up to $17\%\pm3\%$ relative to H$_2$O, validating the earlier interpretation and confirming the presence of abundant CH$_3$OH ice in these red CMZ objects.

However, these follow-up studies consistently reported a relatively lower abundance of CH$_3$OH with respect to H$_2$O in CMZ objects compared to values found in star-forming clouds within the Galactic disk. This discrepancy may reflect intrinsic differences in the chemical composition of CMZ clouds and/or variations in the chemical reaction networks operating under the extreme physical conditions of the CMZ interstellar medium (ISM). To better understand the origin of this systematic difference in the chemical properties of CMZ YSOs, it is useful to investigate the spatial distribution and structure of ice species within YSOs. For example, \citet{pontoppidan:08} used observations of background stars in the Ophiuchus cloud to spatially resolve the distribution of ices within the extended envelope of a YSO. However, at a distance of $\sim8$~kpc from the Sun \citep{reid:14}, direct observations of the detailed structure of individual YSO envelopes in the CMZ are challenging; for example, most CMZ objects appear as point-like sources in the mid-IR bands of Spitzer/IRAC images \citep[FWHM$\sim2\arcsec$;][]{ramirez:08}.

On the other hand, near- and mid-IR spectroscopy can provide a unique opportunity to probe the internal structure of YSO envelopes through absorption features. From our near-IR spectroscopic observations \citep{an:17, jang:22}, we found that many of the YSO candidates previously identified from Spitzer mid-IR spectra \citep{an:11} exhibit strong $2.3\ \mu$m CO band-head absorption, a feature characteristic of (super-)giant stars. The extreme red colours of these objects further suggest that their near-IR fluxes are significantly attenuated by dust. These findings indicate that the observed near- and mid-IR spectra are, in fact, composites of two sources along the line of sight: the mid-IR emission is dominated by the YSO in the early evolutionary stage, while the near-IR absorption originates primarily from a background giant star situated behind the YSO's extended envelope. Such a configuration, in which a background giant star backlights a foreground YSO, is plausible in the CMZ given its high source density. This backlighting scenario naturally could explain both the spectral energy distributions (SEDs) of Spitzer's red point-like sources in the CMZ and their observed absorption features.

Building on the unique backlighting configuration of CMZ YSOs, we aim to expand the sample of such systems and use them as probes of how ice is distributed within YSO envelopes. To address this question, we carried out follow-up near-IR spectroscopic observations of 23 red point-like sources that exhibit a strong CO$_2$ shoulder feature, and measured the column densities ($N$) of H$_2$O and CH$_3$OH ice. By combining these measurements with our previous mid-IR spectral data, we present a comprehensive analysis of these CMZ objects, including extinction estimates from both foreground dust and the envelopes of YSOs. Assuming that our sample represents YSOs at a similar evolutionary stage, we investigated the internal distribution of methanol ice within their envelopes. This work presents the first effort to model an ensemble-averaged radial profile of CH$_3$OH ice abundance as a function of projected radial distance within YSO envelopes in the CMZ.

This paper is organised as follows. In Sect.~\ref{sec:obs}, the Gemini/GNIRS observations of red point-like sources in the CMZ are described, along with the data reduction procedures. In Sect.~\ref{sec:composite}, the GNIRS spectra are supplemented with existing near-IR data and combined with mid-IR spectra to produce the composite SEDs of the CMZ sources. In Sect.~\ref{sec:params}, the modelling of the overall SED from near- to mid-IR spectra is presented, from which the foreground dust extinction is derived. Column densities of H$_2$O and CH$_3$OH measured from the GNIRS spectra are also reported. The ice compositions of the CMZ objects are examined and compared with those in the Galactic disk. In Sec.~\ref{sec:ensemble}, the derived ice abundances and extinction estimates are combined to construct an ensemble-averaged radial profile of methanol ice abundance. A summary of the results and a discussion of the observed trends are given in Sect.~\ref{sec:summary}.

\section{Gemini/GNIRS Observations}\label{sec:obs}

\subsection{Sample Selection}

\begin{table*}
\centering
\caption{Red CMZ objects observed with Gemini/GNIRS}
\label{tab:sample}
\begin{tabular}{cccrcl}
\hline\hline
Object ID & R.A.\ (J2000) & Decl.\ (J2000) & Obs.\ date & Wavelength & Mid-IR \\
(SSTGC) & (h:m:s) & (d:m:s) & (UTC) & range\tablefootmark{a} & classification\tablefootmark{b} \\
\hline
372630 & 17:44:42.76 & -29:23:16.19 & 19 July 2022 & $L$ & Possible YSO\\
425399 & 17:45:02.90 & -29:22:11.36 & 19 June 2017  & $K$& AGB star (background?)\tablefootmark{c} \\
&&& 21 June 2017 &  $L$ &\\
524665 & 17:45:39.86 & -29:23:23.39 & 22 July 2022 & $L$ & YSO \\
563780 & 17:45:54.15 & -28:58:12.39 & 6 August 2022 & $L$ & Possible YSO \\
 &  &  & 9 August 2022 & $L$ &  \\
619522 & 17:46:14.33 & -28:43:18.41 & 21 July 2022 & $L$ & Possible YSO \\
619964 & 17:46:14.48 & -28:36:39.54 & 19 June 2017 & $K$ & OH/IR star (background?)\tablefootmark{c} \\
 & &  & 21 June 2017 & $L$ &  \\
653270 & 17:46:26.55 & -28:18:59.77 & 20 June 2017 & $L$\tablefootmark{d} & Possible YSO \\
679036 & 17:46:35.98 & -28:43:58.19 & 19 August 2022 & $L$ & Possible YSO \\
696367 & 17:46:42.29 & -28:33:26.13 & 20 June 2017 & $L$\tablefootmark{d} & OH/IR star (background?)\tablefootmark{c} \\
718757 & 17:46:50.50 & -28:43:33.38 & 19 August 2022 & $L$ & Possible YSO \\
719445 & 17:46:50.72 & -28:31:24.67 & 2 August 2022 & $L$ & YSO \\
728480 & 17:46:54.13 & -28:29:39.51 & 27 July 2022 & $L$ & YSO \\
770393 & 17:47:11.75 & -28:31:21.90 & 21 July 2022 & $L$ & YSO \\
772981 & 17:47:12.90 & -28:32:05.50 & 5 August 2022 & $L$ & YSO \\
799887 & 17:47:24.80 & -28:15:56.80 & 19 June 2017 & $K$& Possible YSO \\
 & &  & 20 June 2017 &  $L$ &  \\
\hline
\end{tabular}
\tablefoot{
\tablefoottext{a}{Wavelength coverage: $K$ ($2.0$--$2.5\ \mu$m), $L$ ($2.9$--$3.9\ \mu$m).}
\tablefoottext{b}{Mid-IR spectroscopic classifications in \citet{an:11}.}
\tablefoottext{c}{The object is classified as a star in the literature, but this does not exclude the possibility that it is a background (super-)giant seen through a YSO envelope or a dense cloud core.}
\tablefoottext{d}{Supplemented with $K$-band spectra obtained using NASA IRTF/SpeX.}
}
\end{table*}

We used the Gemini Near-IR Spectrograph \citep[GNIRS;][]{elias:06a, elias:06b} on Gemini North to obtain $K$- and $L$-band spectra of 15 CMZ objects, as listed in Table~\ref{tab:sample}. These targets were primarily selected from \citet{an:11}, which reported mid-IR spectroscopic observations with Spitzer/IRS. Their near- and mid-IR colours are extremely red, with $1.6 \leq K - {\rm [3.6]} \leq 4.8$ and $2.4 \leq {\rm [3.6]} - {\rm [8.0]} \leq 5.3$, which was the basis for their inclusion in the original Spitzer/IRS follow-up observations. These colours distinguish them clearly from other stars towards the CMZ, even though typical CMZ stars already appear significantly reddened due to large foreground dust extinction ($A_V \sim 30$~mag).

The GNIRS follow-up targets were primarily selected to maximise the likelihood of detecting high CH$_3$OH ice abundance. The last column of Table~\ref{tab:sample} shows the classification of each object from \citet{an:11}, which was determined based on the strength of the $15.4\ \mu$m shoulder absorption feature of CO$_2$ ice. Since the $15.4\ \mu$m feature is attributed to CO$_2$ ice mixed with CH$_3$OH \citep{an:17, jang:22}, objects classified as YSOs are those for which \citet{an:11} assigned greater weight to the presence of a strong and statistically significant detection of this feature. In contrast, sources classified as possible YSOs show either weaker $15.4\ \mu$m absorption or less significant detections. In total, $12$ of the brightest sources ($L < 12.2$~mag) from \citet{an:11} were observed, including 5 YSOs and 7 possible YSOs.

In addition, three objects (SSTGC~425399, 619964, and 696367) previously classified as known stars in \citet{an:11}, based on a literature search, were also included in this study. Among them, SSTGC~619964 and SSTGC~696367 were identified as OH/IR stars in \citet{lindqvist:92}. However, our understanding of the red point-like objects in \citet{an:11} has since been significantly revised, following the realisation that many of these sources exhibit spectroscopic signatures of both young and evolved stars due to overlapping sightlines \citep{an:17, jang:22}, as is further demonstrated in this work. Although these sources show relatively weak $15.4\ \mu$m shoulder features, their IR colours remain as red as those of the YSO candidates in \citet{an:11}, providing an opportunity to explore the range of physical conditions and geometries of YSOs in the CMZ.

Most of our CMZ objects are likely Class~0 YSOs, i.e.\ in an early stage of evolution \citep{lada:87, andre:93}, as indicated by the large column of ices, including CO$_2$, along the line of sight. Moreover, as discussed in Appendix~\ref{sec:maser}, neither H$_2$O nor CH$_3$OH maser emission was detected from these targets, consistent with them being too young to excite Class~I masers, which are collisionally pumped in shocks and outflows, or Class~II masers, which are radiatively pumped by IR emission from warm dust close to the protostar.

\subsection{Observations and data reduction}

Queue observations were carried out at Gemini North in 2017 and 2022\footnote{Program IDs: GN-2017A-Q-28, GN-2022A-Q-118, GN-2022A-Q-217} to obtain $L$-band spectra for all selected targets. The GNIRS was used in long-slit mode with the 10.44~lines~mm$^{-1}$ grating and the long red camera ($0.05\arcsec$pixel$^{-1}$). The slit width was matched to the seeing during natural seeing guided observations, providing a spectral resolving power of $R = \lambda / \Delta \lambda \sim 400$--$500$ at $\sim3.5\ \mu$m, where the CH$_3$OH absorption band is located. In addition, three of the $L$-band targets were also observed in the $K$-band during the 2017 run (see Table~\ref{tab:sample}). These observations employed the $31.7$~lines~mm$^{-1}$ grating with the short blue camera. For both $L$- and $K$-band observations, a standard ABBA dithering pattern was used to optimize sky subtraction and achieve a sufficient signal-to-noise ratio (SNR). The total on-source exposure time per exposure sequence was $16$--$32$~min in $K$ and $8$--$40$~min in $L$, depending on the source brightness.

Standard data reduction procedures were performed using the IRAF/Gemini package \citep{tody:86, tody:93, fitzpatrick:24}\footnote{NOAO/IRAF version2.16; Gemini/IRAF version1.14. https://www.gemini.edu/observing/phase-iii/reducing-data/gemini-iraf-data-reduction-software.}. Flat-fielding and wavelength calibration frames were acquired with GCAL, the facility calibration unit, as part of the observing sequences. Corrected individual frames obtained in ABBA dithering mode were combined. Wavelength calibration for the $K$-band spectra was performed using arc lamp lines, while the Maunakea atmospheric transmission model (for an airmass of 1.0 and a precipitable water vapour column of 1.0~mm) was employed for the $L$-band. Telluric absorption features were corrected using observations of bright standard stars with spectral types $B3$--$A5$, taken with the same instrumental configuration as the science targets and observed immediately before or after the target to ensure airmass matching. Telluric corrections were applied using the Spextool package (version~4.0; \citealt{cushing:04})\footnote{\url{https://irtfweb.ifa.hawaii.edu/~spex/observer/}}.

Repeat $L$-band observations were obtained for four targets (SSTGC~563780, SSTGC~619964, SSTGC~696367, and SSTGC~799887) at different epochs to assess potential systematics in the observing conditions and data reduction. With the exception of SSTGC~563780, the $L$-band fluxes measured at two epochs differ systematically by up to $\sim$50\%, suggesting uncertainties arising from the extraction and calibration of the spectra. In particular, SSTGC~619964 and SSTGC~799887 lie in very crowded regions, where contamination from nearby sources likely affected the source extraction. Among the targets with only single $L$-band observations, SSTGC~524665 and SSTGC~770393 also have nearby sources that may have contributed to their measured fluxes.

The SNR of each spectrum was estimated by fitting a low-order polynomial around $\sim3.7\ \mu$m, where no strong absorption features are present. The typical SNR is $\sim40$, which is sufficiently high to allow confident measurement of CH$_3$OH column densities above $\sim1\times10^{17}~\mathrm{cm}^{-2}$. From telluric absorption lines in the standard star spectra, the resolving power was measured to be in the range $R \sim 400$--$500$. For the repeat observations, we carried out independent modelling of spectra from each exposure sequence before combining the resulting parameter estimates, including column densities and foreground extinctions (see below).

\section{IR spectra of red CMZ objects}\label{sec:composite}

\subsection{Composite spectra}

\begin{figure*}
\centering
\includegraphics[width=0.32\textwidth]{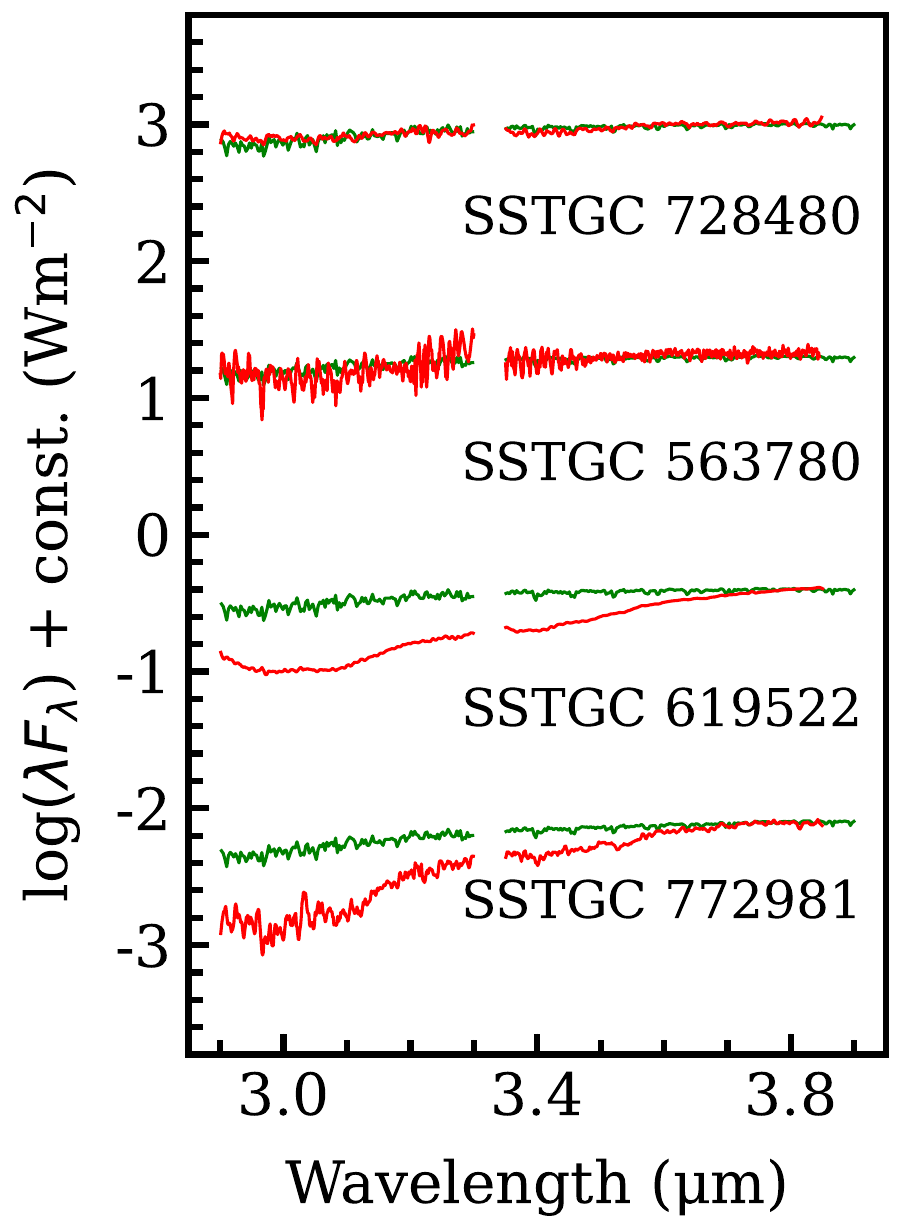}
\includegraphics[width=0.32\textwidth]{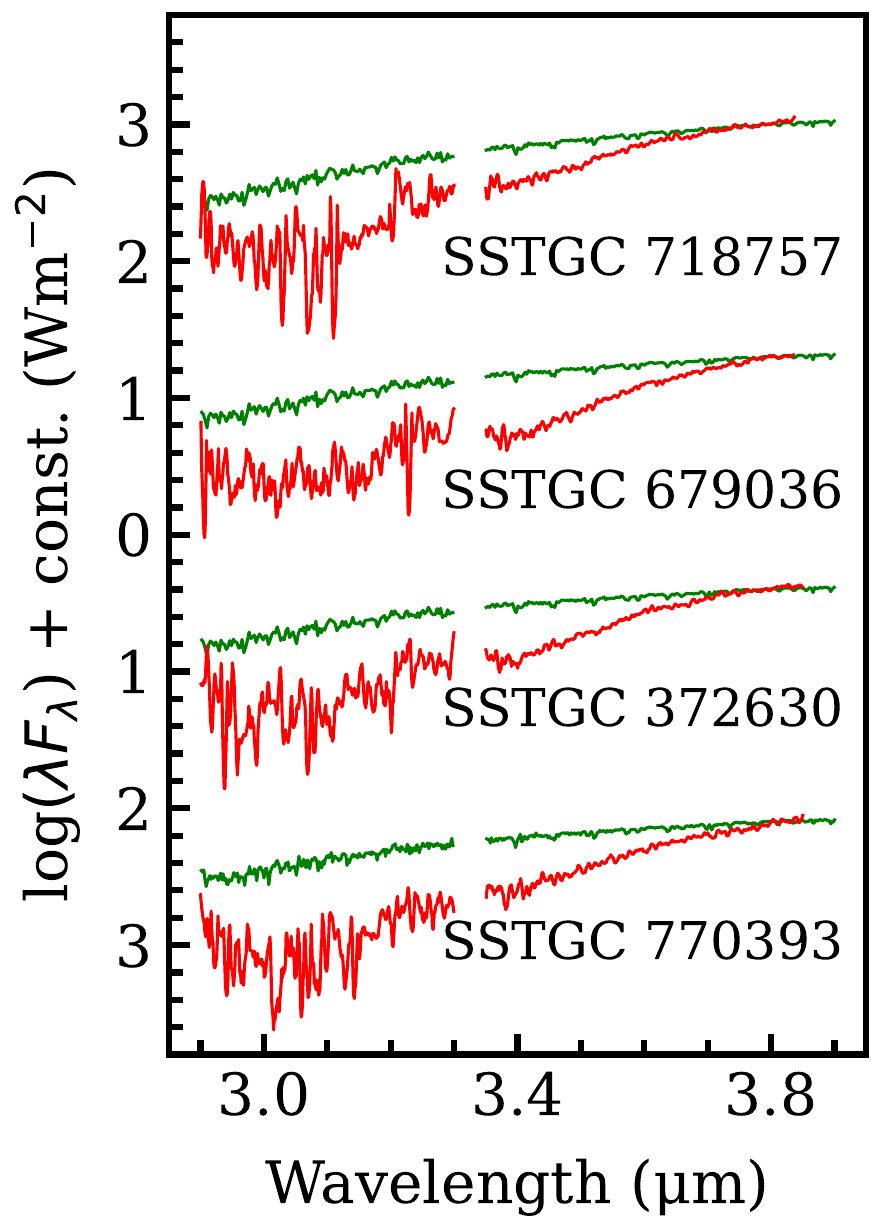}
\includegraphics[width=0.32\textwidth]{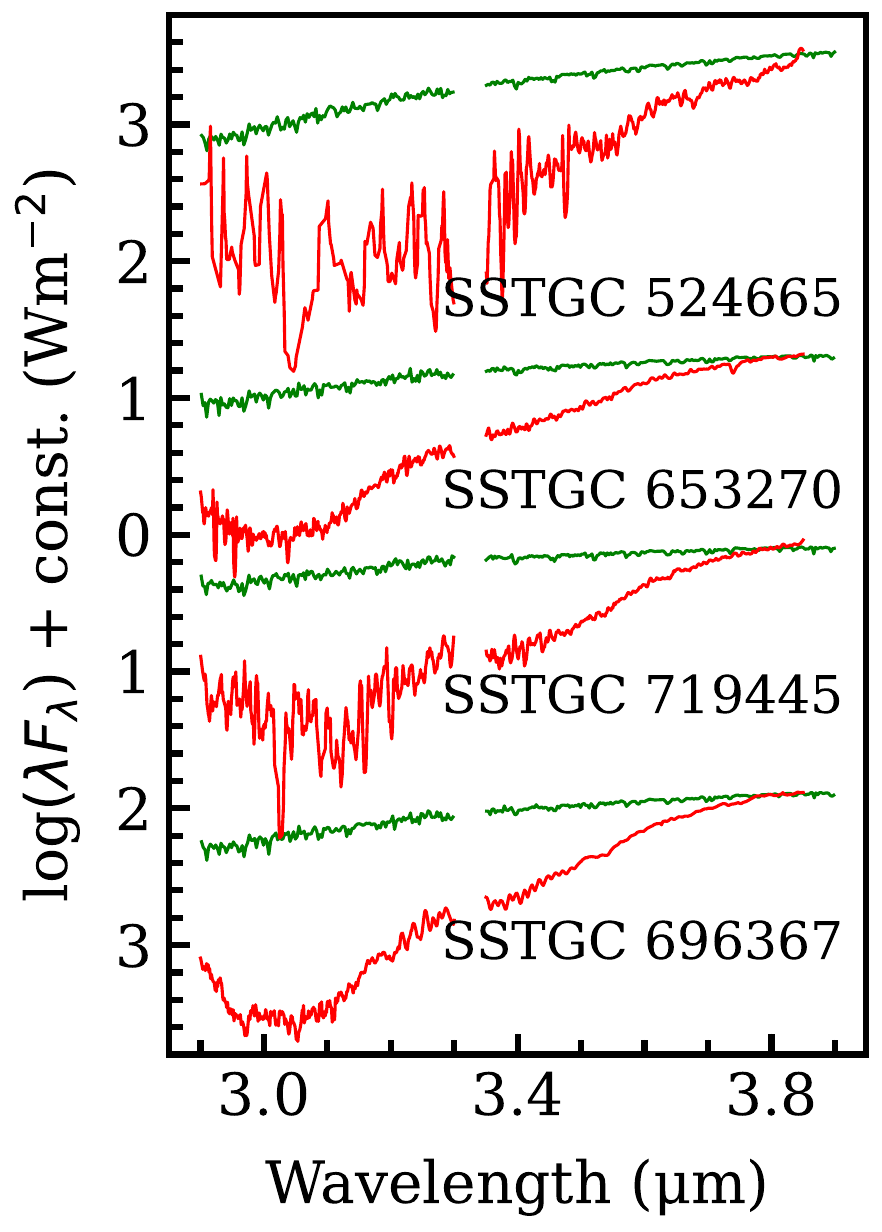}
\caption{GNIRS $L$-band spectra of 12 red sources in the CMZ. The observed spectra (red) are vertically offset for clarity and approximately ordered by increasing strength of the broad H$_2$O ice absorption band centred at $3\ \mu$m. To illustrate the varying depth of this feature, the best-fit IRTF/SpeX M5.5III template spectrum \citep{rayner:09} is shown in green for each source. These templates were reddened using the foreground extinction estimate derived from our spectral energy distribution (SED) fitting procedure (see Sect.~\ref{sec:sed}). Together with the $3\ \mu$m ice absorption profile, they reproduce the overall continuum shape and slope of the observed spectra.}
\label{fig:gnirs_l}
\end{figure*}

\begin{figure*}
\centering
\includegraphics[width=0.55\textwidth]{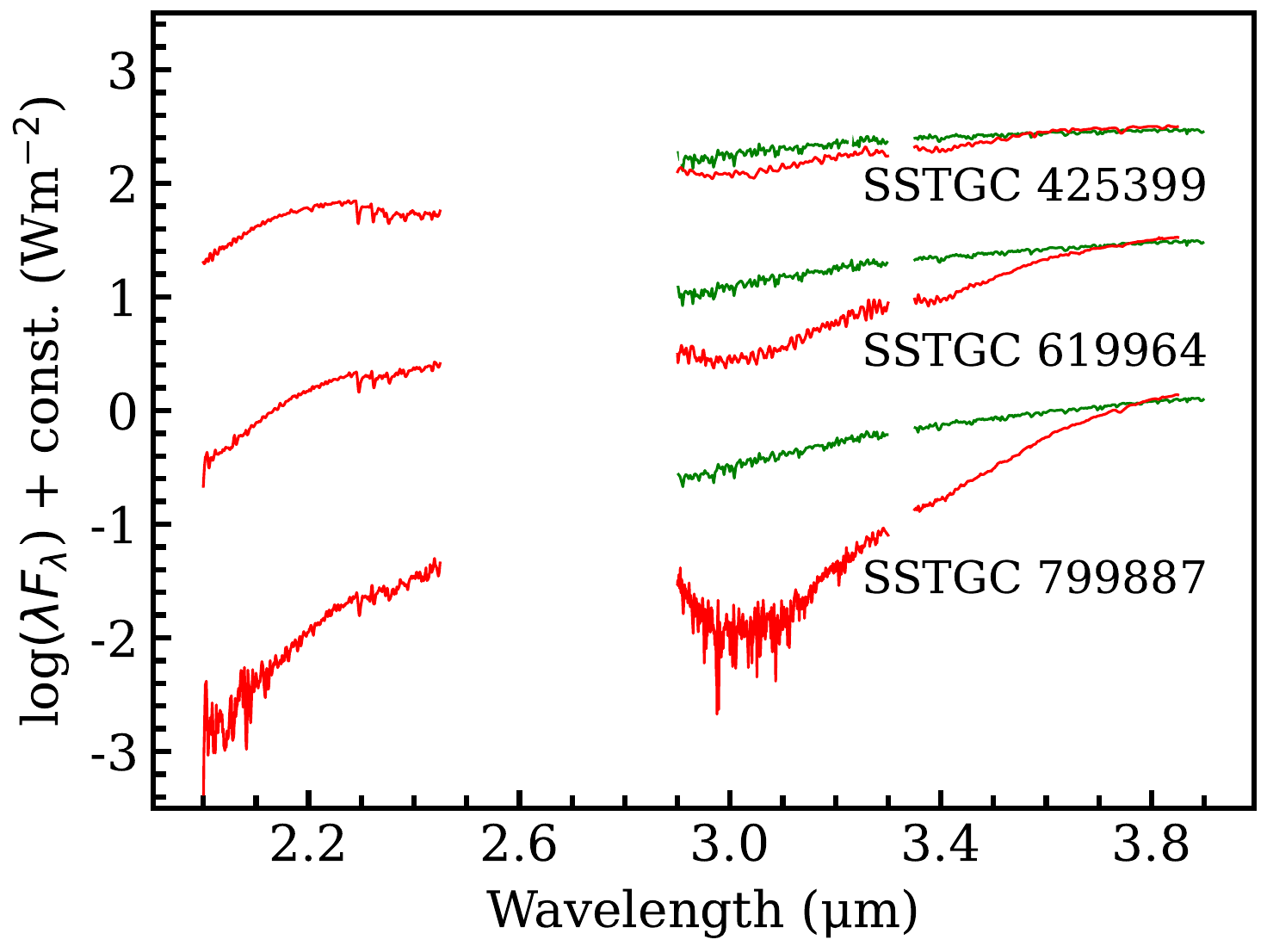}
\caption{Same as Figure~\ref{fig:gnirs_l}, but those containing both $K$- and $L$-band spectra from GNIRS observations.}
\label{fig:gnirs_kl}
\end{figure*}

Figure~\ref{fig:gnirs_l} presents the GNIRS spectra of the 12 red CMZ objects. The spectra are vertically offset with arbitrary zero points and are ordered by increasing strength of the broad H$_2$O ice absorption band centred at $3\ \mu$m, whose wing extends across much of the $L$ band. Some objects, such as SSTGC~728480 and SSTGC~563780, exhibit nearly flat spectra with very weak or absent H$_2$O ice absorption. In contrast, the majority of our sample shows strong band absorption, suggesting significant extinction due to dense molecular clouds along the line of sight. Additionally, Figure~\ref{fig:gnirs_kl} displays the spectra of the three objects for which both $K$- and $L$-band observations were obtained. It reveals CO bandhead absorption features at $2.3$--$2.4\ \mu$m, commonly associated with stellar photospheres. These features are indicative of (super-)giant stars, and their presence, together with the extremely red SED and strong molecular absorption bands, suggests the superposition of such stars behind a YSO envelope or a compact cloud core \citep[see][]{jang:22}.

\begin{figure*}
\centering
\subfigure{
\includegraphics[width=0.6\textwidth]{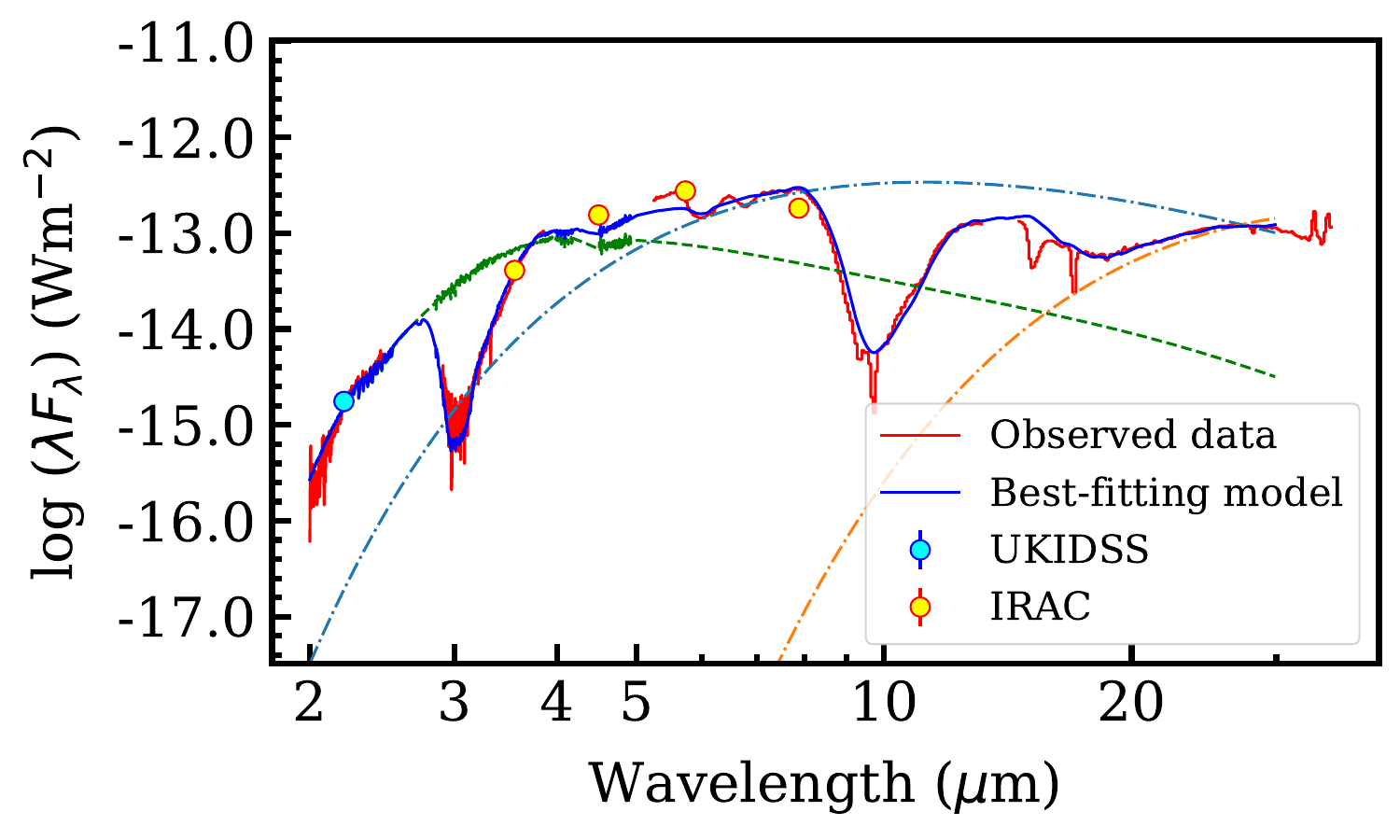}
}
\subfigure{
\includegraphics[width = 0.6\textwidth]{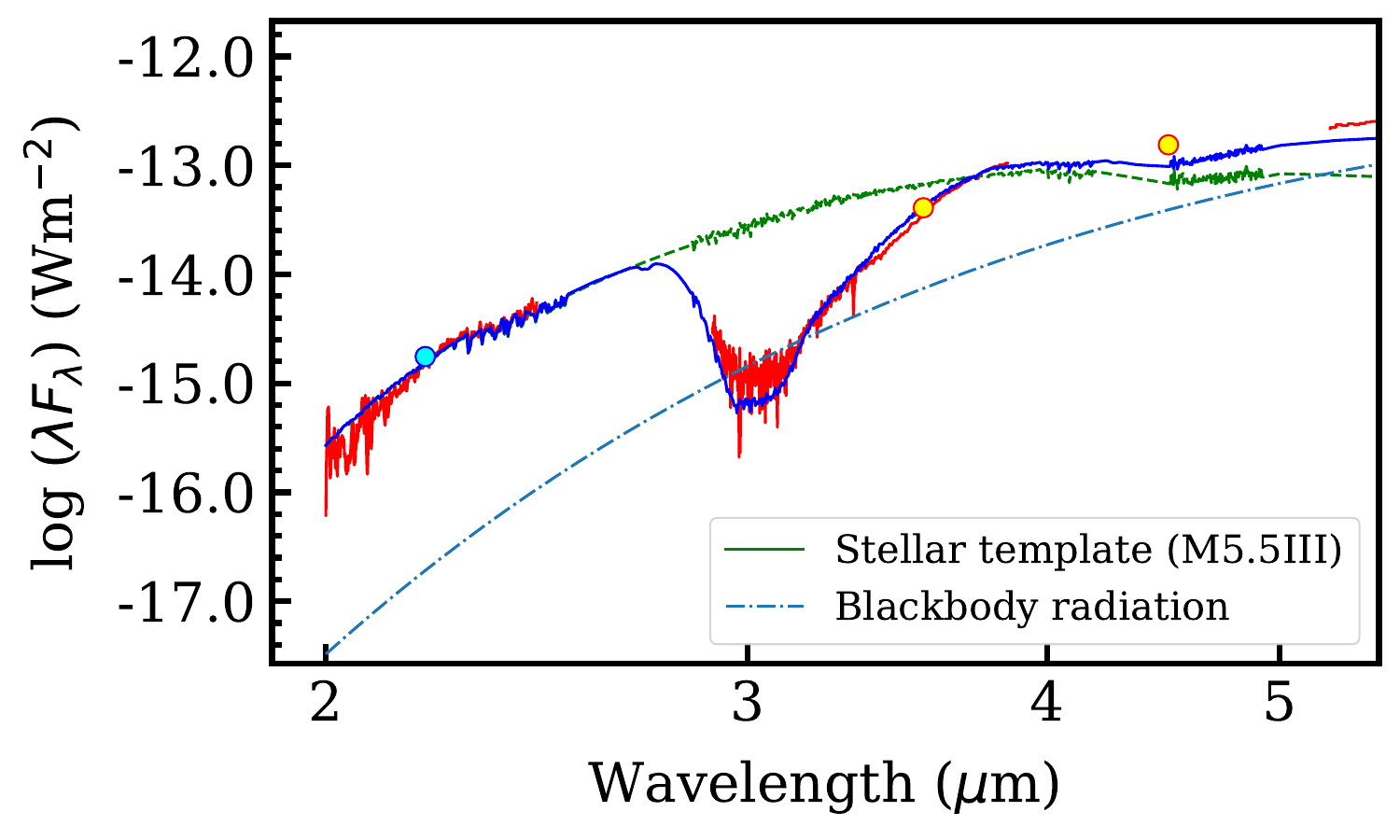}
}
\subfigure{
\includegraphics[width = 0.6\textwidth]{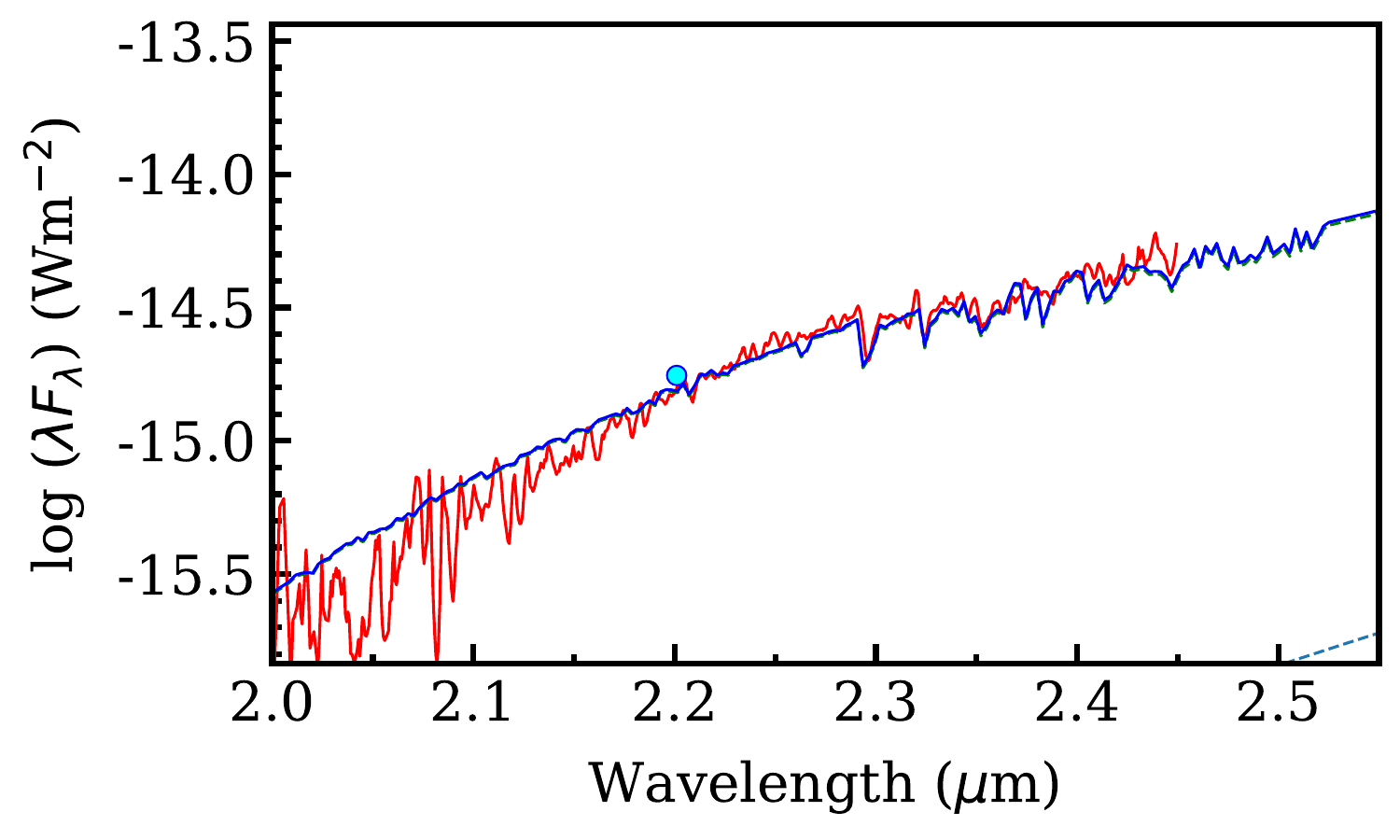}
}
\caption{Composite IR spectrum of a red CMZ object (SSTGC~799887) and modelling of the SED. The observed spectra are shown as a solid red line, constructed by combining near-IR data from Gemini/GNIRS with mid-IR data from Spitzer/IRS. Circles represent photometric measurements from UKIDSS and Spitzer/IRAC, which were used to set the flux scale of the spectra (see text). The blue solid line shows the best-fitting model to the observed spectra, which included the following components: (1) a template of an M5.5-type giant from the IRTF/SpeX spectral library \citep{rayner:09} (green line); (2) a two-component blackbody continuum (dashed blue and orange lines); (3) the $3\ \mu$m band of solid H$_2$O ice, using the template from GC~IRS~7 \citep{jang:22}; and (4) the $10\ \mu$m and $18\ \mu$m silicate bands from the GC~IRS~3 spectrum \citep{kemper:04}. Other weaker absorption features were not included in the modelling. Spectra and their best-fitting models for the additional objects are shown in Appendix~\ref{sec:sed2}.}
\label{fig:sed}
\end{figure*}

In the following analysis, we combined the GNIRS spectra with mid-IR spectra obtained from Spitzer/IRS \citep{an:11} to enable a comprehensive characterisation of the red point-like sources in the CMZ. Figure~\ref{fig:sed} shows an example of such a composite spectrum (SSTGC~799887). The GNIRS spectra cover the wavelength range $2$--$4\ \mu$m, while the low-resolution Spitzer spectra, obtained with the short-low (SL) and long-low (LL) modules, span $5$--$35\ \mu$m. As is detailed below, this multi-wavelength analysis allowed us for the simultaneous estimation of key absorption features, including those of H$_2$O and CH$_3$OH, as well as the dust extinction along the line of sight.

However, the GNIRS spectra occasionally exhibited significant flux discrepancies with the IRS spectra near the $4$--$5\ \mu$m boundary. These offsets likely reflect a combination of observational limitations in the CMZ and variations in spatial resolution and aperture size between instruments. Spectral extraction in this environment is complicated by high source density, where crowding affects background subtraction and source isolation. The large apertures of the IRS modules ($3.7\arcsec$ for SL and $10.7\arcsec$ for LL) are particularly prone to contamination from nearby sources, while GNIRS, with its higher spatial resolution, samples a more localised region. Morphological differences further contribute to these mismatches, as many CMZ sources appear compact in the near-IR but display extended emission at longer wavelengths. In contrast, our ground-based GNIRS observations benefit from higher spatial resolution but are subject to limitations in atmospheric correction, which inevitably lead to systematic uncertainties in the absolute flux calibration. The extended nature of the emission in the $L$ band further increases slit-loss uncertainties, leading to flux mismatches between the $K$- and $L$-band spectra. Overall, both the GNIRS and IRS spectra are affected by distinct sources of systematic uncertainty, which can introduce offsets in the absolute flux calibration and continuum level across the full spectral range.

For these reasons, we anchored all spectra to broadband photometric measurements, both to mitigate source confusion in the large-aperture IRS data and to establish a robust zero-point flux calibration for the GNIRS spectra. To place the GNIRS and IRS spectra on a consistent flux scale, we matched them to broadband photometry from the Galactic plane survey conducted as part of the UKIRT IR Deep Sky Survey \citep[UKIDSS;][]{lucas:08}, and from the Spitzer/IRAC survey of the Galactic centre \citep[GALCEN;][]{ramirez:08}. We used $K$-band photometry to scale the $K$-band spectra, computing synthetic magnitudes using the corresponding filter response function available from the SVO filter profile service\footnote{\url{https://svo2.cab.inta-csic.es/svo/theory/fps/}}. Similarly, the [3.6] photometry was used to set the flux scale for the $L$-band spectra. For the IRS spectra, we adopted the [8.0] photometry for scaling, as the other IRAC bands ([4.5] and [5.8]) fall in wavelength regions that are either only partially covered or entirely missing from the spectra.

Figure~\ref{fig:sed} shows the spectra calibrated in this way, with the broadband photometry over-plotted as filled circles at their respective effective wavelengths. We note that the [8.0] photometric point lies below the observed spectra, owing to the strong silicate absorption feature that affects the long-wavelength side of its passband. Because the $K$-, $L$-, and IRS spectral segments do not overlap, a direct assessment of their relative flux scaling is not possible. Nonetheless, the consistency of the flux calibration can be evaluated through SED modelling, introduced in Sect.~\ref{sec:sed}. The solid blue lines in Fig.~\ref{fig:sed} show the corresponding best-fitting models, and any residual offsets with respect to the data are discussed therein.

As the flux uncertainties in the data products are often underestimated, we scaled them by a multiplicative factor derived from local pseudo-continuum regions free of strong absorption features. Specifically, we fitted a second-order polynomial to $2.10$--$2.25\ \mu$m for the $K$-band spectra and $3.50$--$3.70\ \mu$m for the $L$-band. The scaling factor was chosen to match the median flux uncertainty to the standard deviation of the logarithmic flux in each range. This yielded more realistic random uncertainties of 1--4\% for most $L$-band spectra, although the spectrum of SSTGC~524665 shows uncertainties as high as 17\%. In addition, we applied a second-order \citet{savitzky:64} filter to the spectra to reduce noise while preserving the spectral resolution.

\subsection{NASA IRTF/SpeX}

\begin{table}
\centering
\caption{Red CMZ objects with IRTF/SpeX spectra}
\label{tab:irtf}
\begin{tabular}{cccl}
\hline\hline
Object ID & R.A.\, (J2000) & Decl.\, (J2000) & Mid-IR \\
(SSTGC) & (h:m:s) & (d:m:s) & classification\tablefootmark{a} \\
\hline
300758 & 17:44:14.49 & -29:23:22.04 & Possible YSO\\
348392 & 17:44:33.36 & -29:27:00.70 & Non-YSO \\
388790 & 17:44:48.94 & -29:23:42.36 & Non-YSO \\
404312 & 17:44:54.89 & -29:14:13.10 & Non-YSO \\
405235 & 17:44:55.25 & -29:15:37.80 & Non-YSO  \\
716531 & 17:46:49.69 & -28:36:57.00 & Non-YSO \\
726327 & 17:46:53.31 & -28:32:01.30 & YSO \\
817031 & 17:47:32.96 & -28:34:11.83 & Non-YSO \\
\hline
\end{tabular}
\tablefoot{
\tablefoottext{a}{Mid-IR spectroscopic classifications in \citet{an:11}. The non-YSO classification primarily reflects the weak or absent absorption of the CO$_2$ shoulder component, but does not necessarily rule out the possibility that the source is a YSO envelope or a dense core at an early evolutionary stage.}
}
\end{table}

\citet{jang:22} obtained $K$- and $L$-band spectra of red CMZ objects with a medium-resolution spectrograph SpeX \citep{rayner:03} on the 3.2m NASA IR Telescope Facility (IRTF). Their data are comparable to the GNIRS spectra presented here in both spectral resolution and wavelength coverage. Like our GNIRS sample, the IRTF data combined with IRS observations span the $2$–$35\ \mu$m range. Eleven of these objects are bright enough to be included in \citet{an:11}, and they also exhibit extreme red colours ([3.6]$-$[8.0] $\ga 3.0$). Two of these (SSTGC~653270 and SSTGC~696367) were also observed in this work; for these sources, we supplemented our GNIRS spectra with the $K$-band IRTF/SpeX data. The spectra of the remaining IRTF/SpeX objects, excluding one with very low S/N, are listed in Table~\ref{tab:irtf}, and were combined with the IRS spectra in the same manner as the GNIRS+IRS composite spectra.

The IRTF/SpeX observations were performed simultaneously in the $K$ and $L$ bands using a single instrumental setup, thereby providing a consistency check on the reliability of the flux zero-point adjustments applied to match the photometric measurements. For the eight objects listed in Table~\ref{tab:irtf}, we found that the $L$-band spectra required a scaling factor roughly twice that of the $K$-band to match the photometry. These offsets likely reflect the extended nature of the emission arising from YSO envelopes, which becomes more prominent at longer wavelengths. This interpretation is supported by the fact that the objects appear essentially point-like in the $K$ band, but more extended in the $L$ band \citep{an:17}. It is also possible that the lower spatial resolution of the IRAC [3.6] images used for $L$-band scaling may have included flux contributions from nearby sources, in contrast to the higher-resolution UKIDSS $K$-band images used for the $K$-band scaling; however, this effect is likely less significant than that of the extended emission.

In total, we have 23 objects with $3$--$35\ \mu$m composite spectra assembled from Gemini/GNIRS, IRTF/SpeX, and Spitzer/IRS observations. About half of these also have $K$-band spectra. For the remainder, we used UKIDSS $K$-band photometry to constrain their physical parameters, as is detailed below. The full set of composite spectra, corrected for zero-point flux offsets and including scaled flux uncertainties, is publicly available via Zenodo.\footnote{\url{https://doi.org/10.5281/zenodo.17605042}}

\section{Spectral analysis and derivation of physical parameters}\label{sec:params}

Most of the observed CMZ objects in this study show deep silicate and H$_2$O ice absorption features, indicating the significant amount of molecular clouds along the line of sight to each object. While foreground dust extinction and column densities of ice species cannot be reliably constrained using broadband photometry alone, the SED traced by IR spectra provides a valuable diagnostic for quantifying such parameters. Below, we performed a global modelling of the SED to extract foreground extinctions, along with the column density of solid H$_2$O with a wide absorption band (Sect.~\ref{sec:sed}). We combined this with additional analysis of more localised absorption features from CH$_3$OH ice (Sect.~\ref{sec:ch3oh}) to reveal complex structures of dust and ices along each line of sight towards the red CMZ objects. 

\subsection{Column densities of solid H$_2$O and foreground extinction}\label{sec:sed}

As shown in Fig.~\ref{fig:sed}, the spectra of the red CMZ objects are characterised by several absorption bands. The near-IR absorption feature at $3\ \mu$m is primarily caused by the O--H stretching vibration in solid H$_2$O ice. The broad absorption features centred at $10\ \mu$m and $18\ \mu$m are characteristic of silicate dust and originate from the Si--O stretching and O--Si--O bending modes, respectively, in amorphous silicate grains. Superimposed on a continuum that would otherwise be nearly flat in $\lambda F_\lambda$ space, or slowly rising towards longer wavelengths, these absorption features are sufficiently broad to modify both the continuum level and the overall shape of the SED. In our previous SED modelling \citep{an:17, jang:22}, we found that the overall SED can be well explained by the superposition of two sources of radiation: blackbody emission representing warm dust emission from the extended envelope of a YSO, and the spectrum of a red (super-)giant, both attenuated by significant columns of foreground dust. Given the above considerations, we modelled the composite IR spectra using the following components:

\begin{itemize}

\item{Photospheric component: in the near-IR, many red CMZ objects exhibit CO bandhead absorption at $2.3\ \mu$m, indicating the presence of a red (super-)giant along the line of sight to the mid-IR source. \citet{jang:22} performed detailed spectral fits using the NASA IRTF/SpeX library \citep{rayner:09}, and showed that varying the template spectral type between late-K and late-M giants only mildly affects the parameter estimates, because, beyond $2\ \mu$m, the stellar spectra of late-type giants show little variation in slope. Based on this, we adopted the spectrum of an M5.5III giant (HD~94705) as a representative photospheric component. This choice was also motivated by the fact that $K$-band spectra are available for only 5 of the 15 GNIRS targets.}

\item{H$_2$O ice bands at $3\ \mu$m and $13\ \mu$m: we utilised the observed IRTF/SpeX spectrum of GC~IRS~7 from \citet{jang:22} to model the $3\ \mu$m band, which arises from the O--H stretching mode. In addition, H$_2$O ice exhibits an even broader feature centred at $13\ \mu$m, associated with the librational mode. Since the GC~IRS~7 spectrum is limited to the near-IR region, we complemented it with a laboratory spectrum of pure H$_2$O ice at 15~K from the Leiden Ice Database for Astrochemistry \citep[LIDA;][]{rocha:22} for wavelengths beyond $3.8\ \mu$m. Although the laboratory data also include the $6\ \mu$m feature, we did not incorporate it, as it only partially accounts for the observed absorption in that region.}

\item{Silicate bands at $9.7\ \mu$m and $18\ \mu$m: following \citet{an:09, an:11}, we adopted the mid-IR silicate absorption profile from \citet{kemper:04}. Besides the continuum extinction, the strengths of the silicate features provide a strong constraint on the dust column density along the line of sight. The relative strengths of the two silicate modes were fixed during scaling to match the observed data. The best-fitting peak optical depth at $9.7\ \mu$m was converted to $A_K$ by multiplying it by $0.99$ \citep{roche:85}, which is a factor of two lower than that towards the local ISM \citep{roche:84, madden:22}.}

\item{Warm dust clouds: as demonstrated by \citet{an:17}, the mid-IR continuum of red CMZ objects can be effectively reproduced by emission from two blackbody components, likely originating from the extended envelope of a YSO. According to our modelling (see below), the warmer component, with a temperature of $\sim300$--$400$~K, dominates the spectral range between $\sim5\ \mu$m and $20\ \mu$m, where the photospheric contribution becomes negligible. In contrast, the cooler component, with a temperature of $\sim100$~K, primarily accounts for the rising slope of the spectrum beyond $\sim20\ \mu$m.}

\end{itemize}

We incorporated the components described above into our global SED modelling, assuming the continuum extinction curve from \citet{boogert:11}. In addition to these components, narrower but still significant absorption features are seen in our composite spectra, including the $6\ \mu$m and $6.8\ \mu$m complexes, and the $15\ \mu$m CO$_2$ ice band. Furthermore, the spectrum of SSTGC~799887 in Fig.~\ref{fig:sed} shows a narrower absorption feature at $\sim18.7\ \mu$m, which is in fact due to over-subtraction of background emission from [\ion{S}{iii}]. A similar effect is seen at $\sim10\ \mu$m, where over-subtraction at the bottom of the silicate band creates a spurious narrow absorption feature. On the other hand, the emission lines at $33.5\ \mu$m and $34.8\ \mu$m, from [\ion{S}{iii}] and [\ion{Si}{ii}], respectively, could trace the localised emission from the YSO itself. For our modelling, we selected spectral regions free of these features, using only clean continuum segments across the $K$-, $L$-, and mid-IR bands (see Appendix~\ref{sec:sed2}).

The CH$_3$OH ice bands at $3.535\ \mu$m and $9.74\ \mu$m were not included in our global SED analysis and were instead masked, because both features are highly sensitive to the shape of the local continuum. The $3.535\ \mu$m band lies within the broader $3\ \mu$m H$_2$O and hydrocarbon absorption complex, while the $9.74\ \mu$m band is embedded in the deep $10\ \mu$m silicate absorption. Because a global SED fit provides only a coarse approximation to the continuum in these wavelength ranges, the continuum level in both regions is particularly uncertain. We therefore analysed the CH$_3$OH bands separately, fitting a third-order polynomial to more accurately trace the local baseline (see Sects.~\ref{sec:ch3oh} and \ref{sec:97}).

In modelling the water-ice absorption, we tied the strength of the $13\ \mu$m libration band to the peak optical depth of the $3\ \mu$m stretching mode. This choice was motivated by the strong and broad silicate absorption at $10\ \mu$m in our CMZ targets, which overlaps with the libration feature and prevents a reliable decomposition. In addition, while some studies isolated the $13\ \mu$m band using simplified polynomial continua \citep[e.g.,][]{boogert:11}, we adopted a more physically motivated continuum model based on blackbody dust emission. As a result, the $13\ \mu$m feature could not be independently constrained in our fits and was instead anchored to the better-defined $3\ \mu$m absorption. We note, however, that the two bands may have probed slightly different lines of sight (see below).

The solid blue lines in Fig.~\ref{fig:sed} show the best-fitting model for SSTGC~799887, derived as described above. The model included eight free parameters: a flux normalisation for the near-IR stellar template, temperatures and normalisations for the two blackbody components, foreground extinction affecting the near-IR source, foreground extinction affecting the mid-IR source (including both continuum and silicate absorption), and the column density of solid H$_2$O ice. To derive global parameters that reproduce the overall SED shape, we used {\tt curve\_fit}, the non-linear least-squares optimisation routine from the SciPy library, applying weights based on the scaled uncertainties in logarithmic flux space. Results for the remaining objects are presented in Appendix~\ref{sec:sed2}.

Overall, the models reproduce the observed SEDs of the red CMZ sources well, capturing the spectral curvature shaped by the major broad absorption bands. The $2.3\ \mu$m CO bandheads are also well matched. However, some local mismatches remain. For instance, the model slightly overestimates the depth of the $3\ \mu$m band (middle panel) and underestimates the flux on the short-wavelength side of the $9.7\ \mu$m silicate feature. The continuum at wavelengths shorter than $2.2\ \mu$m also appears systematically brighter in the model. Most notably, the model underpredicts the IRS flux between $5.5$ and $7\ \mu$m. This mismatch may partly arise from source contamination in the IRAC photometry used to scale the IRS spectra, as the coarse spatial resolution of IRAC ($\sim2\arcsec$) increases the risk of source contamination in crowded regions. Taken together, these deviations likely reflect a combination of factors, including uncertainties in spectral extraction in the dense CMZ environment, residual uncertainties in flux calibration and background subtraction, and possible limitations in the adopted templates for the $3\ \mu$m H$_2$O and $9.7\ \mu$m silicate bands in capturing the true continuum and band shapes. Nevertheless, as the model captures the global shape of the SED and the key spectral features, the derived parameters provide a reasonable approximation to the intrinsic spectral properties of the sources.

\begin{table*}
\centering
\caption{Estimates on foreground extinction and H$_2$O ice absorption}
\label{tab:sed}
\begin{tabular}{cccccc}
\hline\hline
Object ID & $\akstar$ & $\akmid$ & $\tau_{3.0}$ & $\nwater$ \\
(SSTGC) & (mag) & (mag) & & ($10^{17}$\ cm$^{-2}$) \\
\hline
\multicolumn{5}{c}{Gemini/GNIRS} \\
\hline
372630 & 9.4 $\pm$ 1.4 & 4.3 $\pm$ 0.1 & 1.2 $\pm$ 0.3 & 21 $\pm$ 6 \\
425399 & 6.5 $\pm$ 1.2 & 2.6 $\pm$ 0.7 & $<0.6$ & $<11$ \\
524665 & 13.3 $\pm$ 1.8 & 3.0 $\pm$ 0.1 & 2.1 $\pm$ 0.6 & 34 $\pm$ 11 \\
563780 & 4.8 $\pm$ 0.9 & 4.1 $\pm$ 0.2 & $<0.01$ & $<0.1$ \\
619522 & 4.6 $\pm$ 0.9 & 3.0 $\pm$ 0.1 & 1.0 $\pm$ 0.2 & 17 $\pm$ 4 \\
619964 & 10.2 $\pm$ 1.6 & 4.5 $\pm$ 1.9 & 1.6 $\pm$ 0.3 & 26 $\pm$ 6 \\
653270 & 8.2 $\pm$ 1.4 & 2.2 $\pm$ 1.2 & 2.2 $\pm$ 0.4 & 36 $\pm$ 7 \\
679036 & 10.0 $\pm$ 1.4 & 4.5 $\pm$ 0.1 & 1.2 $\pm$ 0.3 & 20 $\pm$ 5 \\
696367 & 9.0 $\pm$ 1.5 & 4.1 $\pm$ 1.3 & 3.1 $\pm$ 0.3 & 51 $\pm$ 7 \\
718757 & 12.1 $\pm$ 1.8 & 3.4 $\pm$ 0.2 & 1.2 $\pm$ 0.4 & 19 $\pm$ 6 \\
719445 & 7.0 $\pm$ 1.2 & 4.9 $\pm$ 0.1 & 2.6 $\pm$ 0.3 & 43 $\pm$ 6\\
728480 & 4.7 $\pm$ 0.9 & 4.2 $\pm$ 0.1 & $<0.3$ & $<4$ \\
770393 & 9.6 $\pm$ 1.4 & 3.1 $\pm$ 0.1 & 1.4 $\pm$ 0.3 & 24 $\pm$ 6 \\
772981 & 6.5 $\pm$ 1.0 & 4.5 $\pm$ 0.1 & 1.1 $\pm$ 0.3 & 17 $\pm$ 5 \\
799887 & 14.4 $\pm$ 1.8 & 4.1 $\pm$ 0.7 & 4.0 $\pm$ 0.4 & 65 $\pm$ 10 \\
\hline
\multicolumn{5}{c}{IRTF/SpeX} \\
\hline
300758 & 5.9 $\pm$ 1.1 & 3.3 $\pm$ 0.01 & 0.6 $\pm$ 0.2 & 10 $\pm$ 4 \\
348392 & 7.2 $\pm$ 1.4 & 2.1 $\pm$ 0.1 & $<0.5$ & $<8$ \\
388790 & 5.7 $\pm$ 1.2 & 2.6 $\pm$ 0.2 & $<0.3$ & $<6$ \\
404312 & 7.4 $\pm$ 1.3 & 2.3 $\pm$ 0.2 & $<0.2$ & $<4$ \\
405235 & 8.2 $\pm$ 1.4 & 2.0 $\pm$ 0.1 & $<0.7$ & $<12$ \\
716531 & 6.1 $\pm$ 1.2 & 2.0 $\pm$ 0.2 & $<0.3$ & $<5$ \\
726327 & 5.8 $\pm$ 1.2 & 4.4 $\pm$ 0.2 & 1.1 $\pm$ 0.2 & 18 $\pm$ 4 \\
817031 & 4.7 $\pm$ 1.1 & 2.7 $\pm$ 0.2 & $<0.6$ & $<11$ \\
\hline
\end{tabular}
\tablefoot{$\akstar$ denotes the foreground extinction in $K$ derived by matching a near-IR template spectrum, whereas $\akmid$ refers to the foreground extinction in $K$ estimated from the shape of the mid-IR spectrum, including the $10\ \mu$m and $18\ \mu$m silicate features.}
\end{table*}

Results from the global SED fitting are summarised in Table~\ref{tab:sed} for both the Gemini/GNIRS and IRTF/SpeX samples. Only the three key parameters are reported. In the second and third columns, $\akstar$ denotes the foreground extinction in the $K$ band associated with the near-IR photospheric component, while $\akmid$ represents the $K$-band extinction inferred from the mid-IR spectrum, including the contribution from the silicate absorption bands. The fourth column lists the peak optical depth of the $3\ \mu$m H$_2$O ice feature ($\tau_{3.0}$). The corresponding column density is shown in the final column, computed by multiplying $\tau_{3.0}$ by a fixed FWHM of $330$~cm$^{-1}$ for amorphous H$_2$O ice and using a band strength of $A = 2 \times 10^{-16}~\mathrm{cm\,molecule^{-1}}$ \citep{hagen:81}.

The uncertainties in Table~\ref{tab:sed} represent the quadrature sum of fitting errors, derived from the covariance matrix, and various systematic uncertainties. The latter include the effect of varying the stellar template between M0IIIb (HD~213893) and M9III (IRAS~15060+0947), with the difference from the default case taken as an effective $2\sigma$ uncertainty. To evaluate the impact of the extinction law, we repeated the SED fitting with alternative curves from \citet{indebetouw:05} and \citet{fritz:11}. The latter is notably steeper in the near-IR, leading to shifts in the derived parameters. The deviations from the default curve \citep{boogert:11} were treated as an effective $2\sigma$ uncertainty and included in the total error budget for both $A_K$ and $\tau_{3.0}$. For $\nwater$, we adopted a $10\%$ uncertainty in the band strength and included it in the total error budget. For the targets with repeated observations, the differences in $\akstar$ or $\tau_{3.0}$ between individual spectra exceed the statistical uncertainties but remain well below the combined systematic uncertainties. In such cases, half of the observed differences was added in quadrature to the total error budget. Upper limits are quoted at the $3\sigma$ level, calculated from the total error budget.

Several objects in this study overlap with \citet{jang:22}. Specifically, in addition to the eight sources listed as the IRTF/SpeX sample in Table~\ref{tab:sed}, three objects (SSTGC~653270, SSTGC~696367, and SSTGC~719445) were also observed in the $L$-band with Gemini/GNIRS. The unweighted mean and standard deviation of the differences are $2.5\pm0.7$~mag in $\akstar$ and $-0.08\pm0.44$ in $\tau_{3.0}$, indicating systematically higher extinction and lower optical depth in this study. These offsets primarily reflect our revised modelling approach, which incorporates the extended wavelength coverage up to $35\ \mu$m. Nevertheless, the dispersion remains well below the level of the total systematic uncertainties.

\begin{figure}
\centering 
\includegraphics[width = 0.42\textwidth]{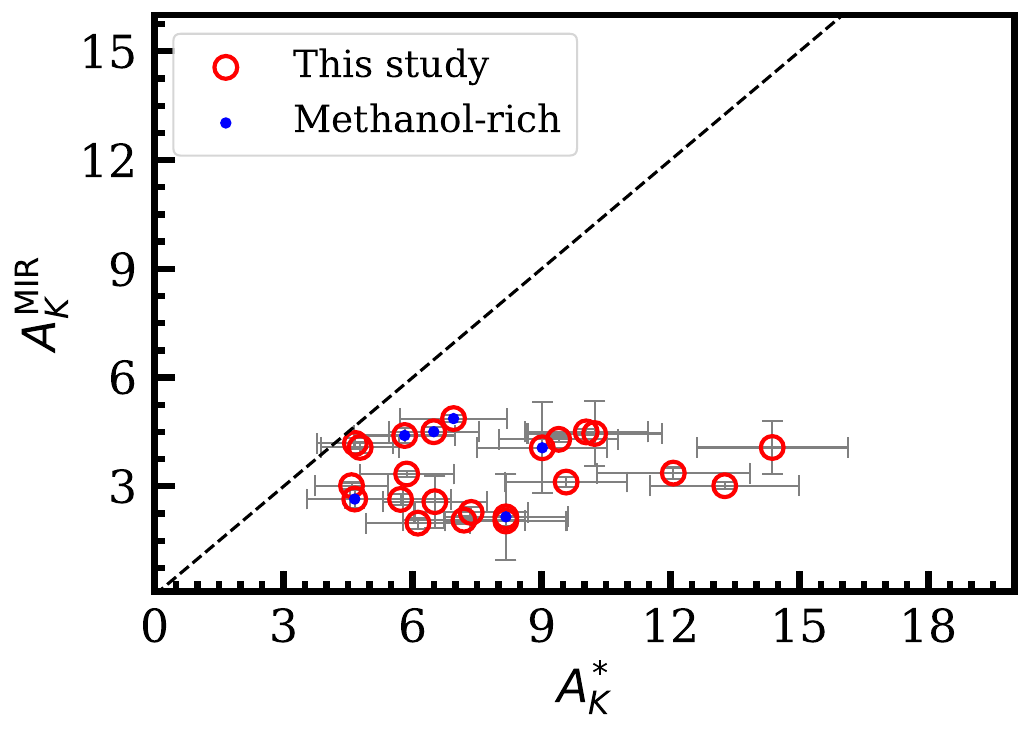}
\includegraphics[width = 0.42\textwidth]{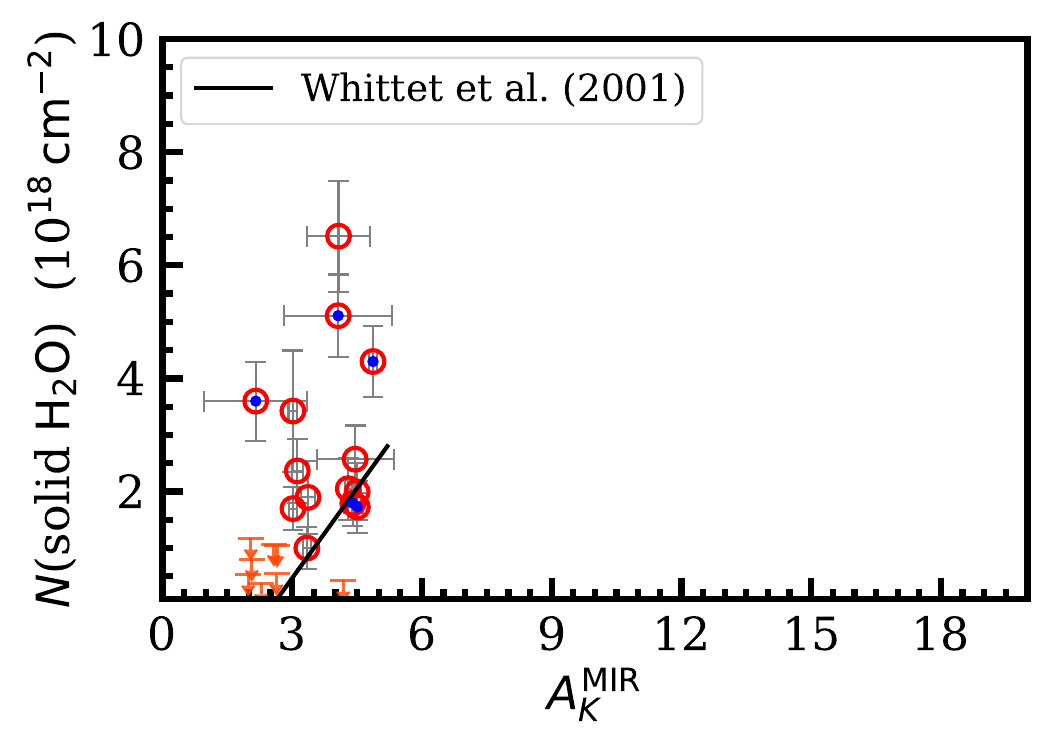}
\includegraphics[width = 0.42\textwidth]{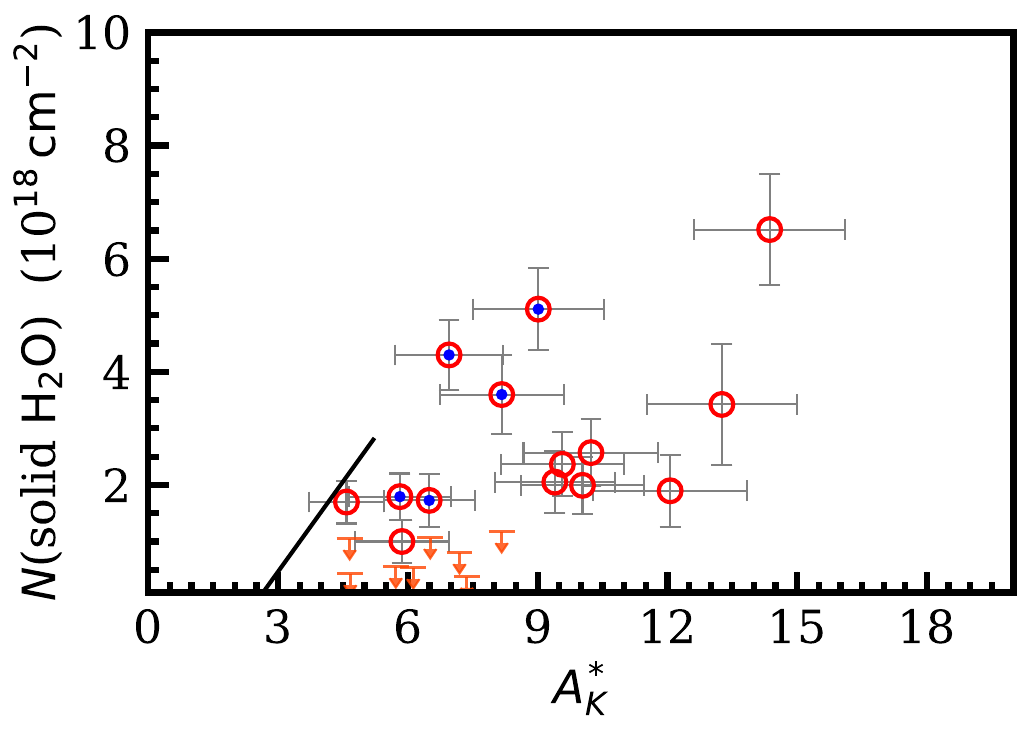}
\caption{Distribution of extinction and H$_2$O ice column densities derived from the global fits to the composite spectra. The two extinction estimates, $\akmid$ and $\akstar$, come from our composite spectra but rely on different wavelength regimes: the mid-IR spectral shape for $\akmid$ and near-IR template matching for $\akstar$. Objects with large solid CH$_3$OH column densities ($N > 3 \times 10^{17}\, \mathrm{cm}^{-2}$; see below) are marked with blue open circles. The dashed line in the top panel indicates unity. For comparison, the solid lines in the middle and bottom panels show the empirical relation derived for the Taurus molecular cloud by \citet{whittet:01}, shifted horizontally by $A_K = 2.2$~mag to account for the average foreground extinction towards the CMZ.}
\label{fig:akh2o}
\end{figure}

As shown in the top panel of Fig.~\ref{fig:akh2o}, $\akmid$ ranges from $2$ to $5$~mag, corresponding to $A_V \sim 18$--$45$~mag. In contrast, $\akstar$ spans a substantially larger range of $4$--$15$~mag. The systematic excess of $\akstar$ over $\akmid$ indicates that the near-IR stellar template requires stronger attenuation than implied by the mid-IR silicate bands. Together with the presence of CO absorption bandheads in many of our CMZ spectra, this enhanced extinction supports the interpretation that our composite spectra combine two sources along nearly the same line of sight \citep{jang:22}.

\begin{figure}
\centering 
\includegraphics[width = 0.48\textwidth]{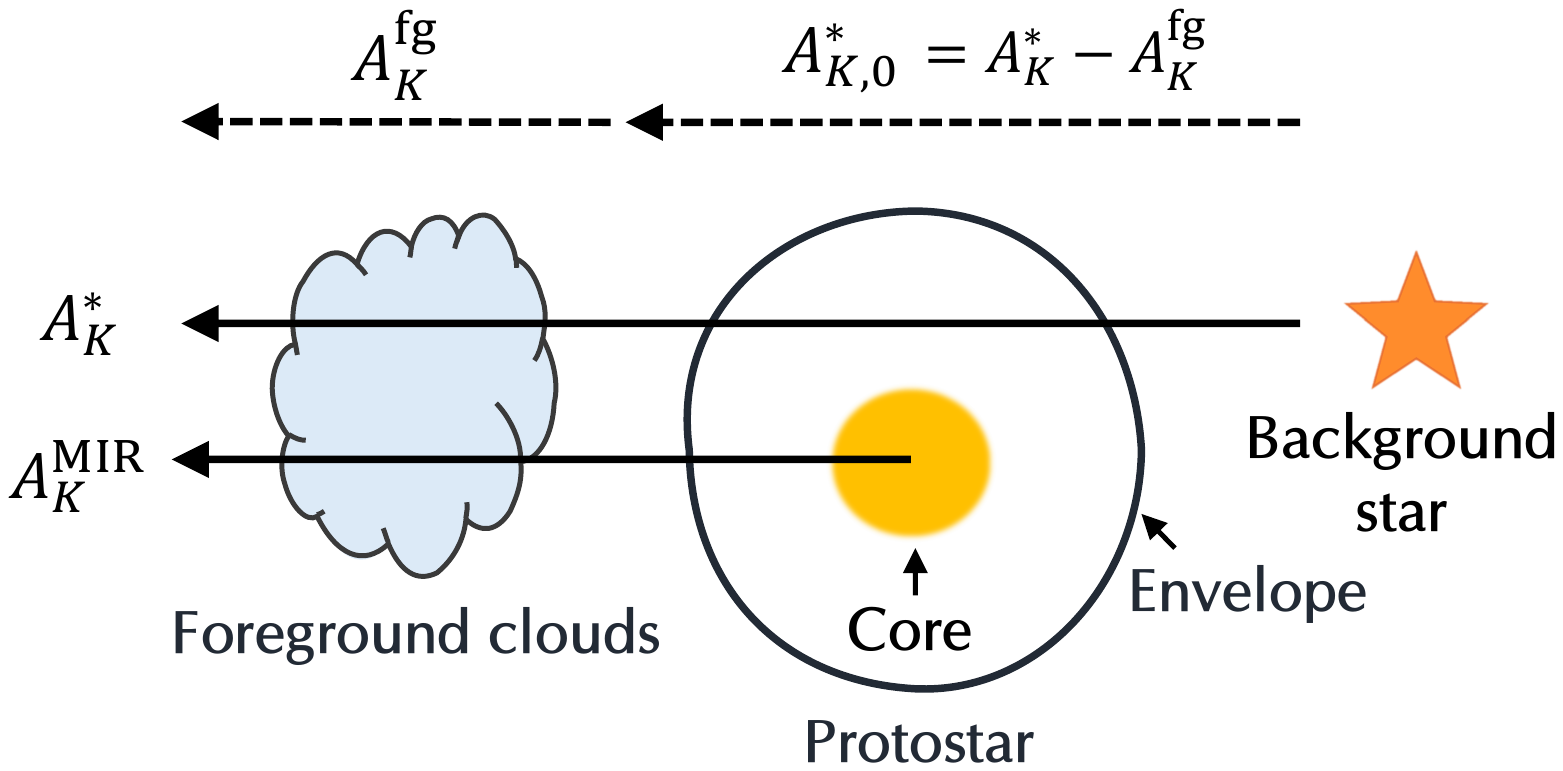}
\caption{Schematic illustration of the light-ray geometry probed by the composite spectra. $\akstar$ reflects extinction constrained mainly by the near-IR, while $\akmid$ is determined primarily from the mid-IR. The quantity $A_K^{\mathrm{fg}}$ represents the foreground extinction estimated from nearby source-free regions, which accounts for dust located along the line of sight in the Galactic disk and within the CMZ. The envelope-only extinction, $\akred$, is defined as $\akstar - A_K^{\mathrm{fg}}$, isolating the extinction produced by material within the protostellar envelope along the sightline to the background star.}
\label{fig:schematic}
\end{figure}

Figure~\ref{fig:schematic} illustrates the light-ray geometry probed by our composite spectra. The photospheric near-IR emission from a background giant star passes through the extended dusty envelope of a protostar and therefore encounters a deeper column of material than the mid-IR emission emerging from the protostar’s core. At the distance of the CMZ, the two light rays are unresolved and their emissions blend after undergoing the same foreground extinction by dust in the Galactic disk and/or within the CMZ. Because the near-IR part of the spectrum is dominated by photospheric emission from the background star, while the mid-IR is dominated by thermal emission from the cooler protostellar core, this configuration naturally produces the distinct extinction estimates $\akstar$ and $\akmid$ obtained from the near- and mid-IR spectral regions, respectively. The absence of H$_2$O and Class I CH$_3$OH maser emission towards nearly all sources (Appendix D) is consistent with the early, envelope-dominated evolutionary stage required for this backlighting geometry.

The middle and bottom panels in Figure~\ref{fig:akh2o} show the H$_2$O column density from the $3\ \mu$m band plotted against the foreground extinction estimates. Of the 23 targets in our sample, eight exhibit strong H$_2$O ice absorption with $\nwater > 2 \times 10^{18}\ \mathrm{cm}^{-2}$, while six show moderate absorption ($1$--$2 \times 10^{18}\ \mathrm{cm}^{-2}$). The remaining sources display only weak features or yield upper limits. No clear correlation is found between $\nwater$ and the YSO classifications of \citet[][see Table~\ref{tab:sample}]{an:11}: for example, two OH/IR stars (SSTGC~619964 and SSTGC~696367) show significant detections, whereas another star known in the literature has only an upper limit. As discussed further below, in the context of near-/mid-IR source superposition, the near-IR classifications of these sources as (super-)giants do not necessarily preclude them from being YSOs.

For sources with significant detections of H$_2$O ice, the measured optical depths are $\tau_{3.0} \approx 0.5$--$4.0$. A minor contribution could arise from intervening Galactic disk clouds, but this effect is expected to be negligible, with \citet{moultaka:15} finding $\tau_{3.0} \approx 0.03$ towards stars in the vicinity of Sgr~A$^*$. Even if this full amount were present along our sightlines, it would account for only $\sim6\%$ of the weakest detection and $<1\%$ of the strongest. We therefore consider the foreground H$_2$O ice absorption negligible and apply no correction.

For comparison, the solid lines in Fig.~\ref{fig:akh2o} show the empirical relation for Taurus derived by \citet{whittet:01}, who found that H$_2$O ice appears in dense clouds once the extinction exceeds $A_K \approx 0.35$~mag \citep[see also][]{whittet:11}. Since the CMZ sightlines include foreground diffuse ISM clouds contributing $A_K \sim 2.2$~mag ($A_V \sim 20$~mag), we shifted the Taurus relation horizontally by this amount. The shifted curve thus reflects the extinction--ice column density relation expected for CMZ sources. However, the shifted Taurus relation provides a poor match to either of the observed distributions. In the middle panel of Fig.~\ref{fig:akh2o}, most objects with detected H$_2$O ice (open circles) lie above the $\nwater$ values predicted from their $\akmid$. In the bottom panel, $\akstar$ extends to much higher values, emphasising the exceptionally large extinctions along the near-IR path. The \citeauthor{whittet:01} relation lies between the two distributions defined by $\akmid$ and $\akstar$, indicating that a single Taurus-like relation does not adequately describe our CMZ sample. This is not surprising, since CMZ sightlines are considerably more complex than those in nearby molecular clouds, as illustrated by the following effects:

\begin{itemize}

\item The threshold extinction for the onset of H$_2$O ice formation is not consistently reproduced by either $\akmid$ or $\akstar$. In the middle panel, sources with upper limits in $\nwater$ broadly define a CMZ threshold near $\akmid \approx 2.55$ at the minimum detected $\nwater$, yet one object (SSTGC~653270) already shows substantial ice at this value. In the bottom panel, the apparent threshold in $\akstar$ is nearly twice as large. This behaviour is naturally explained by the highly patchy foreground extinction towards the CMZ, which varies on $\sim1\arcmin$ scales and shifts the effective extinction at which ice appears along a given line of sight.

\item Even where the threshold extinction appears roughly consistent in the middle panel, the shifted Taurus relation still fails to reproduce the observed $\akmid$--$\nwater$ distribution. Nearly half of the sources lie above the Taurus prediction in $\nwater$ at a given $\akmid$, which would imply either that $\akmid$ underestimates the relevant dust column or that H$_2$O ices form more efficiently in the CMZ than in Taurus. The latter scenario seems unlikely in the strongly UV-irradiated, warmer CMZ environment, suggesting instead that $\akmid$ does not trace the same column of cold dust that dominates the ice absorption.

\item Despite the measurement uncertainties, an overall increase of $\nwater$ with $\akstar$ is evident in the bottom panel. Some of the scatter may again be driven by patchy foreground extinction. However, because our $\nwater$ estimates are mainly constrained by the near-IR 3~$\mu$m ice band, while the librational band provides only a weak additional constraint, $\nwater$ is expected to correlate more tightly with $\akstar$ than with $\akmid$, in agreement with the observed trend.

\item This behaviour reflects the different optical paths that define $\akmid$ and $\akstar$ through the same YSO envelope (Fig.~\ref{fig:schematic}): the mid-IR path ($\akmid$, middle panel) samples a shorter route through warmer, more deeply embedded regions, whereas the near-IR path ($\akstar$, bottom panel) traverses cooler, more extended parts of the envelope. If $\akstar$ and $\akmid$ therefore probe different dust columns, then thermal processing of ices and radial gradients in ice abundance within the envelope naturally lead to departures from a simple, single-column Taurus-like dust–$\nwater$ relation.

\end{itemize}

\subsection{Column densities of solid CH$_3$OH}\label{sec:ch3oh}

In addition to the SED fitting, we analysed the $3.535\ \mu$m absorption feature from solid CH$_3$OH \citep{hudgins:93,schutte:96}. The band is sufficiently narrow that the local continuum was fitted with a third-order polynomial over the $3.40$--$3.50\ \mu$m and $3.60$--$3.70\ \mu$m intervals. The feature was modelled in optical-depth space with the laboratory spectrum of pure CH$_3$OH ice at 10~K \citep{hudson:24}, fitted using the {\tt curve\_fit} routine from the SciPy library. As it lies on the long-wavelength wing of a stronger, broader band centred at $3.42\ \mu$m, it was extracted using the same polynomial fit and wavelength ranges adopted for the optical-depth spectra. The model spectrum was then smoothed to match the spectral resolution of our observations.

\begin{table}
\centering
\caption{Peak optical depths and column densities of solid CH$_3$OH}
\label{tab:ch3oh}
\begin{tabular}{ccc}
\hline\hline
Object ID & $\tau_{3.535}$ & $\nmethanol$ \\ 
(SSTGG) & & ($10^{17}\ \mathrm{cm}^{-2}$) \\
\hline
\multicolumn{3}{c}{Gemini/GNIRS} \\
\hline
372630 & $<0.06$ & $<5.4$ \\
425399 & $<0.02$ & $<1.3$ \\
524665 & $<0.68$ & $<42.6$ \\
563780 & $<0.07$ & $<4.6$ \\
619522 & $<0.02$ & $<1.2$ \\
619964 & $<0.01$ & $<0.9$ \\
653270 & 0.03 $\pm$ 0.01 & 2.1 $\pm$ 0.4 \\
679036 & $<0.02$ & $<1.3$ \\
696367 & 0.08 $\pm$ 0.03 & 4.7 $\pm$ 1.9 \\
718757 & $<0.02$ & $<1.2$ \\
719445 & 0.05 $\pm$ 0.02 & 2.9 $\pm$ 1.5 \\
728480 & $<0.10$ & $<6.2$ \\
770393 & $<0.02$ & $<1.6$ \\
772981 & 0.08 $\pm$ 0.01 & 5.2 $\pm$ 1.1 \\
799887 & $<0.04$ & $<3.8$ \\
\hline
\multicolumn{3}{c}{IRTF/SpeX} \\
\hline
300758 & $<0.12$ & $<7.6$ \\
348392 & $<0.08$ & $<5.0$ \\
388790 & $<0.03$ & $<1.9$ \\
404312 & $<0.05$ & $<2.9$ \\
405235 & $<0.08$ & $<5.2$ \\
716531 & $<0.04$ & $<2.7$ \\
726327 & 0.08 $\pm$ 0.02 & 4.8 $\pm$ 1.3 \\
817031 & 0.13 $\pm$ 0.04 & 8.1 $\pm$ 2.7 \\
\hline
\end{tabular}
\end{table}

Table~\ref{tab:ch3oh} summarises the peak optical depth of the $3.535\ \mu$m absorption in the best-fit model ($\tau_{3.535}$) for both the Gemini/GNIRS and IRTF/SpeX samples. The column density $\nmethanol$ was computed using a band strength of $5.19 \times 10^{-18}\ \mathrm{cm\,molecule}^{-1}$ \citep{hudson:24}. The value is close to $5.3 \times 10^{-18}\ \mathrm{cm\,molecule}^{-1}$ in \citet{hudgins:93}; however, a conservative 12\% uncertainty in the band strength was adopted to account for the impact of the $3.42\ \mu$m feature. We further assessed systematic effects by re-fitting the spectra with a second-order polynomial continuum and by modelling the feature with a Gaussian profile to evaluate model dependence in $\tau_{3.535}$. The uncertainties listed in Table~\ref{tab:ch3oh} are the quadrature sum of these contributions and the statistical errors derived from the covariance matrix of the fit. Upper limits are quoted at the $3\sigma$ level, calculated from the total error budget.

Of the targets with repeated observations, only SSTGC~696367 exhibits a significant detection of the $3.535\ \mu$m band, and its individual $\tau_{3.535}$ measurements are consistent within the statistical uncertainty. The three objects (SSTGC~653270, SSTGC~696367, and SSTGC~726327) with significant detections were also included in \citet{jang:22}; the differences in $\tau_{3.535}$ are well within the measurement uncertainties ($<0.01$). Since the comparison for SSTGC~653270 and SSTGC~696367 was made based on independent observations, it demonstrates the robustness of our $\tau_{3.535}$ estimates. In case of SSTGC~726327, the IRTF/SpeX spectra have been re-analysed in this work, which also yields a result consistent with the previously reported value of $0.10\pm0.01$ in \citet{an:17}.

\begin{figure*}
\centering
\includegraphics[width=0.32\textwidth]{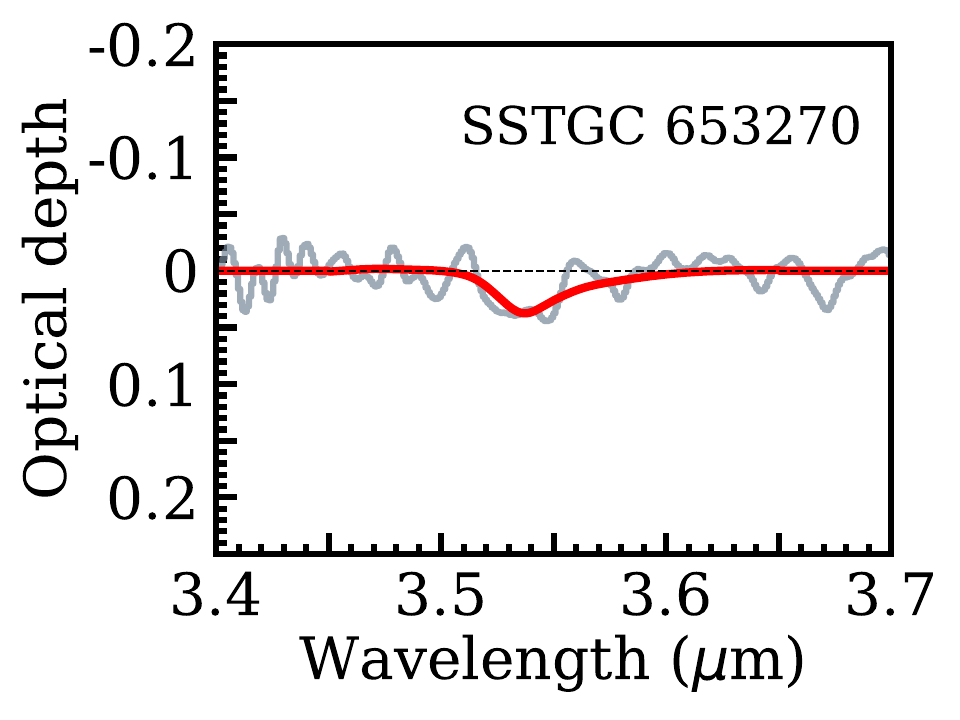}
\includegraphics[width=0.32\textwidth]{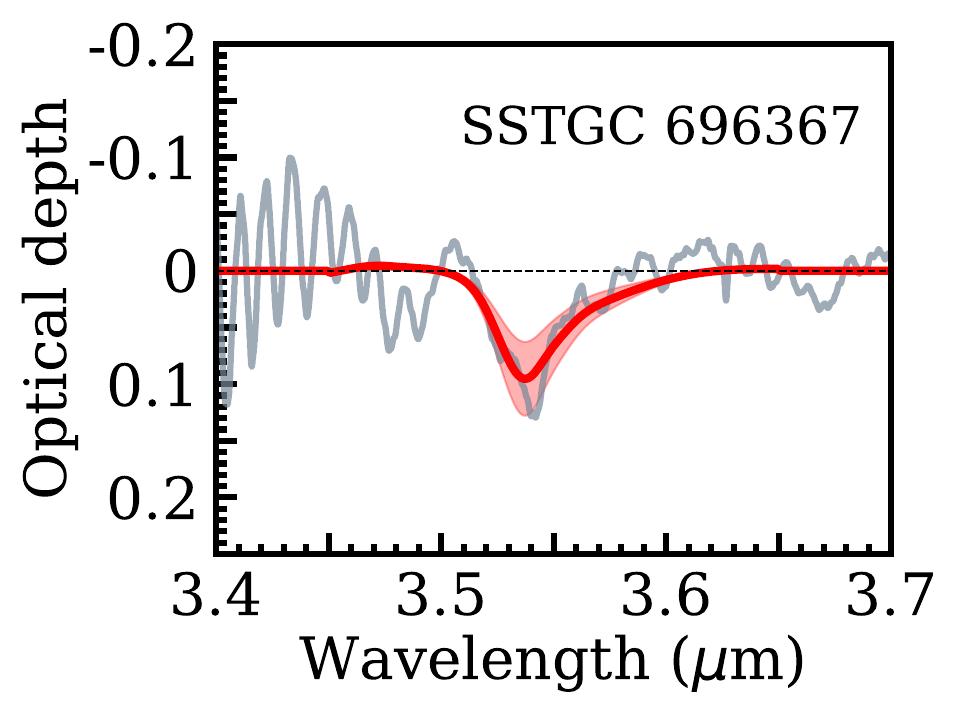}
\includegraphics[width=0.32\textwidth]{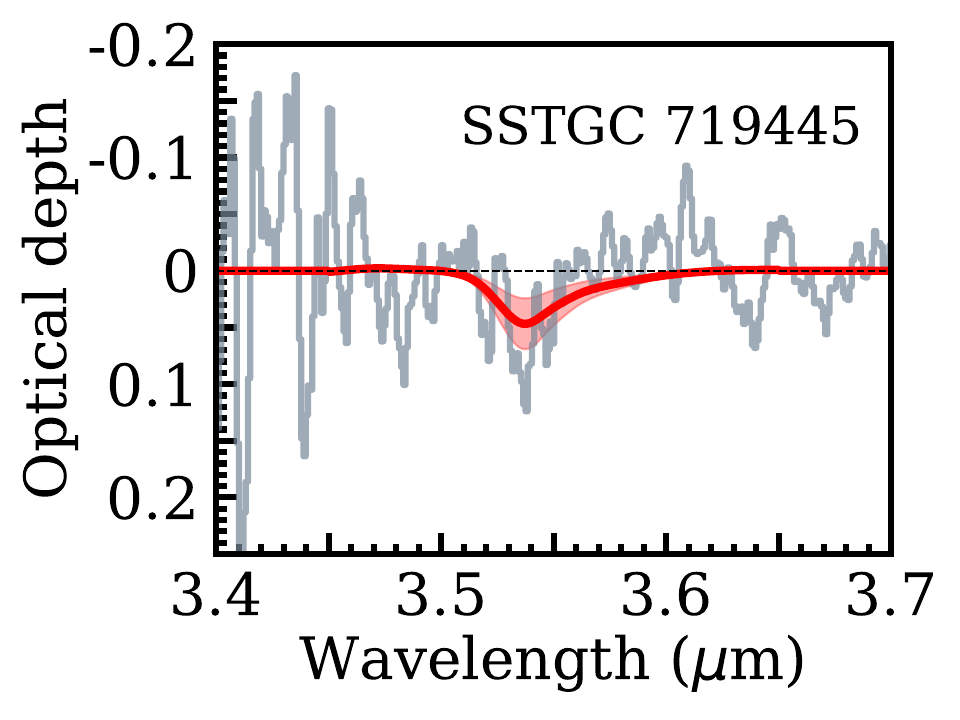}
\includegraphics[width=0.32\textwidth]{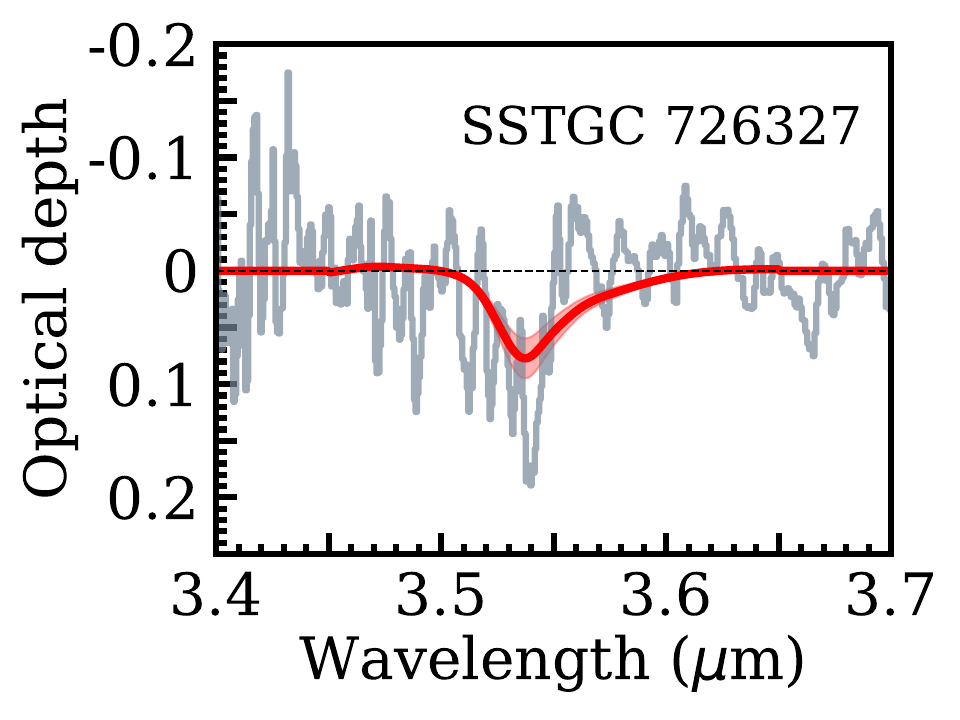}
\includegraphics[width=0.32\textwidth]{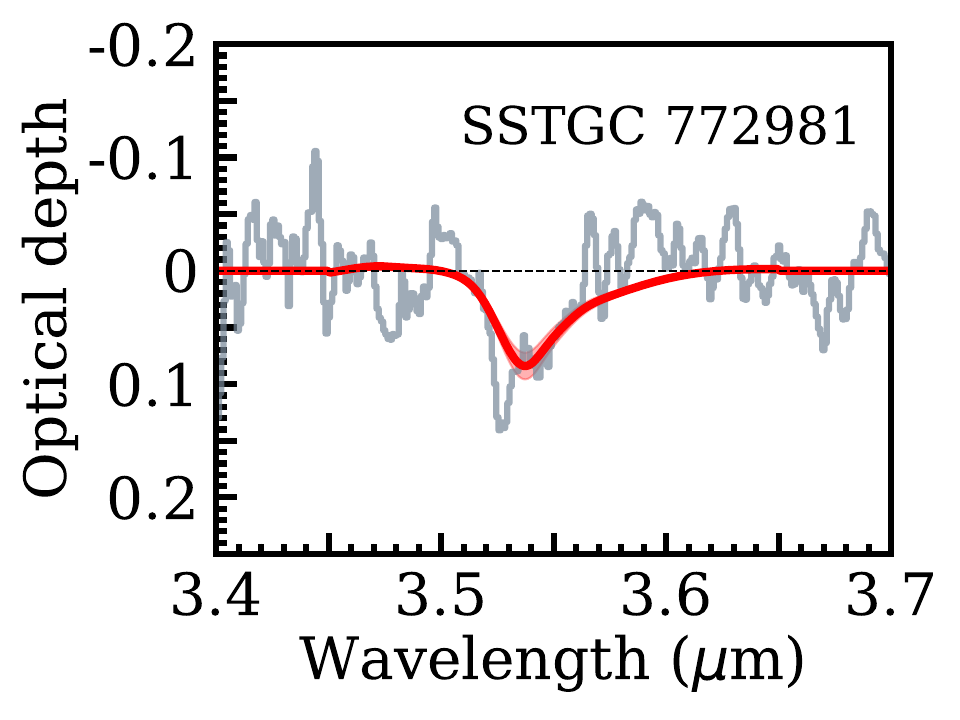}
\includegraphics[width=0.32\textwidth]{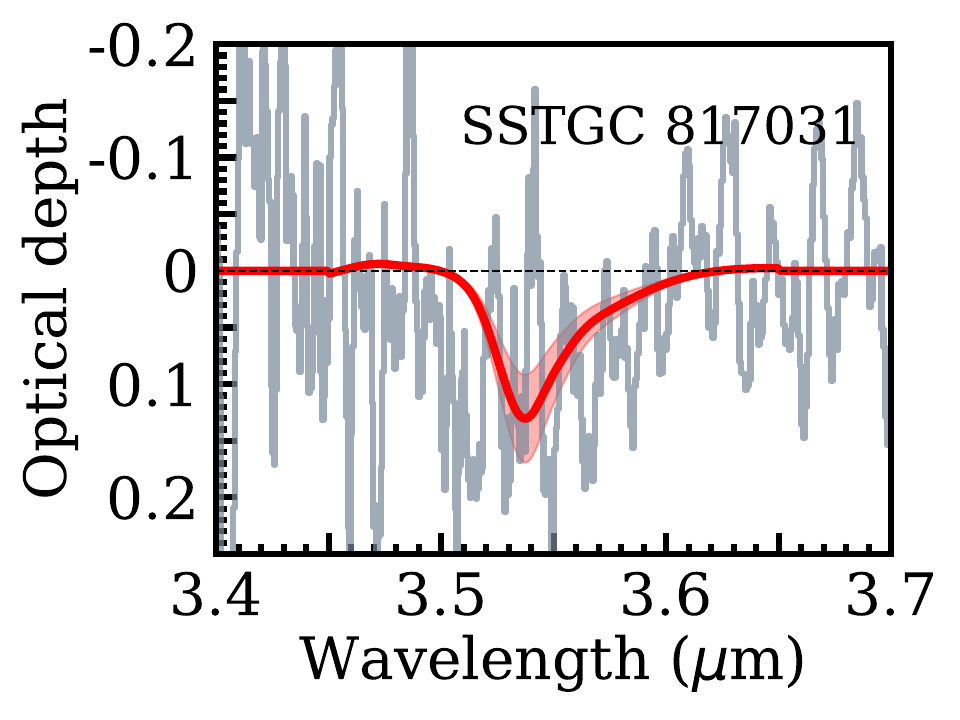}
\caption{Optical depth spectra of CMZ sources exhibiting $3.535\ \mu$m solid CH$_3$OH absorption at $>2.5\sigma$ significance. The red line shows the best-fit model for pure solid CH$_3$OH ice, and the shaded area denotes the $\pm1\sigma$ confidence interval. Spectra for the remaining sources are provided in Appendix~\ref{sec:ch3oh2}.}
\label{fig:ch3oh}
\end{figure*}

Figure~\ref{fig:ch3oh} shows the optical-depth spectra of objects with detections at $>2.5\sigma$ significance, including four with $>3\sigma$ confidence. Spectra of the remaining objects are presented in Appendix~\ref{sec:ch3oh2}. The red shading denotes the $\pm1\sigma$ range of the total uncertainty. Among these six objects, SSTGC~726327 and SSTGC~817031 have previously been reported to show CH$_3$OH \citep{jang:22}, while the $3.535\ \mu$m detections for the remaining four objects are new from this work. As is noted above for H$_2$O ice, the detection of CH$_3$OH ice is not correlated with the mid-IR classifications in \citet{an:11}: for example, the previously known stars (SSTGC~696367 and SSTGC~817031) exhibit moderate to significant detections.

\begin{figure}
\centering
\includegraphics[width = 0.44\textwidth]{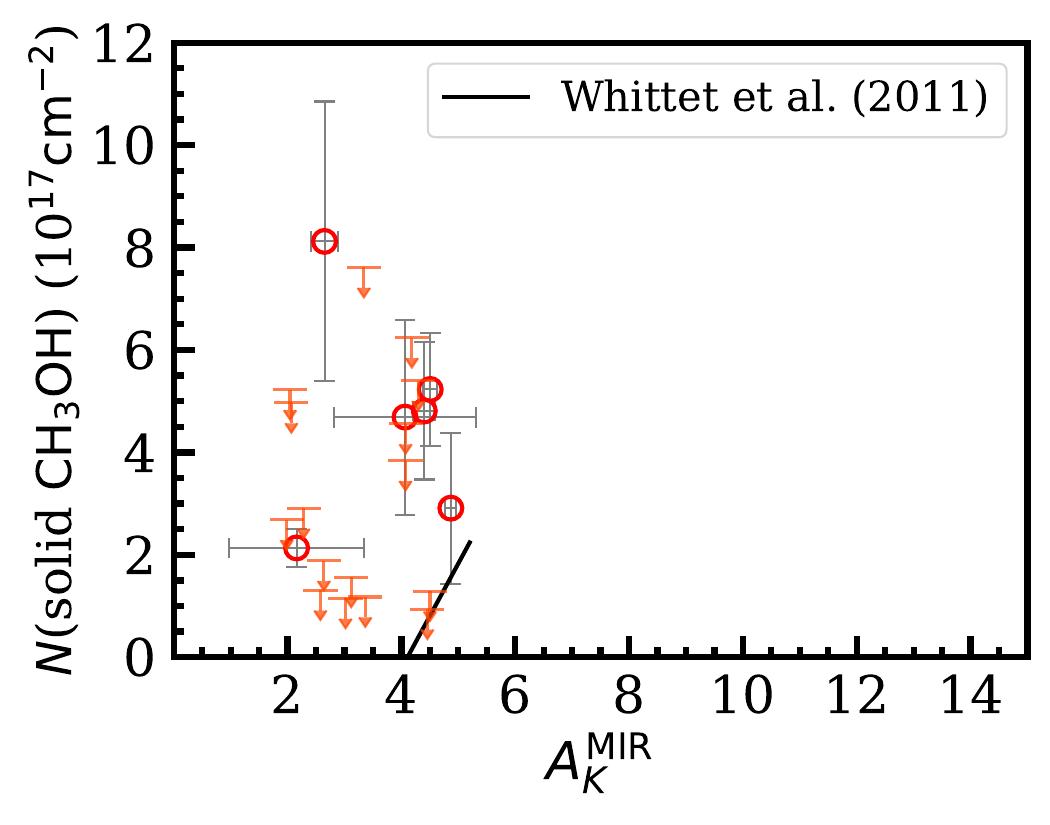}
\includegraphics[width = 0.44\textwidth]{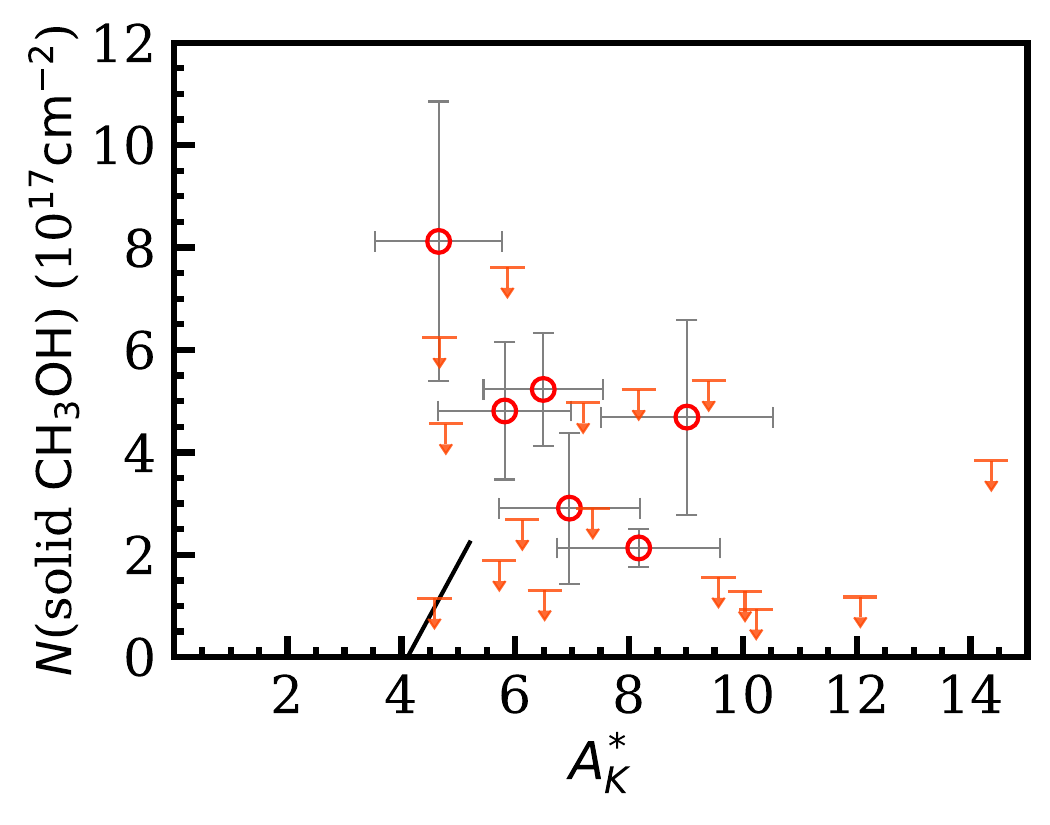}
\includegraphics[width = 0.44\textwidth]{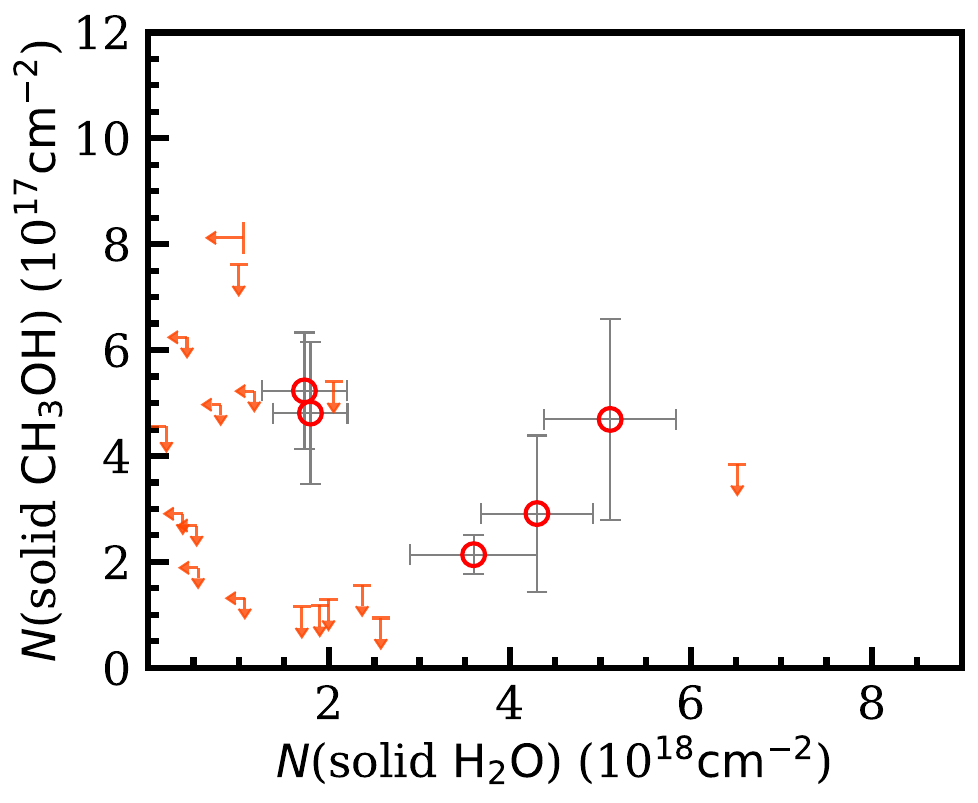}
\caption{Methanol-ice column densities against foreground extinction estimates ($\akmid$ and $\akstar$ in the top and middle panels) and water-ice column densities (bottom panel). The solid line in the top and middle panels shows the empirical relation for local clouds from \citet{whittet:11}, shifted horizontally by the average CMZ foreground extinction ($\Delta A_K = 2.2$~mag).}
\label{fig:akch3oh}
\end{figure}

Figure~\ref{fig:akch3oh} shows $\nmethanol$ plotted against our foreground extinction measurements: $\akmid$ in the top panel and $\akstar$ in the middle panel. As for water ice in Fig.~\ref{fig:akh2o}, $\nmethanol$ measurements show a large upward scatter above the local relation with respect to $\akmid$ (top panel). The scatter of $\nmethanol$ is more pronounced against $\akstar$ (middle panel). For reference, the local relation from \citet{whittet:11} is shown as a solid line, shifted by $A_K = 2.2$~mag to account for the extinction of the foreground diffuse clouds in front of the CMZ. Methanol ice tends to form in denser regions and deeper layers of clouds, where the extinction is generally higher than in the environments where H$_2$O ice forms. Thus, the extinction threshold for detecting CH$_3$OH ice ($A_K \approx 1.87 \pm 0.50$) is higher than that for water ice \citep[see][]{boogert:11}. However, the threshold extinction of $\nmethanol$ in the top panel is less apparent than that for water ice in Fig.~\ref{fig:akh2o}.

In the bottom panel of Fig.~\ref{fig:akch3oh}, $\nmethanol$ is shown against $\nwater$. The absence of a clear correlation reflects the significant scatter observed in the top and middle panels. Nevertheless, in local clouds, both $\nwater$ and $\nmethanol$ typically increase with extinction \citep{boogert:11}, suggesting higher $\nmethanol$ at higher $\nwater$. Therefore, the lack of a positive trend in our sample, particularly considering upper limits in these measurements, may indicate that the ice abundances have been modified by environmental effects, such as thermal processing within YSO envelopes. These possibilities and their implications for ice chemistry in the CMZ are explored in the following section.

\subsection{Mixing ratio of CH$_3$OH and CO$_2$}\label{sec:co2}

The primary feature distinguishing our sample objects from the other CMZ sources is the strong, broad $15\ \mu$m absorption from CO$_2$ ice. As described in Sect.~\ref{sec:intro}, CH$_3$OH-rich CO$_2$ ice grains are responsible for the long-wavelength shoulder of this absorption band. The greater shoulder absorption compared to the polar (H$_2$O-rich) and apolar (CO-rich) components of CO$_2$ ice is uniquely seen towards massive YSOs in the CMZ, suggesting that distinct chemical networks operate in CMZ star-forming clouds, producing unique chemical mixtures.

\begin{figure*}
\centering
\includegraphics[width=0.4\textwidth]{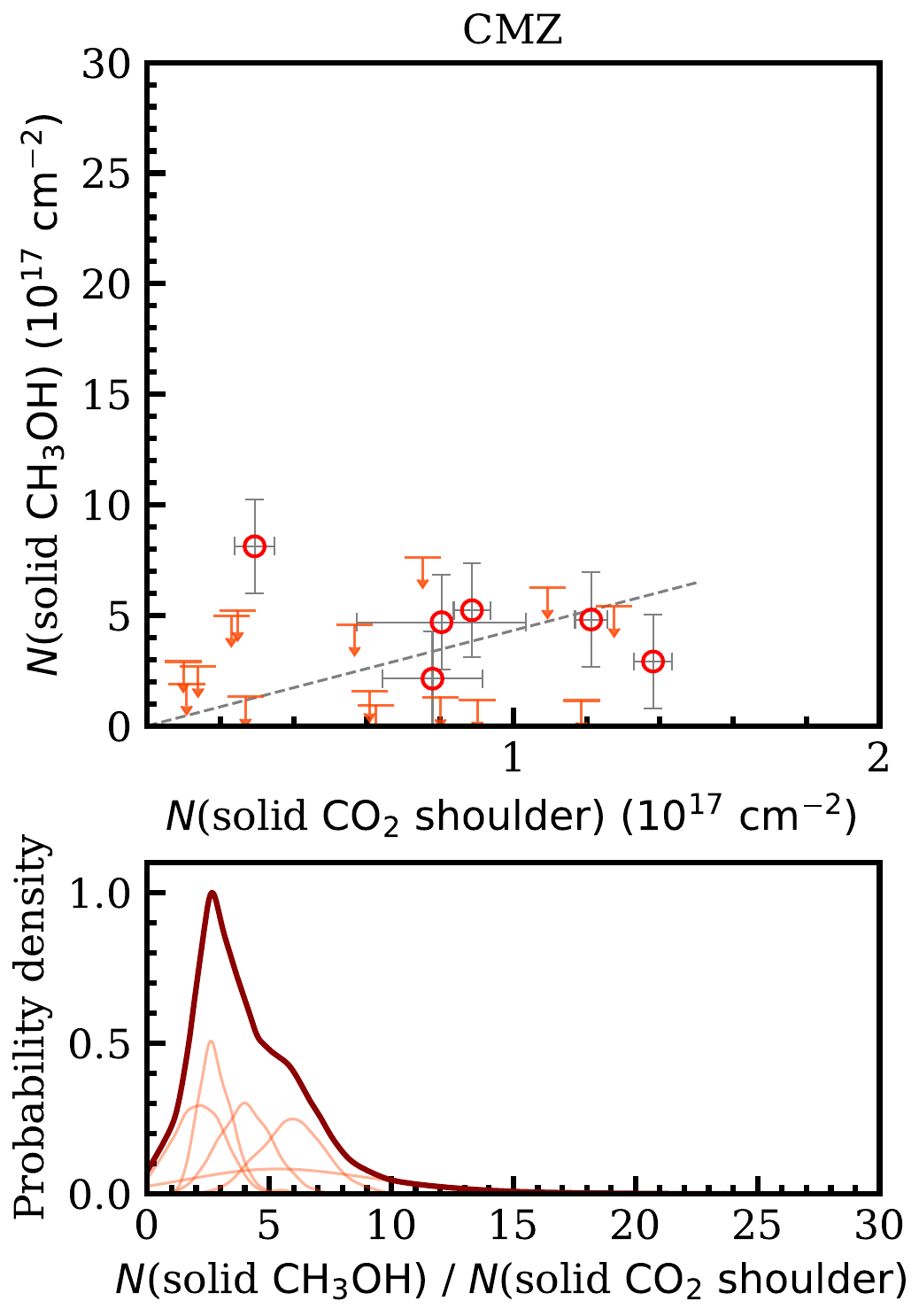}
\includegraphics[width=0.4\textwidth]{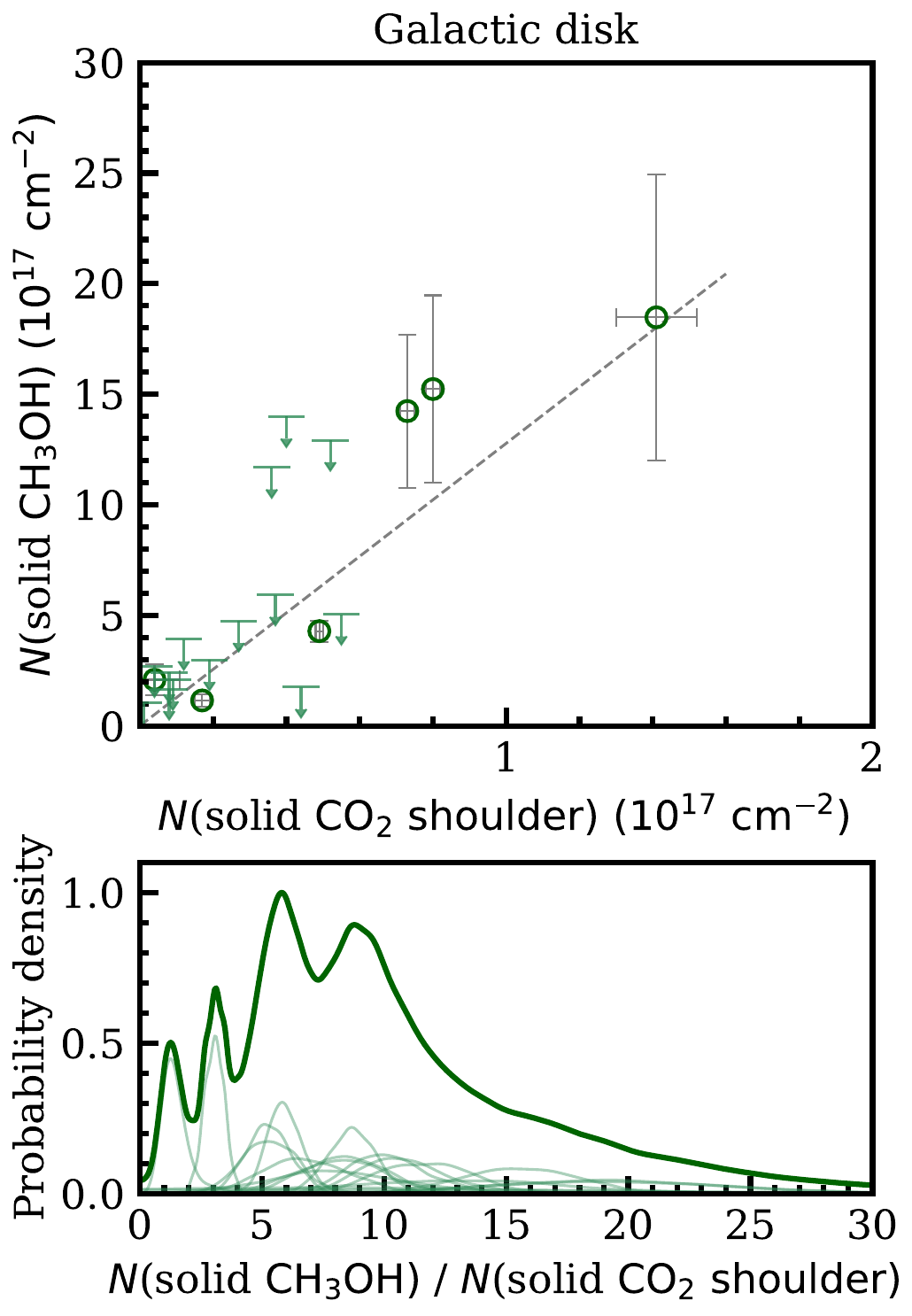}
\caption{Mixing ratios of CH$_3$OH ice to the shoulder component of CO$_2$ ice. Circles and arrows in the top panels indicate measurements and $3\sigma$ upper limits, respectively, for the red CMZ objects in this study (left) and for YSOs in the Galactic disk \citep[][right]{pontoppidan:08}. In the bottom two panels, probability density functions of the mixing ratio $\nmethanol$/$N$(CO$_2$ shoulder) are shown for the CMZ and disk YSOs, respectively. Thin lines correspond to individual measurements, assuming a bivariate normal distribution for the measured column densities and their $1\sigma$ uncertainties; the thick line shows the sum of all distributions.}
\label{fig:co2}
\end{figure*}

The upper left panel of Figure~\ref{fig:co2} shows $\nmethanol$ plotted against the CO$_2$ shoulder column densities from \citet{an:11} for all 23 objects in our CMZ sample, with red circles indicating detections and arrows marking $3\sigma$ upper limits. The ratio of these quantities represents the CH$_3$OH:CO$_2$ mixing ratio in ice grains, spanning $0.7$--$5.9$ for the six objects with useful $3.535\ \mu$m detections ($>2.5\sigma$). A linear fit to the CMZ detections (excluding upper limits) yields a slope of $\nmethanol / N$(CO$_2$ shoulder) $= 3.3 \pm 0.7$ (dashed line), consistent with $2.9 \pm 0.6$ reported by \citet{jang:22}. For comparison, the upper right panel shows measurements for YSOs in the Galactic disk \citep{boogert:08, pontoppidan:08}, which span a broader range ($1.3$--$52.2$) and have a mean ratio of $12.8 \pm 2.4$ (dashed line), significantly higher than the CMZ value. This contrast is also evident in the lower panels, where individual probability density functions (thin lines) are modelled as bivariate normal distributions using the measured column densities and their $1\sigma$ uncertainties; their sum (thick line) represents the overall distribution.

According to the LIDA laboratory experiment of CO$_2$ ice absorptions \citep{ehrenfreund:99}, even a relatively low abundance of CH$_3$OH in CH$_3$OH:CO$_2$ mixtures can produce the shoulder absorption of solid CO$_2$ centred at $15.4\ \mu$m at temperatures below $130$~K. The band strength extracted from the laboratory data is nearly the same as in more methanol-rich mixtures, such as those appropriate for the Galactic disk population. On the other hand, the shoulder component could also arise in mixtures with species other than CH$_3$OH. In that case, the presence of additional species would enhance the strength of the $15.4\ \mu$m shoulder, while $\nmethanol$ would remain unchanged because the $3.535\ \mu$m band originates primarily from CH$_3$OH. Thus, if other species that act as bases in Lewis acid--base interactions, such as ethanol or related molecules, contribute to the CO$_2$ shoulder \citep[e.g.,][]{dartois:99}, the inferred CH$_3$OH:CO$_2$ ratio would decrease, since part of the $15.4\ \mu$m absorption would not be due to CH$_3$OH. Such a scenario would make the CMZ objects distinct from those in the Galactic disk, but only if the CMZ uniquely hosts a large reservoir of these more complex molecules.

If the $15.4\ \mu$m shoulder arises entirely from CO$_2$ ice mixed with CH$_3$OH, the relatively low abundance ratios shown in Fig.~\ref{fig:co2} may point to a distinctive chemical environment in the CMZ. In particular, they could suggest that CH$_3$OH formation is less efficient relative to CO$_2$ in CMZ clouds, even though both species are produced through CO reactions involving hydrogenation or oxidation \citep[see][]{jang:22}. However, the CH$_3$OH:CO$_2$ mixing ratio in the CMZ displays substantial scatter. Among the six CMZ sources with significant detections, SSTGC~817031 shows the highest ratio ($\sim 27.7$), whereas SSTGC~719445 has the lowest ($\sim 2.1$). Including upper limits for the other 17 objects broadens the range even further, implying large variations across the sample. Such a wide spread in abundance ratios would, in principle, require unusually high chemical diversity among CMZ clouds.

Instead, we argue in the next section that the observed scatter primarily reflects variations in the lines of sight, with each spectrum probing different regions of the YSO envelope. This geometry offers a rare opportunity to investigate the internal chemical structure of CMZ YSOs: CO$_2$ absorption is measured from mid-IR spectra, whereas $\nmethanol$ is derived from near-IR data. By contrast, the consistently higher abundance ratios found for YSOs in the Galactic disk may stem from tracing the same mid-IR sightlines. In that case, $\nmethanol$ estimates were based on the $9.7\ \mu$m C--O stretching mode \citep{skinner:92} in the same mid-IR spectra \citep{pontoppidan:08}. We revisited this feature in our own data in Appendix~\ref{sec:97}, but the CMZ sources are considerably fainter, leading to much lower S/N at the bottom of the deep $9.7\ \mu$m silicate bands.

\section{Ensemble average of CMZ YSOs}\label{sec:ensemble}

\subsection{Reduced foreground extinction}\label{sec:reduced}

Our CMZ objects appear as point-like sources in Spitzer/IRAC mid-IR images (FWHM $\sim 2\arcsec$). At a distance of $8.3$ kpc to the Galactic centre \citep{reid:14}, this corresponds to $\lesssim 8300$~au in projection, implying that the near-IR sightlines towards the background giant through the YSO envelope are unresolved within this scale and can be regarded as the near-IR counterparts of the mid-IR sources. Consequently, the projected separation between each near-IR sightline and the YSO centre cannot be directly measured. As an alternative, foreground extinction can serve as a proxy for projected separation, as the integrated dust column density within the YSO envelope generally increases towards its centre. However, the complex nature of foreground extinction fields of the CMZ --- arising from patchy extinction in intervening Galactic disk clouds, additional obscuration by CMZ clouds, and dust within the YSO envelope --- means that such an approach requires a careful understanding of the foreground dust properties.

\begin{table}
\centering
\caption{Reduced foreground extinction}
\label{tab:sed2}
\begin{tabular}{cccc}
\hline\hline
Object ID & $\langle A_K^{\mathrm{fg}} \rangle$ & $\akred$ & $\akmid - \langle A_K^{\mathrm{fg}} \rangle$ \\
(SSTGC) & (mag) & (mag) & (mag) \\
\hline
\multicolumn{4}{c}{Gemini/GNIRS} \\
\hline
372630 & $2.9 \pm 0.1$ & 6.6 $\pm$ 1.4 & 1.5 $\pm$ 0.1 \\
425399 & $2.3 \pm 0.3$ & 4.2 $\pm$ 1.3 & 0.2 $\pm$ 0.8 \\
524665 & $0.9 \pm 0.3$ & 12.4 $\pm$ 1.8 & 2.2 $\pm$ 0.3\\
563780 & $3.8 \pm 0.4$ & 1.0 $\pm$ 1.0 & 0.3 $\pm$ 0.4\\ 
619522 & $3.7 \pm 0.6$ & 0.9 $\pm$ 1.1 & -0.7 $\pm$ 0.6\\
619964 & $3.2 \pm 1.0$ & 7.1 $\pm$ 1.9 & 1.3 $\pm$ 1.3\\
653270 & $4.5 \pm 3.2$ & 3.7 $\pm$ 3.5 & -2.4 $\pm$ 3.4\\
679036 & $4.7 \pm 0.5$ & 5.4 $\pm$ 1.5 & -0.2 $\pm$ 0.6\\
696367 & $2.6 \pm 0.3$ & 6.4 $\pm$ 1.6 & 1.4 $\pm$ 1.3\\
718757 & $4.0 \pm 0.5$ & 8.0 $\pm$ 1.8 & -0.7 $\pm$ 0.5\\
719445 & $3.7 \pm 1.2$ & 3.2 $\pm$ 1.7 & 1.1 $\pm$ 1.2\\
728480 & $2.9 \pm 0.8$ & 1.8 $\pm$ 1.2 & 1.3 $\pm$ 0.8\\
770393 & $5.2 \pm 0.1$ & 4.4 $\pm$ 1.4 & -2.1 $\pm$ 0.2\\
772981 & $4.6 \pm 0.5$ & 1.9 $\pm$ 1.2 & -0.1 $\pm$ 0.5\\
799887 & $2.3 \pm 0.7$ & 12.0 $\pm$ 1.9 & 1.8 $\pm$ 1.0\\
\hline
\multicolumn{4}{c}{IRTF/SpeX} \\
\hline
300758 & $3.3 \pm 0.5$ & 2.6 $\pm$ 1.2 & 0.0 $\pm$ 0.6\\
348392 & $3.6 \pm 0.5$ & 3.6 $\pm$ 1.5 & -1.6 $\pm$ 0.5\\
388790 & $2.4 \pm 0.3$ & 3.3 $\pm$ 1.2 &  0.3 $\pm$ 0.3\\
404312 & $3.2 \pm 0.2$ & 4.2 $\pm$ 1.3 &  -0.9 $\pm$ 0.2\\
405235 & $2.8 \pm 0.1$ & 5.3 $\pm$ 1.4 &  -0.8 $\pm$ 0.1\\
716531 & $2.9 \pm 1.7$ & 3.2 $\pm$ 2.1 &  -0.9 $\pm$ 1.7\\
726327 & $3.8 \pm 0.6$ & 2.0 $\pm$ 1.3 &  0.6 $\pm$ 0.7\\
817031 & $2.8 \pm 0.2$ & 1.9 $\pm$ 1.1 &  -0.1 $\pm$ 0.3\\
\hline
\end{tabular}
\tablefoot{$\akred$ is the reduced extinction, obtained by subtracting the foreground component ($\langle A_K^{\mathrm{fg}} \rangle$) from $\akstar$, and thus represents the extinction through the YSO envelope itself.}
\end{table}

In order to isolate the extinction contributed by a YSO envelope from the rest of the foreground extinction, we defined a reduced extinction as
\begin{equation}
\akred \equiv \akstar - \langle A_K^{\mathrm{fg}} \rangle
\end{equation}
where $\langle A_K^{\mathrm{fg}} \rangle$ is the mean foreground extinction from dust lying between the Sun and the YSO's outer surface. Since $\akstar$ measures the total $A_K$ from the Sun to the background giant, subtracting $\langle A_K^{\mathrm{fg}} \rangle$ yields a reduced $A_K$ that more closely represents the extinction through the YSO envelope alone. For $A_K^{\mathrm{fg}}$, we adopted the foreground extinction estimates from \citet{an:13}, who measured the depth of the $9.7\ \mu$m silicate feature in spectra of the diffuse ISM, deliberately selected to exclude point sources. For most of our sample, such measurements are available at four orthogonal positions, each located approximately $1\arcmin$ (equivalent to $2.4$ pc in projection) from the target. We averaged the $A_K$ values converted from the silicate depths and incorporated their standard deviation into the total $\akred$ uncertainty budget. These results are listed in the second and third columns of Table~\ref{tab:sed2}.

Alternatively, the fourth column of Table~\ref{tab:sed2} shows the differential extinction, obtained by subtracting the foreground extinction (second column) from the mid-IR extinction $\akmid$. Although $\akmid$ does not trace the column density towards a background giant but rather probes the line of sight into the YSO itself, one might expect that subtracting the foreground ISM extinction from the total mid-IR extinction would yield a positive residual, corresponding to the local envelope material. The presence of prominent mid-IR absorption features, such as the $15\ \mu$m CO$_2$ band, supports this expectation. However, even in sources that exhibit significant CH$_3$OH ice absorption, the residual extinction listed in the fourth column often appears consistent with zero or even slightly negative values. This result does not contradict the observed mid-IR absorption but rather reflects the limitations imposed by the measurement uncertainties.

\begin{figure}
\centering 
\includegraphics[width = 0.46\textwidth]{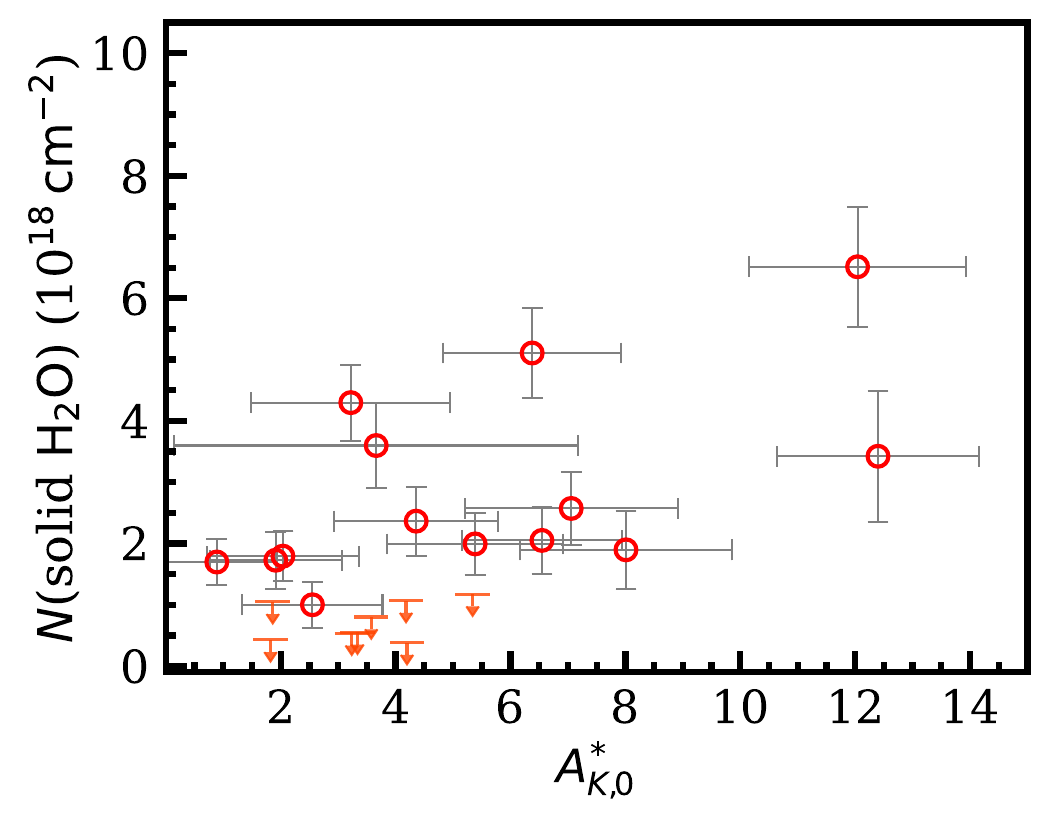}
\includegraphics[width = 0.46\textwidth]{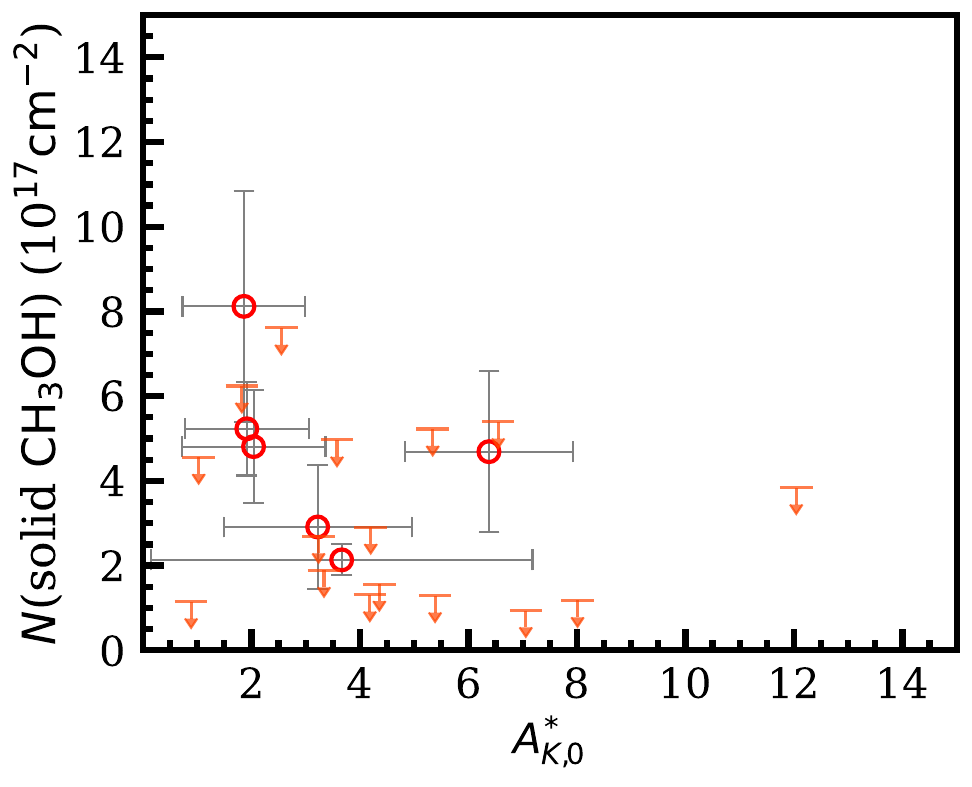}
\caption{Column densities of H$_2$O ice (top) and CH$_3$OH ice (bottom) as a function of reduced extinction, $\akred$. Since $\akred$ traces the dust column integrated along the line of sight within a YSO envelope, it serves as a proxy for projected distance from the centre of the system, with higher $\akred$ corresponding to smaller impact parameters.}
\label{fig:akred}
\end{figure}

Figure~\ref{fig:akred} shows the column densities of H$_2$O and CH$_3$OH ices as a function of $\akred$. As in the case of $\akstar$ (Fig.~\ref{fig:akh2o}), $\nwater$ increases with extinction (Pearson's correlation coefficient of $r = 0.565$), but the zero-point shift in $\akred$ lowers the minimum extinction to $\sim$1~mag. The decline of $\nmethanol$ with $\akred$ closely parallels that with $\akstar$ (Fig.~\ref{fig:akch3oh}), although the trend is less significant than in the H$_2$O case ($r = -0.367$). Interpreting $\akred$ as a proxy for projected distance from the source centre, these trends suggest that $\nwater$ rises and $\nmethanol$ falls towards the source centre. The linear fits yield dispersions of $\Delta N = 1.2\, {\rm cm}^{-2}$ and $1.8\, {\rm cm}^{-2}$ in the top and bottom panels, respectively, close to the expected uncertainties of $1.0\, {\rm cm}^{-2}$ and $1.4\, {\rm cm}^{-2}$. The small excess may reflect intrinsic variations in ice abundances due to differing temperature structures among YSOs of varying masses or evolutionary stages.

Nonetheless, $\akred$ extends from $\sim1$~mag to nearly $12$~mag, indicating a wide range of dust column densities along the various lines of sight in our sample. To assess whether dust extinction can be quantitatively linked to the projected distance from the source centre, we constructed a spherically symmetric gas model representing the envelope in the early phase of YSO evolution. Assuming a density power-law index of $1.5$ \citep[e.g.][]{jorgensen:02} and a total mass of $10$--$20\ M_\odot$ for our CMZ YSOs, as inferred from multi-wavelength SED fitting \citep{an:11}, and adopting the gas-to-dust ratio from \citet{bohlin:78}, we estimated that the dust extinction through the envelope is $4 \lesssim A_K \lesssim 8$~mag at $1\arcsec$ from the centre. At $0.5\arcsec$, the extinction increases to $6 \lesssim A_K \lesssim 11$~mag. This range of $A_K$ is broadly consistent with $\akred$ in Fig.~\ref{fig:akred}, supporting the view that our observed near-IR sightlines probe different regions of the extended envelopes of the CMZ YSOs.

\subsection{Average chemical structure of YSOs}

\begin{figure}
\centering
\includegraphics[width =0.46\textwidth]{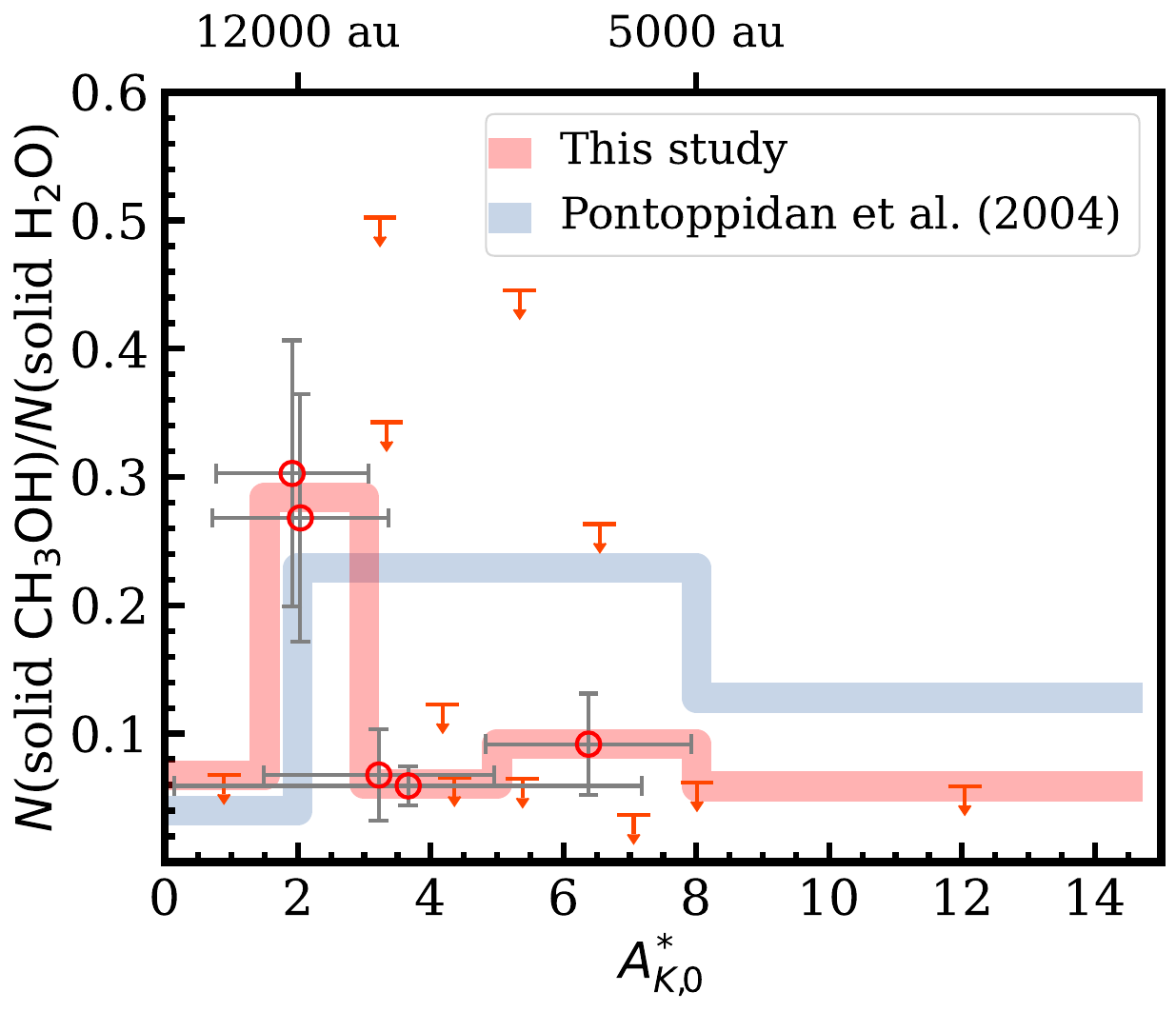}
\includegraphics[width =0.46\textwidth]{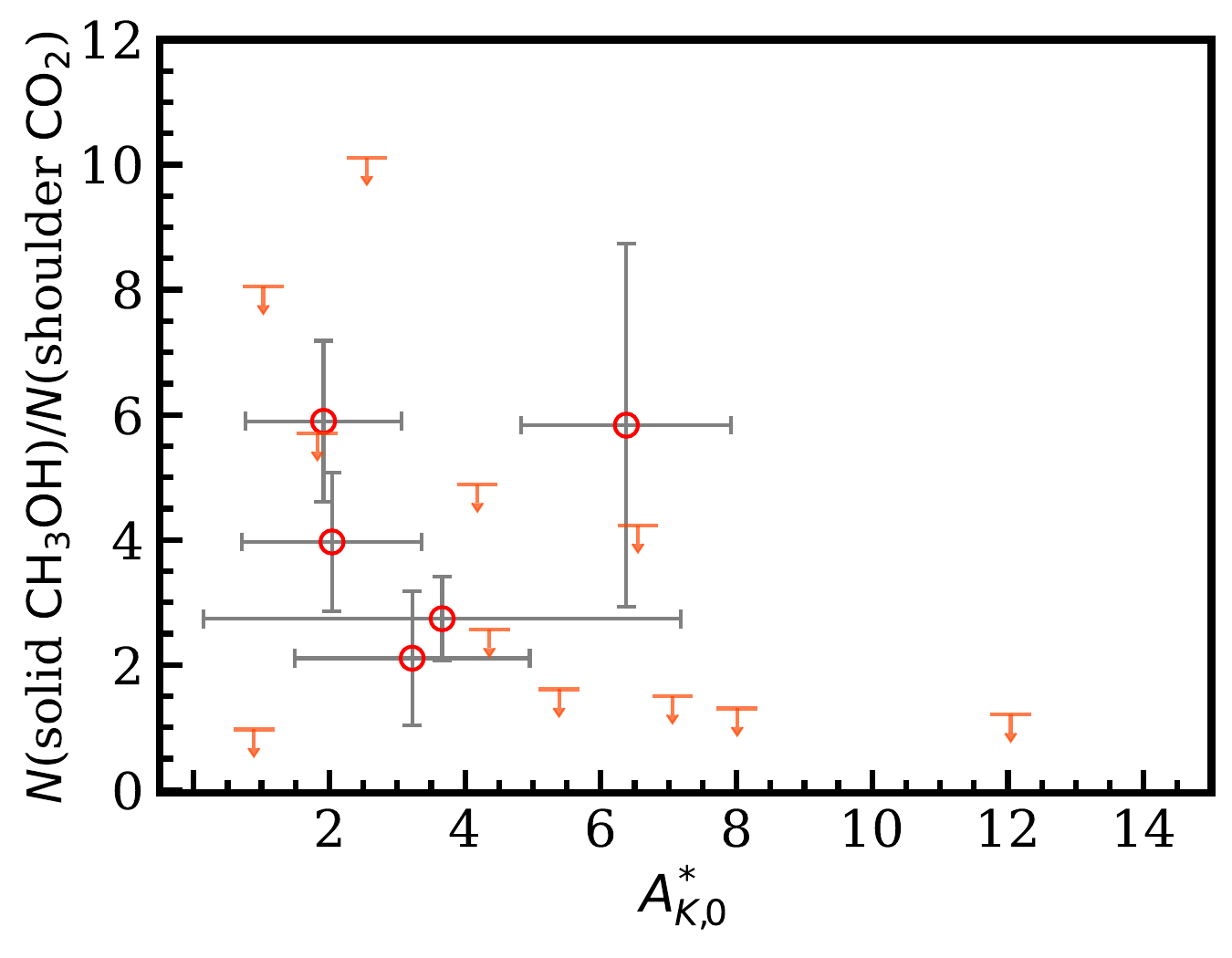}
\caption{Relative abundance of CH$_3$OH ice in the CMZ sources, shown with respect to H$_2$O ice (top panel) and to the CO$_2$ shoulder component (bottom panel), as a function of reduced extinction $\akred$. The fractional abundance is defined as the column density of CH$_3$OH ice divided by the sum of $\nmethanol$ and either H$_2$O or CO$_2$ shoulder column densities, respectively. Circles show our measurements, and arrows mark $3\sigma$ upper limits. Only sources with significant detections ($>2.5\sigma$) of $3\ \mu$m H$_2$O ice are included. In the top panel, the thick red line shows the weighted mean abundance in four adaptive intervals, while the blue thick line shows the abundance measured in the envelope of SMM~4, a low-mass YSO in the Serpens cloud \citep{pontoppidan:04}. The approximate length scale shown at the top corresponds to the projected distance from the centre of SMM~4.}
\label{fig:ensemble}
\end{figure}

The top panel of Figure~\ref{fig:ensemble} shows the fractional abundance of CH$_3$OH ice, defined as $\nmethanol$ divided by $\nwater$, for sources in our sample. The ratios are shown as a function of reduced extinction $\akred$, which serves here as a proxy for the projected distance from the source centre. The ratio peaks at $\sim0.3$ near $\akred \approx 2$~mag, but generally stays below $0.1$ for $\akred > 3$~mag. To highlight the overall trend, the thick red line represents the weighted mean abundance ratio in four adaptive extinction intervals, including the single upper limit at $\akred \sim 1$~mag. When viewed as an ensemble, the CMZ objects show a trend in which a typical CMZ YSO exhibits a sharp rise in methanol ice abundance in the outer envelope, while the abundance remains low ($<0.1$) in the inner region.

For comparison, the thick blue line in the top panel of Fig.~\ref{fig:ensemble} shows the methanol abundance profile towards SMM~4, a nearby, well-characterized low-mass Class~0 YSO projected against the young cluster SVS~4 \citep{pontoppidan:04}. To construct this profile, we used their ice abundance measurements, which include H$_2$O column densities along ten sightlines towards background stars spanning projected distances of $\sim$4500--19,000~au from the centre of SMM~4. The H$_2$O abundance relative to H$_2$ was found to be nearly constant across most of this range, except for a single data point at $\sim$4700~AU that showed a significantly higher abundance. CH$_3$OH ice was detected in four sightlines within $\sim$12,000~au, and the inferred abundance remained nearly constant across these positions, with a strong upper limit placed at the outermost radius ($\sim$19,000~au). In constructing the CH$_3$OH/H$_2$O ratio profile, we adopted their abundance values and assumed a constant CH$_3$OH abundance inward of the innermost detection, based on their best-fitting envelope model. This results in a broad peak in the CH$_3$OH/H$_2$O ratio at intermediate radii, caused by the combination of flat CH$_3$OH abundance and a centrally increasing H$_2$O abundance.

As in SMM~4, the CH$_3$OH abundance of the CMZ ensemble exhibits a peak at intermediate radii within the envelope. In the cold and less dense outer envelope, methanol formation is suppressed due to the reduced efficiency of CO hydrogenation, likely caused by the limited mobility and availability of atomic hydrogen at low temperatures \citep[e.g.][]{watanabe:03}. While water ice can still accumulate in these conditions, methanol production becomes less efficient, leading to a declining $\nmethanol/\nwater$ ratio with increasing distance from the source centre. 

The peak CH$_3$OH abundance in the CMZ ensemble ($\approx 0.3$) is comparable to that observed towards SMM~4, suggesting that methanol formation via CO hydrogenation on grain surfaces may impose a common upper limit on CH$_3$OH ice production in star-forming cores. However, the abundance peak in the CMZ sources spans a narrower $A_K$ range than in SMM~4, potentially reflecting their higher total masses ($\sim$10--20\,$M_\odot$; \citealt{an:11}) compared to $\sim$5.8\,$M_\odot$ for SMM~4 \citep{pontoppidan:04}. The shorter evolutionary timescales of high-mass YSOs likely result in hotter internal structures, which in turn enhance CH$_3$OH ice sublimation in the inner envelope. As a consequence, methanol ice is efficiently thermally desorbed near the protostar, producing a characteristic abundance peak at intermediate depths.

This interpretation can further explain the observed low $\nmethanol$/$N(\mathrm{CO}_2\ \mathrm{shoulder})$ values of CMZ YSOs (Fig.~\ref{fig:co2}). The bottom panel of Fig.~\ref{fig:ensemble} shows the trend of $\nmethanol$ relative to the CO$_2$ shoulder component, with the ratio peaking at lower $\akred$ (i.e.\ in the outer envelope regions) and gradually decreasing towards higher $\akred$ (i.e.\ the warmer, inner regions of the YSOs). The declining ratio with increasing $\akred$ suggests that CH$_3$OH is more efficiently depleted than CO$_2$ from the ice mantles in the inner envelopes. This trend is somewhat counter-intuitive, given the higher sublimation temperature of CH$_3$OH compared to CO$_2$. However, CH$_3$OH may be efficiently destroyed or chemically transformed into more complex organic species as the ices are subjected to UV irradiation followed by thermal processing. UV photons dissociate CH$_3$OH into radicals, which then become mobile and recombine during warm-up to form complex molecules. This combined processing reduces the CH$_3$OH abundance relative to CO$_2$, which is more photo-stable and less reactive under similar conditions \citep[e.g.][]{gerakines:96, oberg:09}. In addition, desorption driven by increasing gas temperature may remove both CO$_2$ and CH$_3$OH; however, CH$_3$OH may appear preferentially depleted, as it is a minor component in the ice mixture. In contrast, CO$_2$ is more abundant and remains detectable through its prominent $15.4\ \mu$m shoulder feature.

Previously, \citet{jang:22} proposed that the reduced CH$_3$OH abundance could result from elevated oxygen abundance in the CMZ, which favours oxidation of CO over hydrogenation, thereby limiting CH$_3$OH production. However, the present result suggests that the lower methanol ice abundance may not directly arise from suppressed CH$_3$OH formation pathways. Instead, it may reflect a sample bias towards higher-mass YSOs, in which methanol ice has been more significantly depleted.

\section{Summary}\label{sec:summary}

We have presented a spectroscopic study of 23 extremely red, point-like CMZ sources, combining new $L$-band data from Gemini/GNIRS (15 objects) with previously published IRTF/SpeX $L$-band data (8 objects), and supplementing these with archival \textit{Spitzer}/IRS spectra to trace ice chemistry and foreground attenuation from $2$ to $35~\mu$m. Many targets show $K$-band CO band heads consistent with (super-)giant photospheres, while their mid-IR continua and strong ice bands are characteristic of embedded YSOs \citep{an:17, jang:22}. This supports a backlighting geometry in which a background giant star illuminates a foreground YSO envelope (or dense core), enabling us to probe the envelope material through absorption features. We modelled each composite SED and derived both near- and mid-IR extinction estimates, along with robust ice column densities.

Water ice absorption is strong and ubiquitous among our sources, with $\nwater \sim (0.1$--$6.5) \times 10^{19}\,\mathrm{cm}^{-2}$. The mid-IR extinction ($\akmid$) is typically $2$--$5$~mag, whereas the extinction towards the background star ($\akstar$) spans $4$--$15$\,mag, with $\akstar$ systematically exceeding $\akmid$ at fixed $\nwater$. This offset is consistent with the background star being behind additional envelope material. Solid CH$_3$OH is detected at $>2.5\sigma$ towards six objects; in the remainder we report meaningful upper limits.

To compare $\nmethanol$ measured in the near-IR with the CO$_2$:CH$_3$OH mixtures traced in the mid-IR \citep{an:09, an:11}, we revisited the $15.4~\mu$m CO$_2$ ``shoulder'' component. In the CMZ, the bulk ratio $N(\mathrm{CH_3OH})/N(\mathrm{CO}_2\ \mathrm{shoulder})$ is $\sim$2–5\%, with a mean of $3.3\pm0.7\%$, confirming earlier findings \citep{jang:22} that it is systematically lower than typical Galactic disk values (5–15\%). This indicates either intrinsically reduced CH$_3$OH production or preferential depletion/processing of CH$_3$OH in CMZ envelopes. Exploiting the backlighting geometry, we introduced a reduced extinction ($\akred$) as a proxy for projected distance through an envelope. In this ensemble framework, $N(\mathrm{CO}_2\ \mathrm{shoulder})$ increases with $\akred$, while $\nmethanol$ shows a declining trend; the ice abundance ratio $\mathrm{CH_3OH/CO_2}$ is $\sim$10\% in inner lines of sight (large $\akred$) and rises steeply to $\sim$30\% in the outer envelope (small $\akred$). As our CMZ sample is likely biased towards more massive YSOs, the trends suggest that internal heating preferentially sublimates and/or chemically alters CH$_3$OH relative to CO$_2$ in the inner envelope, lowering the line-of-sight CH$_3$OH/CO$_2$ while leaving high methanol abundances in the outer regions. This thermal/UV processing scenario offers an alternative to purely chemical-network explanations (e.g.\ distinct elemental ratios in the ISM) for the CMZ's low integrated CH$_3$OH/CO$_2$ ratios.

Our results demonstrate that (i) red CMZ point sources often comprise superposed near- and mid-IR components; (ii) joint modelling of near- and mid-IR spectra enables an estimate of the foreground extinction towards the backlit giant and thus the line-of-sight integrated column through the YSO envelope; and (iii) backlit spectroscopy can reveal ensemble-averaged radial abundance trends that would otherwise be inaccessible at the distance of the CMZ. Our conclusions are tempered by systematic uncertainties in cross-instrument flux scaling and by crowding/background-subtraction effects; moreover, limited S/N at the base of the $9.7\ \mu$m silicate band precludes robust use of the CH$_3$OH C--O stretching mode. High-S/N, higher-resolution mid-IR spectroscopy and spatially resolved mapping (e.g.\ with JWST) of additional backlit sightlines will enable tests of the inferred abundance gradients and quantify the roles of sublimation, photochemistry, and mixing in setting the distinct ice chemistry of CMZ YSOs.

\begin{acknowledgements}

We thank the anonymous referee for constructive comments that helped improve the clarity and presentation of this manuscript.

D.A.\ thanks Kris Sellgren and Solange Ram\'{\i}rez for their advice and support in the preparation of the Gemini observations. D.A.\ acknowledges the hospitality and support provided by the staff at the Gemini Observatory's Hilo office during a research visit in 2017, which contributed to the development of this work. D.A.\ also thanks Minjin Kim for his assistance with the data acquisition during that visit. Y.K.\ and D.A.\ are grateful to Young-Jun Kim for discussions on YSO modelling. We acknowledge support provided by the National Research Foundation (NRF) of Korea grant funded by the Ministry of Science and ICT (No.\ 2021R1A2C1004117).\\

Based on observations obtained at the international Gemini Observatory, a program of NSF NOIRLab, which is managed by the Association of Universities for Research in Astronomy (AURA) under a cooperative agreement with the U.S. National Science Foundation on behalf of the Gemini Observatory partnership: the U.S. National Science Foundation (United States), National Research Council (Canada), Agencia Nacional de Investigaci\'{o}n y Desarrollo (Chile), Ministerio de Ciencia, Tecnolog\'{i}a e Innovaci\'{o}n (Argentina), Minist\'{e}rio da Ci\^{e}ncia, Tecnologia, Inova\c{c}\~{o}es e Comunica\c{c}\~{o}es (Brazil), and Korea Astronomy and Space Science Institute (Republic of Korea). This work was enabled by observations made from the Gemini North telescope, located within the Maunakea Science Reserve and adjacent to the summit of Maunakea. We are grateful for the privilege of observing the Universe from a place that is unique in both its astronomical quality and its cultural significance.\\

NOIRLab IRAF is distributed by the Community Science and Data Center at NSF NOIRLab, which is managed by the Association of Universities for Research in Astronomy (AURA) under a cooperative agreement with the U.S.\ National Science Foundation.

\end{acknowledgements}

\begin{appendix}
\onecolumn

\section{Additional spectra of the sample}\label{sec:sed2}

Figure~\ref{fig:sed2} presents the full set of observed composite spectra for all sources in our sample. Each panel follows the same format as the top panel of Fig.~\ref{fig:sed}, which illustrates the representative case of SSTGC~799887. The observed data are shown in red, and the coloured curves represent the best-fitting model components as described in Sect.~\ref{sec:sed}, including the stellar continuum, the foreground extinction, and thermal dust emission, as well as major molecular ice absorption features. To avoid strong molecular features and ensure reliable continuum estimation, the model fitting was performed over the following wavelength intervals: 1.90--3.30\ $\mu$m, 3.65--5.5\ $\mu$m, 3.65--6.5\ $\mu$m, 7.8--9.3\ $\mu$m, 10.2--12.4\ $\mu$m, 14.0--14.9\ $\mu$m, 15.8--16.7\ $\mu$m, 17.8--18.4\ $\mu$m, 19.0--27.5\ $\mu$m, and 28.8--30.0\ $\mu$m. These plots are provided to illustrate the consistency and diversity of spectral shapes across the sample and to complement the parameter values listed in Table~\ref{tab:sed2}.

\begin{figure*}[h]
\centering
\includegraphics[width=0.45\textwidth]{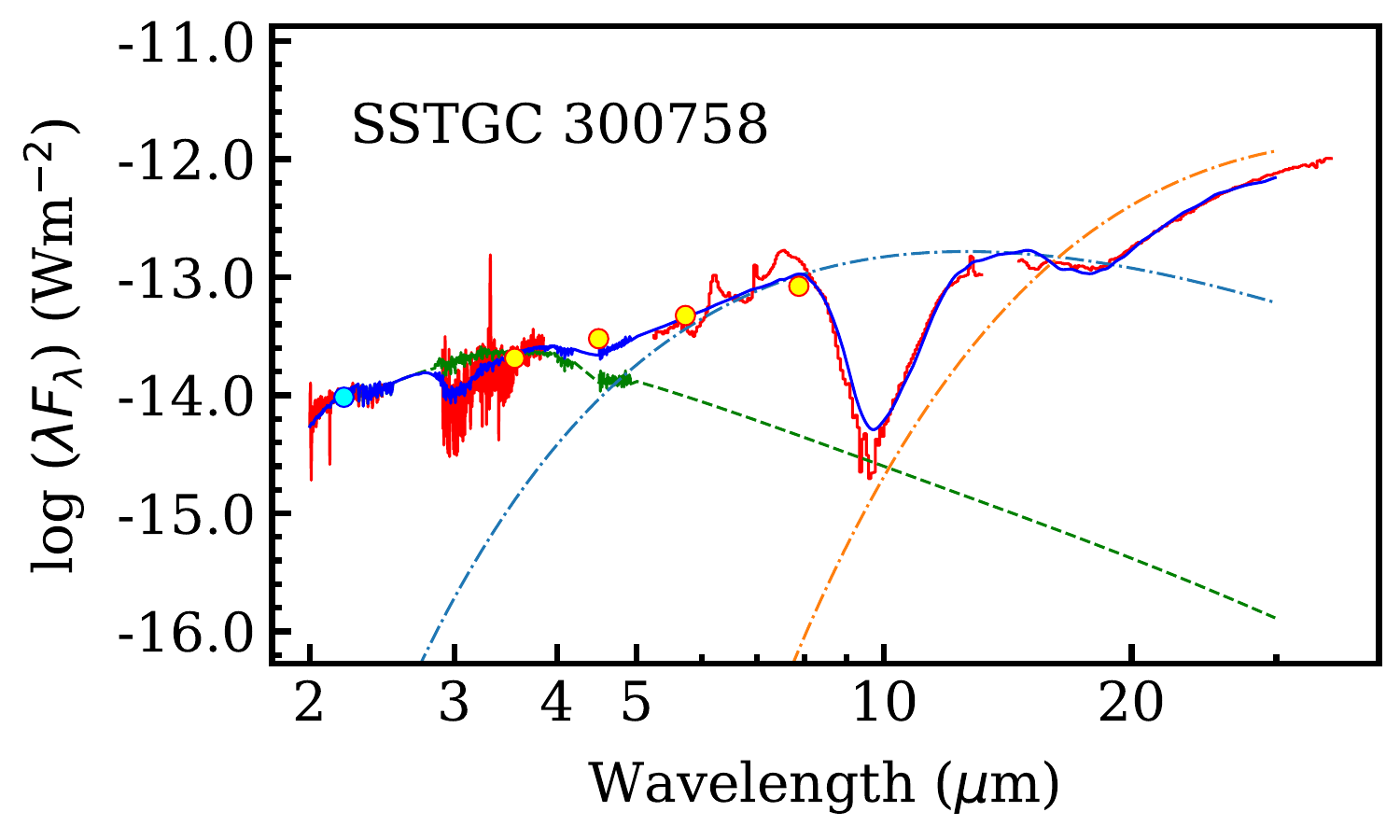}
\includegraphics[width=0.45\textwidth]{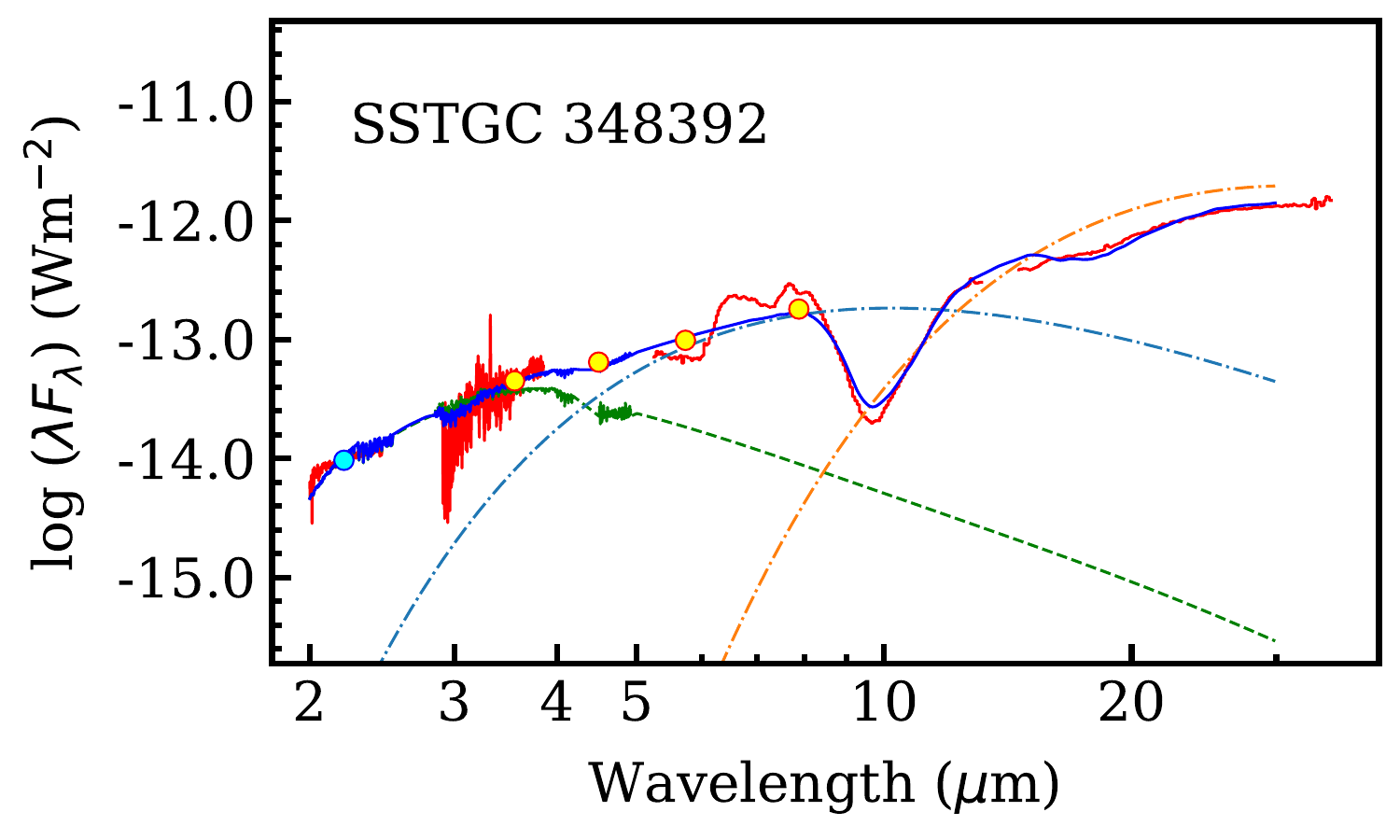}
\includegraphics[width=0.45\textwidth]{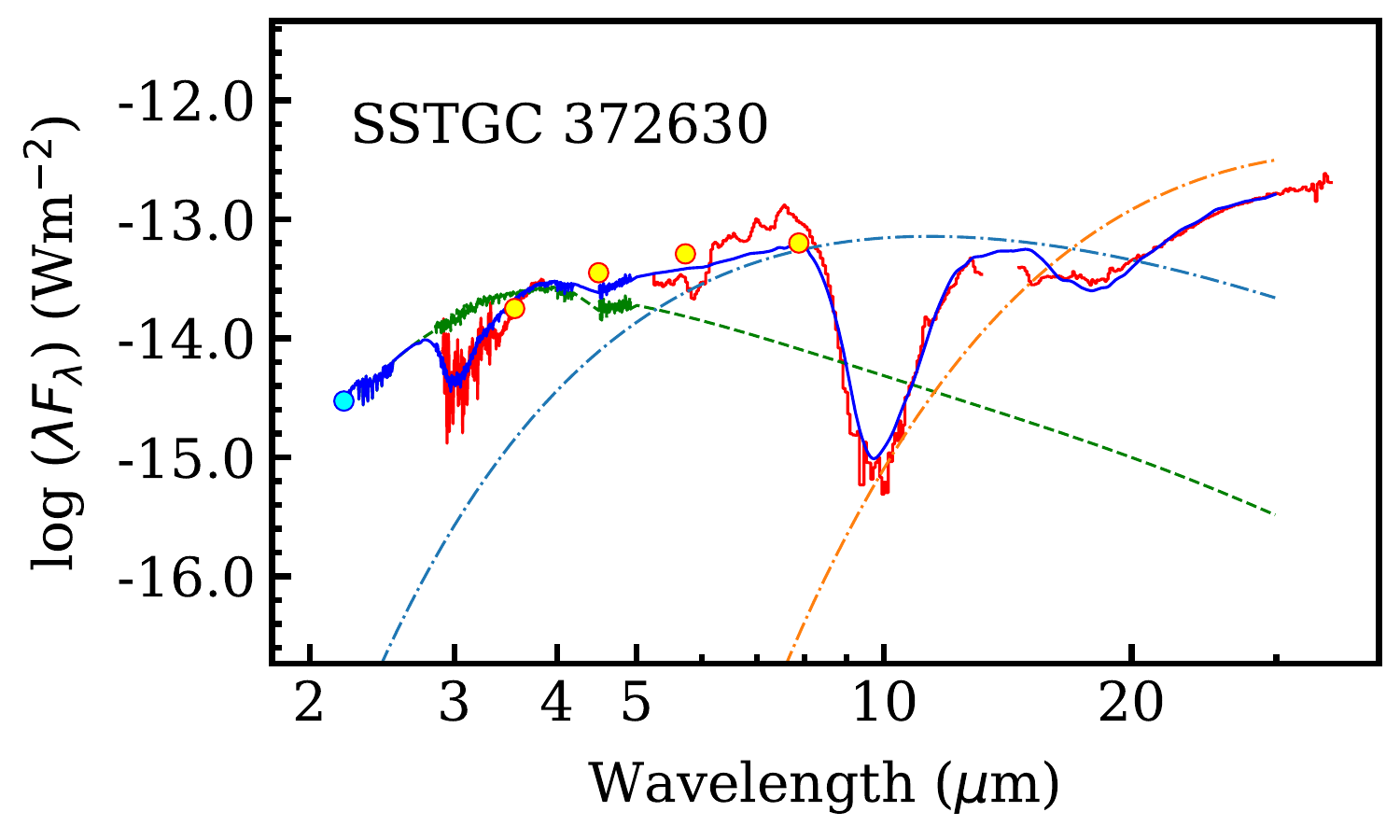}
\includegraphics[width=0.45\textwidth]{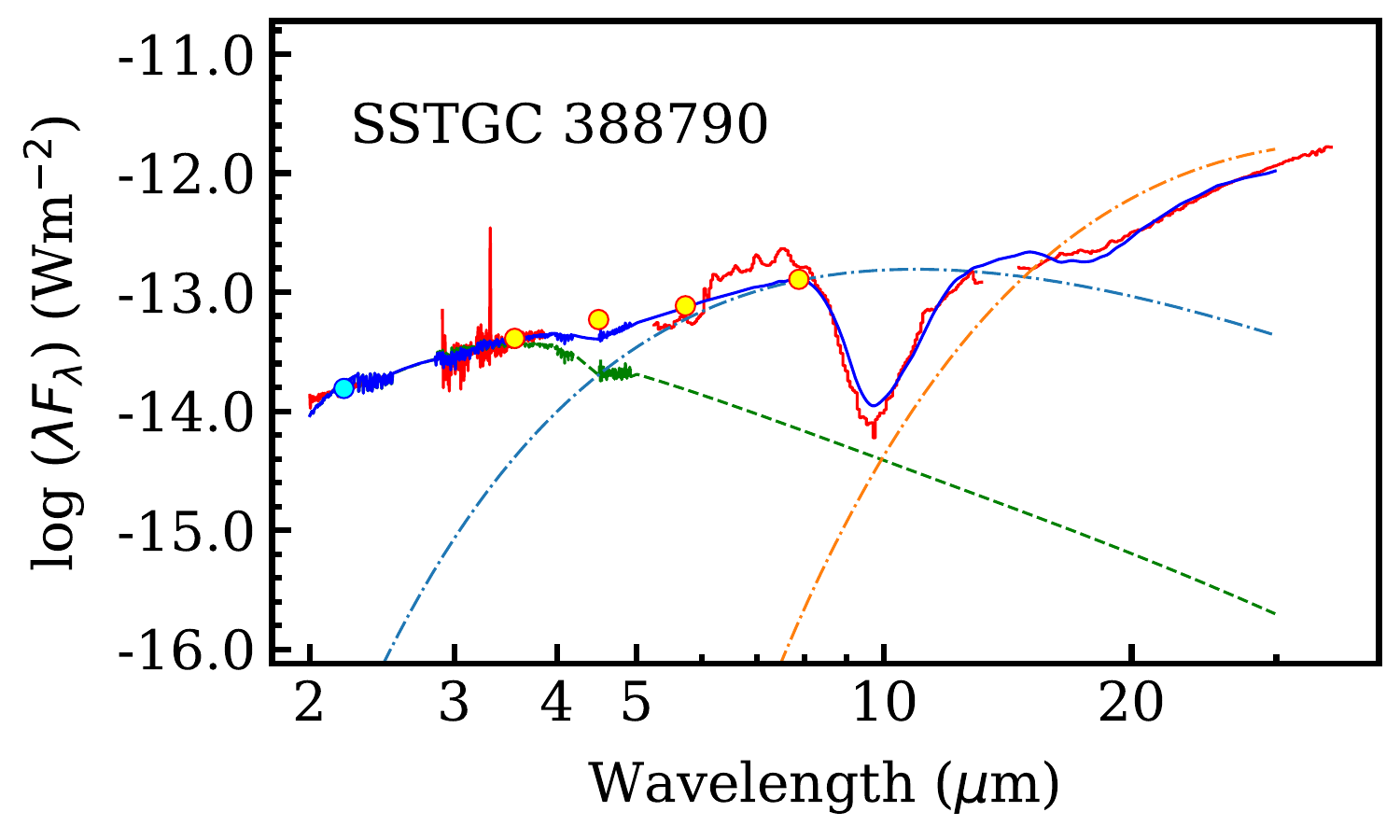}
\includegraphics[width=0.45\textwidth]{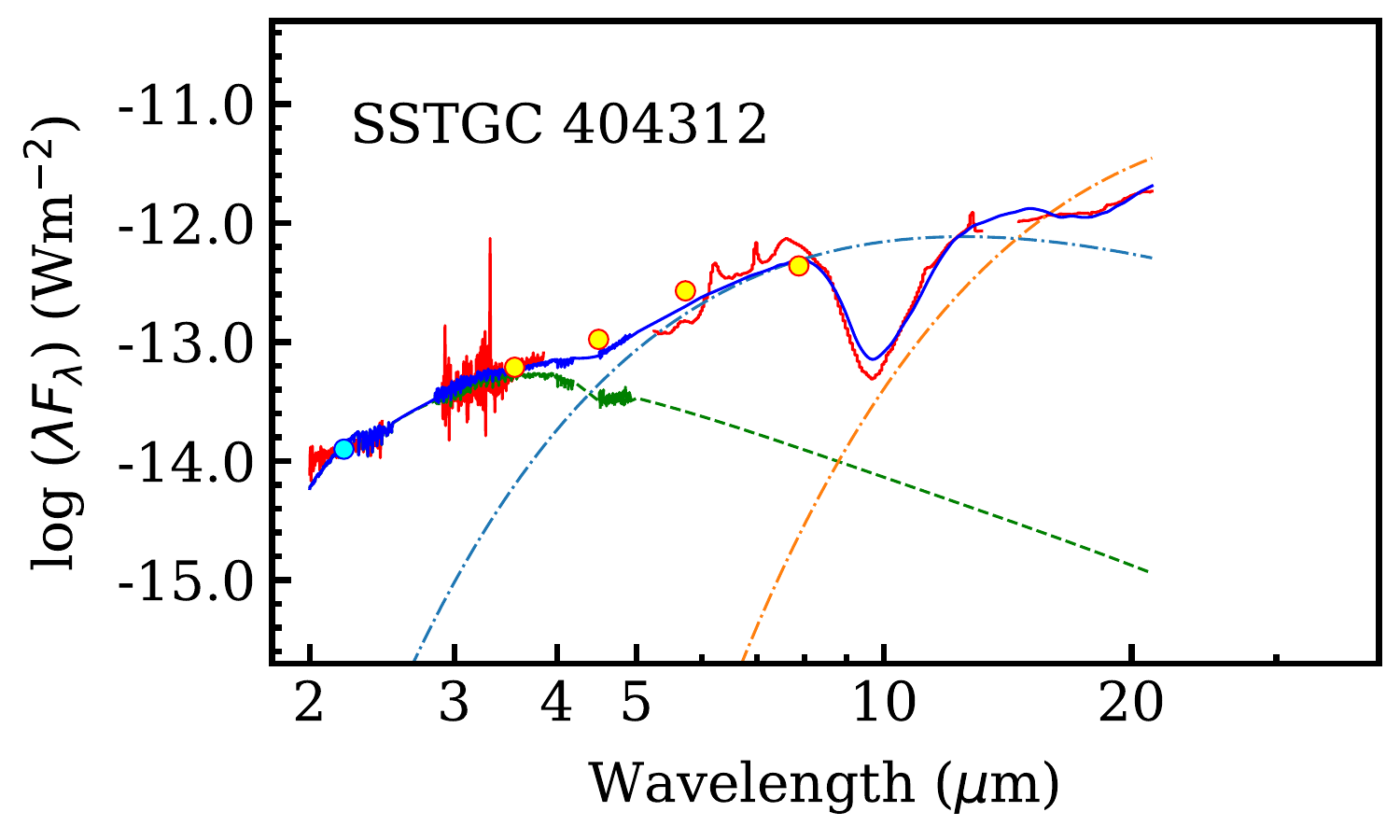}
\includegraphics[width=0.45\textwidth]{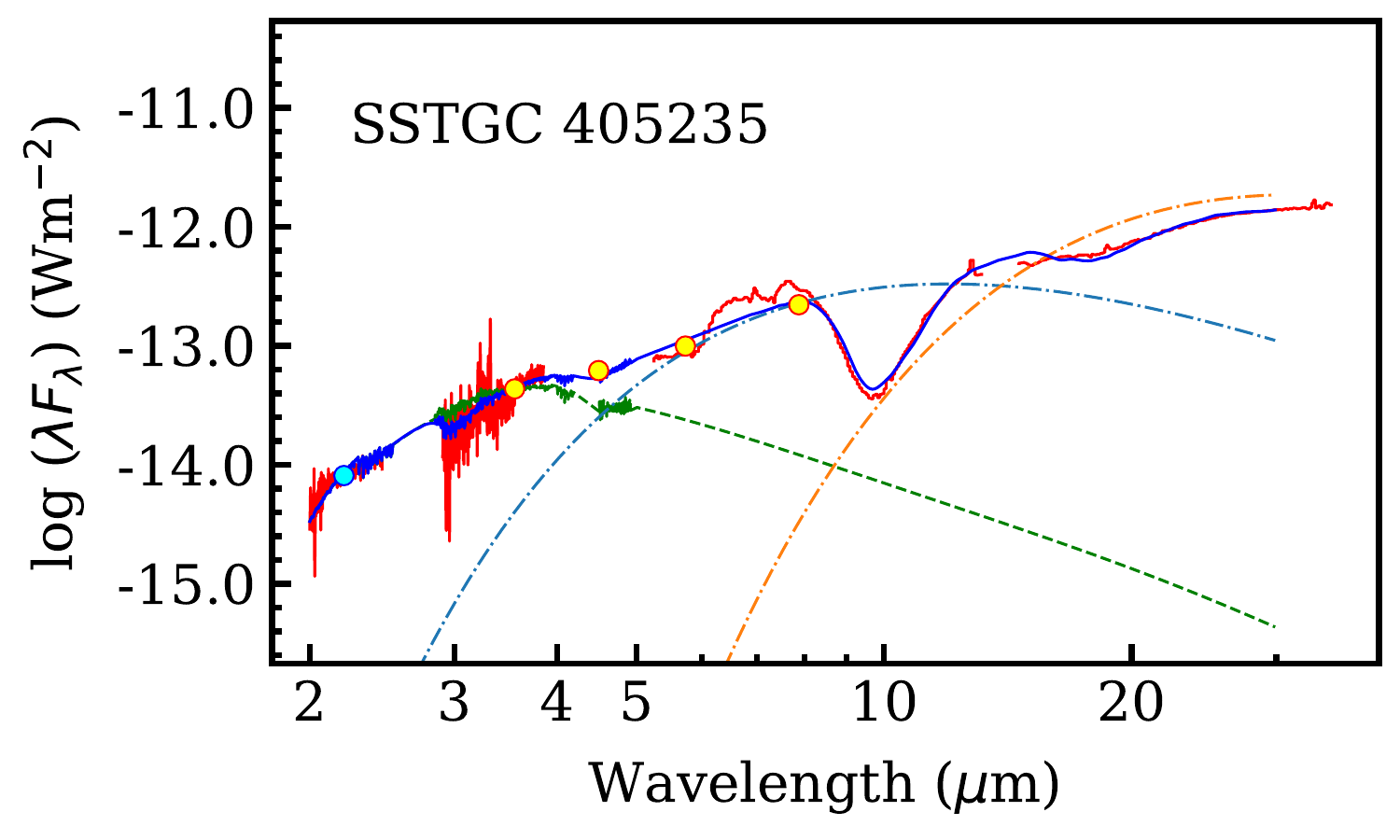}
\caption{Same as Fig.~\ref{fig:sed}, but showing the composite spectra of the remaining red CMZ sources.}
\label{fig:sed2}
\end{figure*}

\begin{figure*}[h]
\centering
\ContinuedFloat
\includegraphics[width=0.45\textwidth]{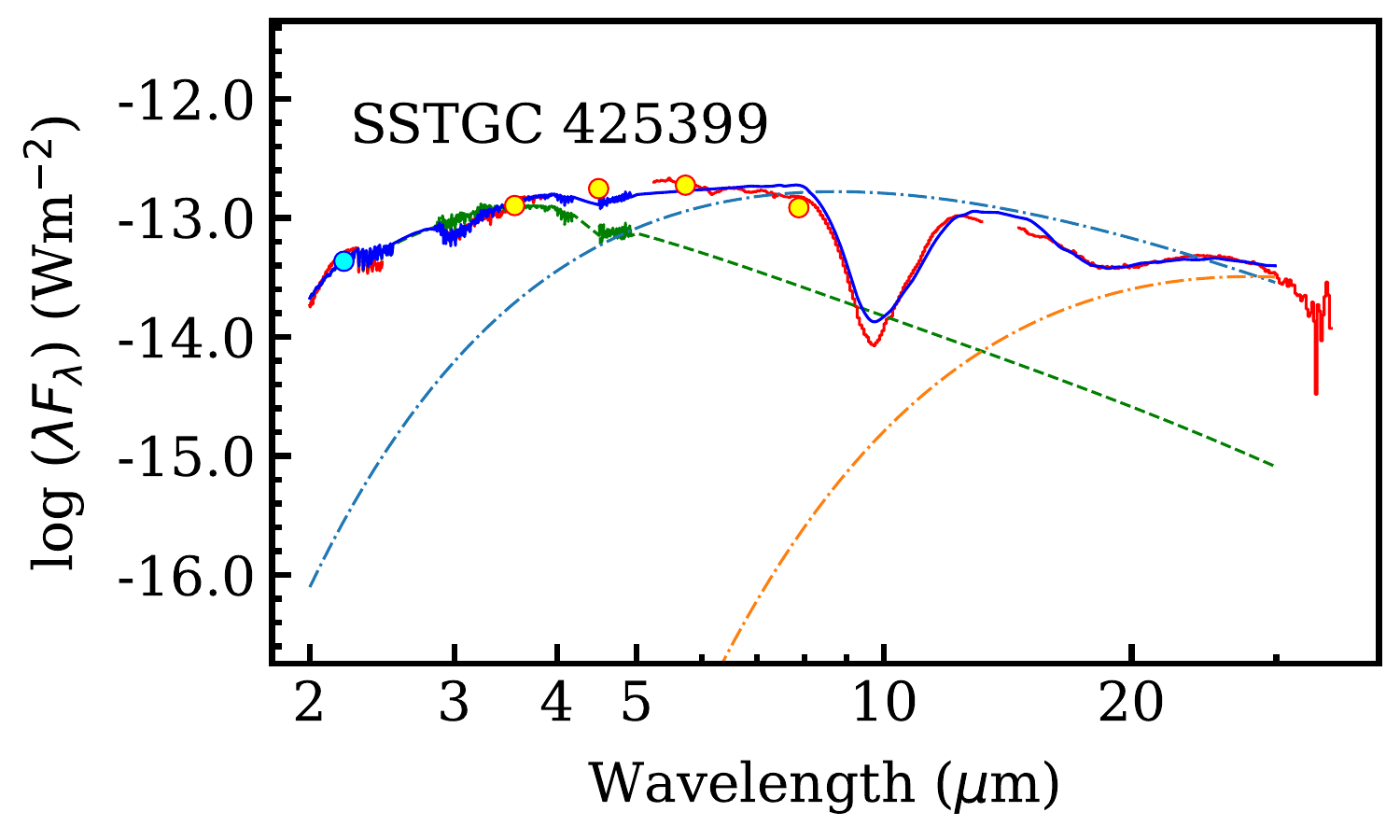}
\includegraphics[width=0.45\textwidth]{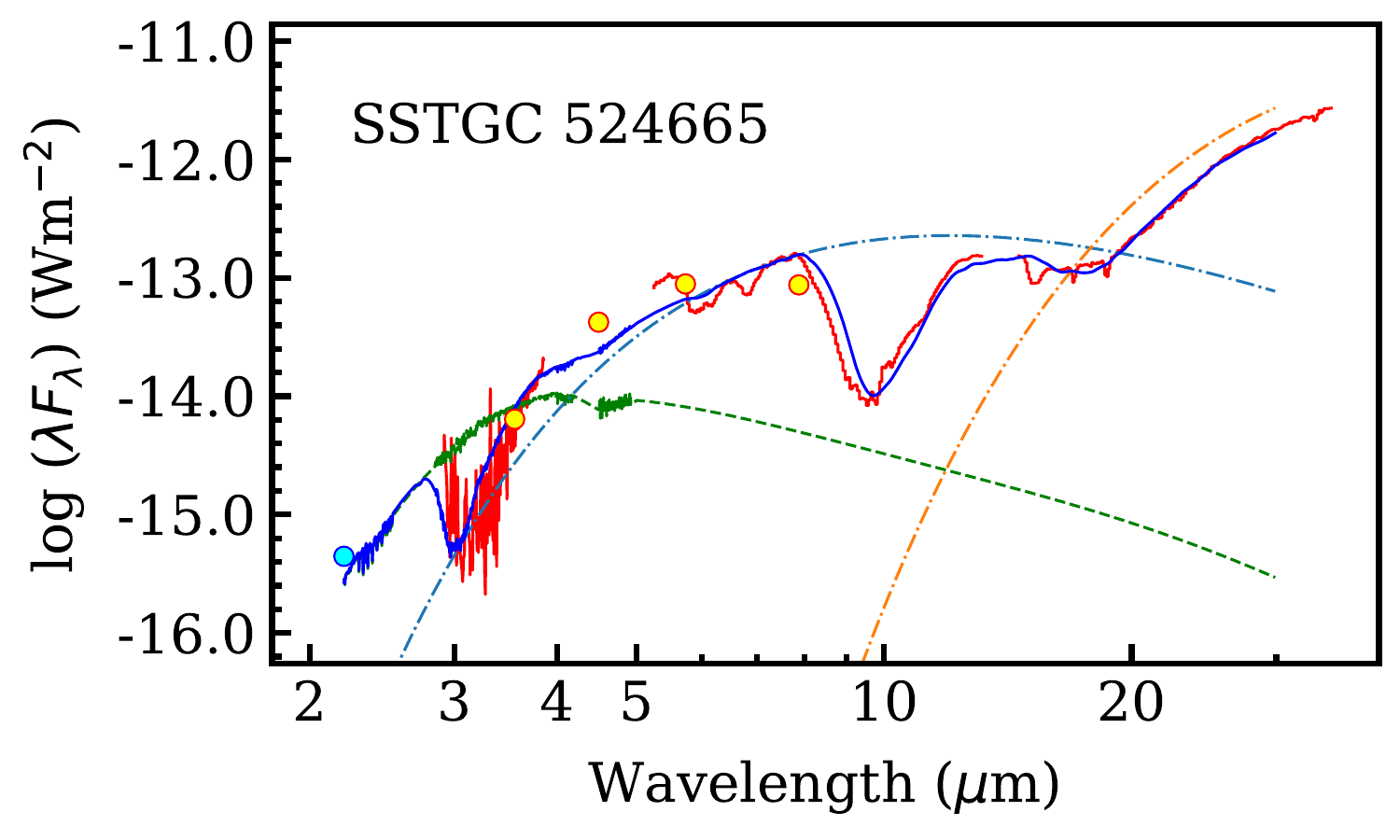}
\includegraphics[width=0.45\textwidth]{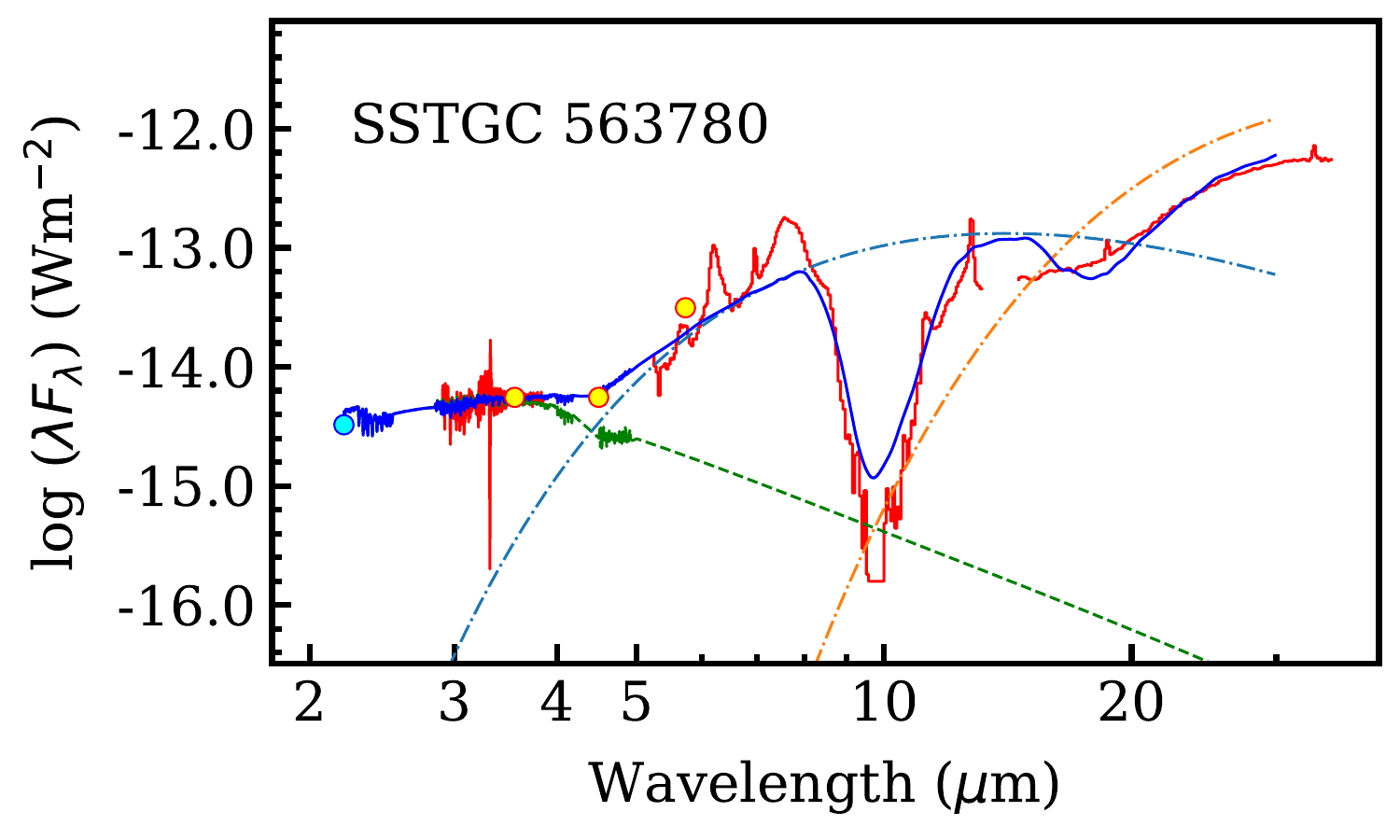}    
\includegraphics[width=0.45\textwidth]{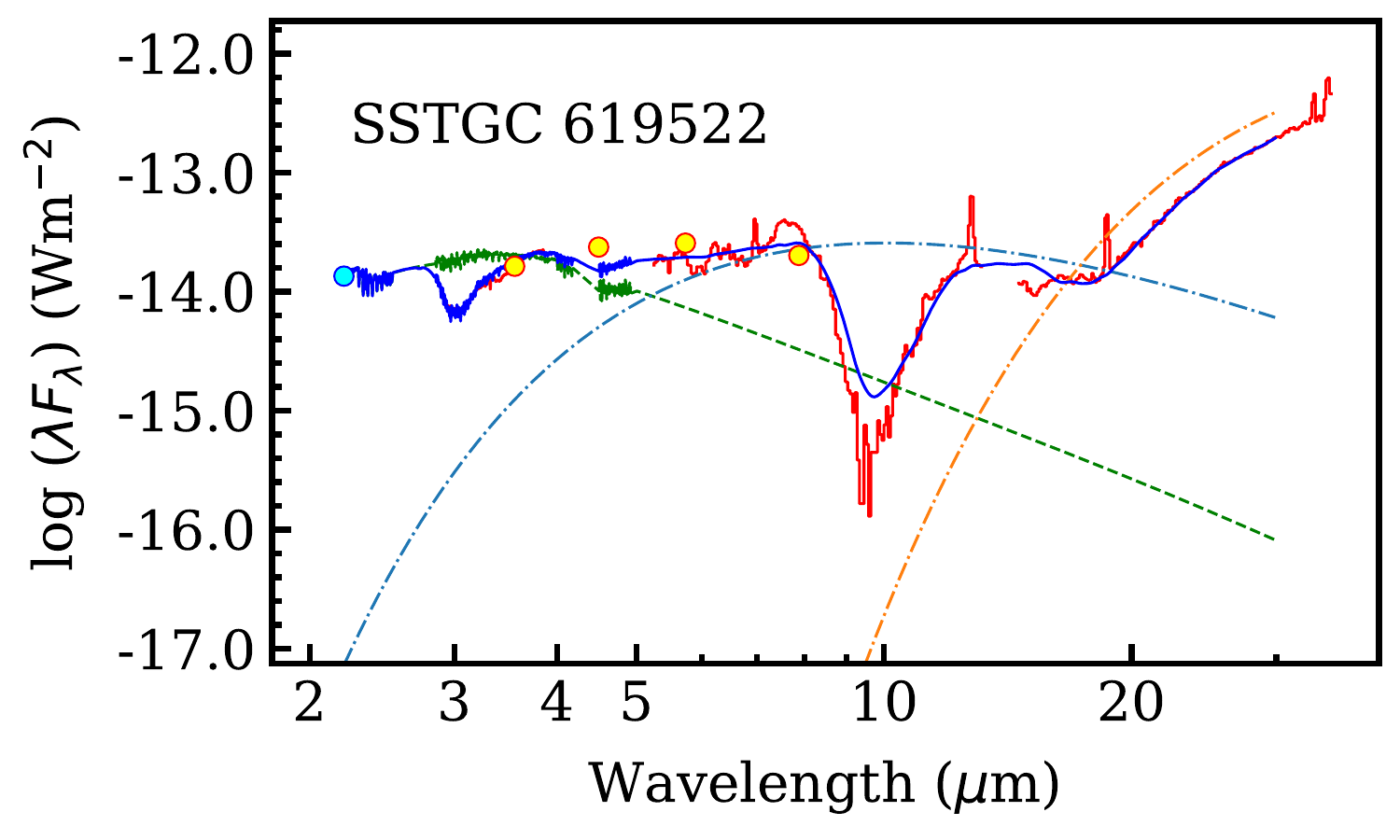}    
\includegraphics[width=0.45\textwidth]{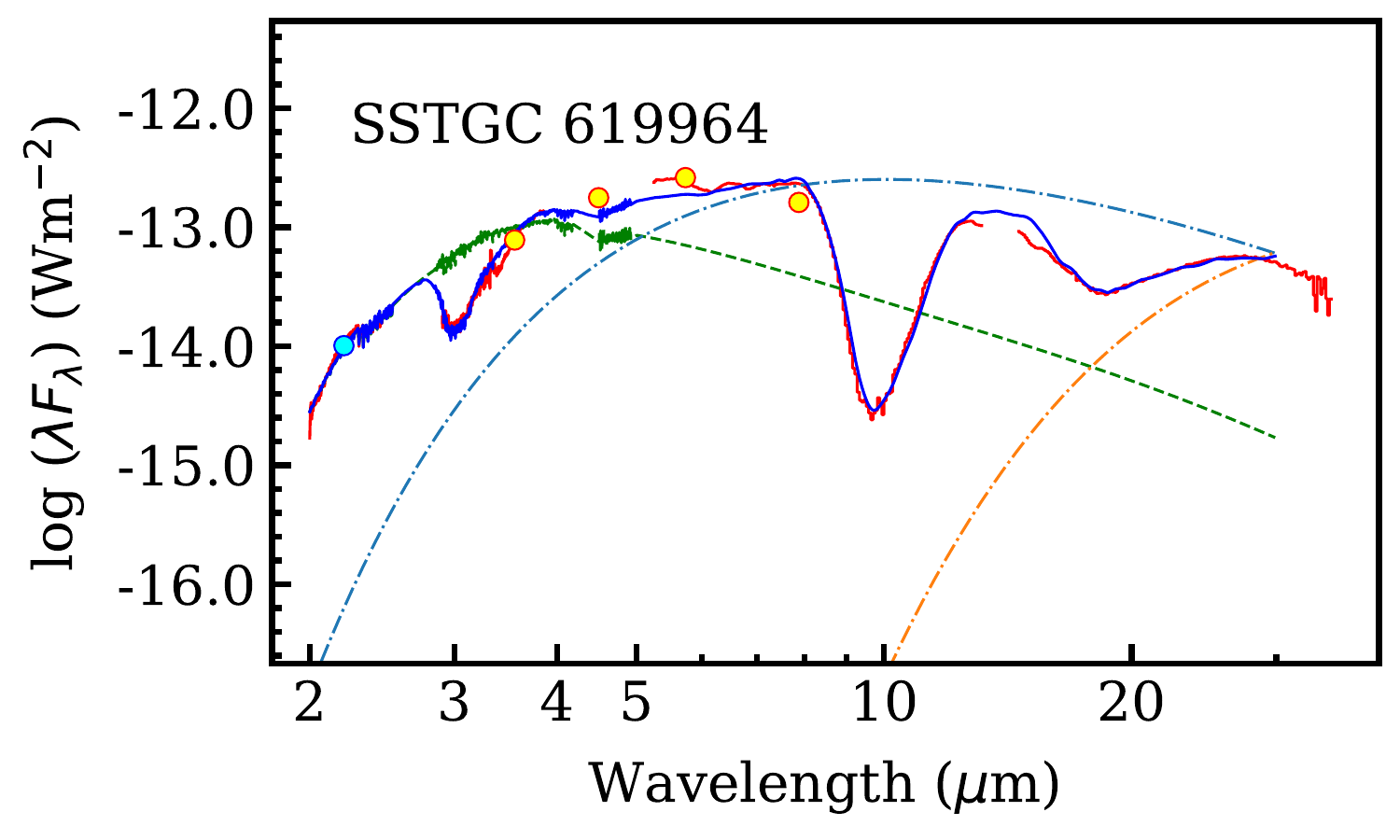}
\includegraphics[width=0.45\textwidth]{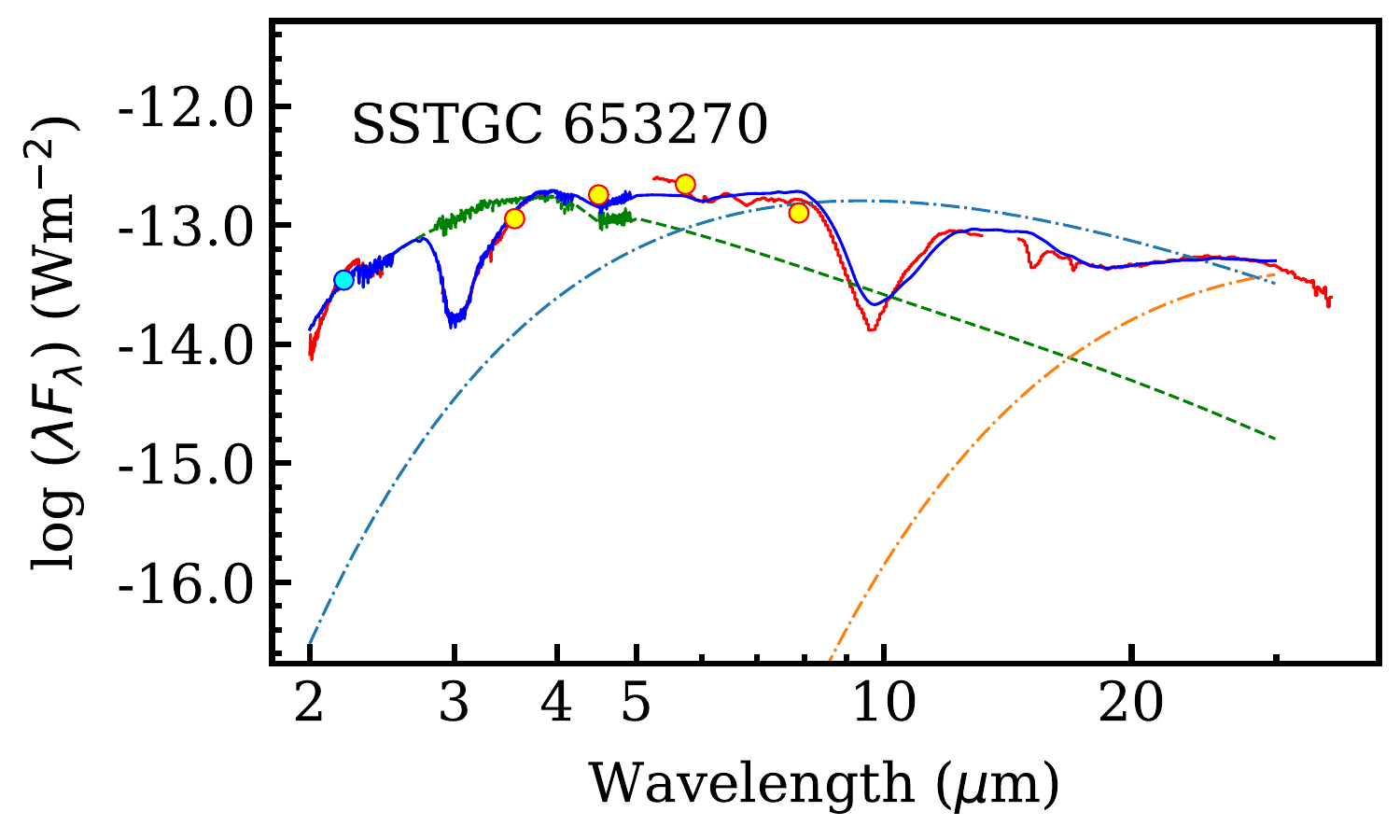}
\includegraphics[width=0.45\textwidth]{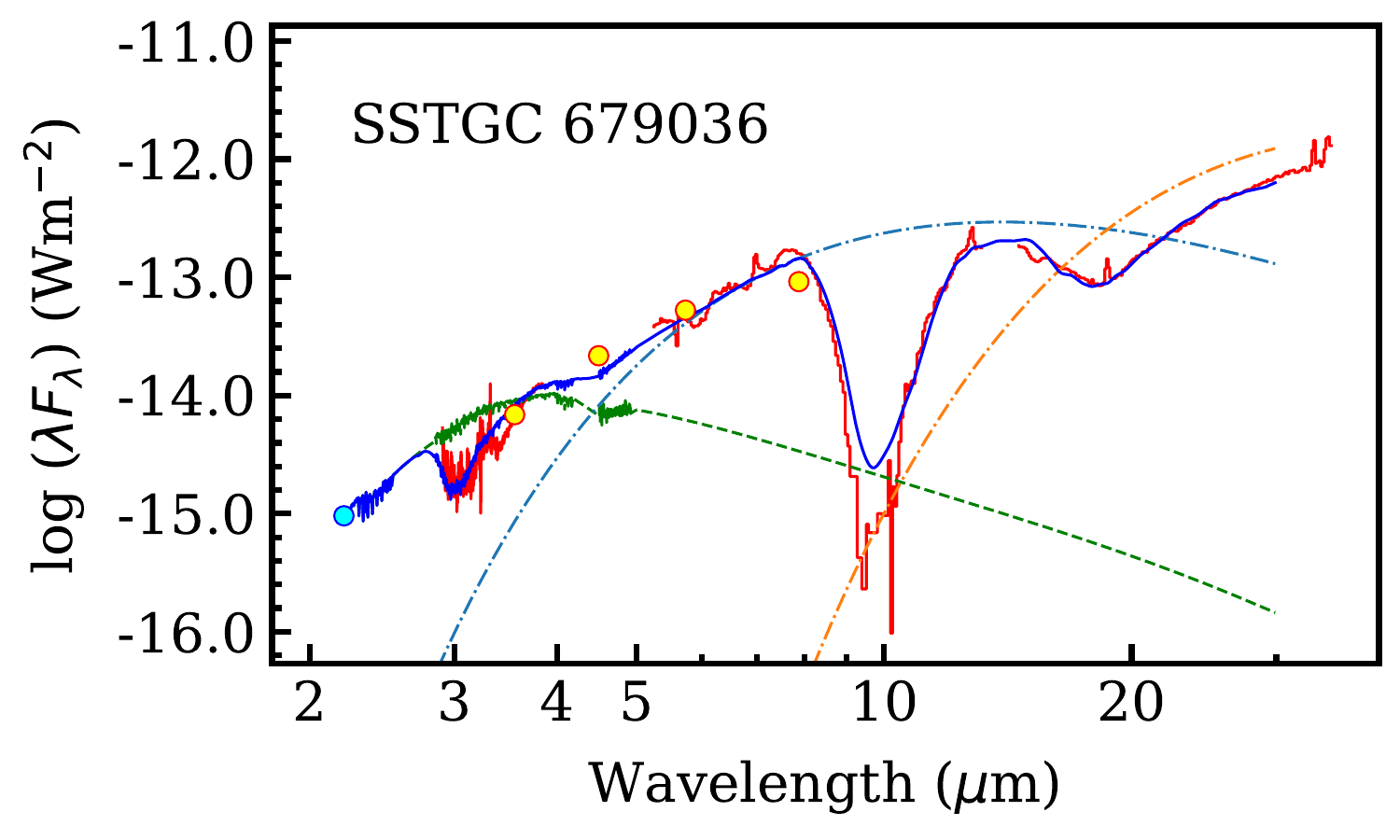}
\includegraphics[width=0.45\textwidth]{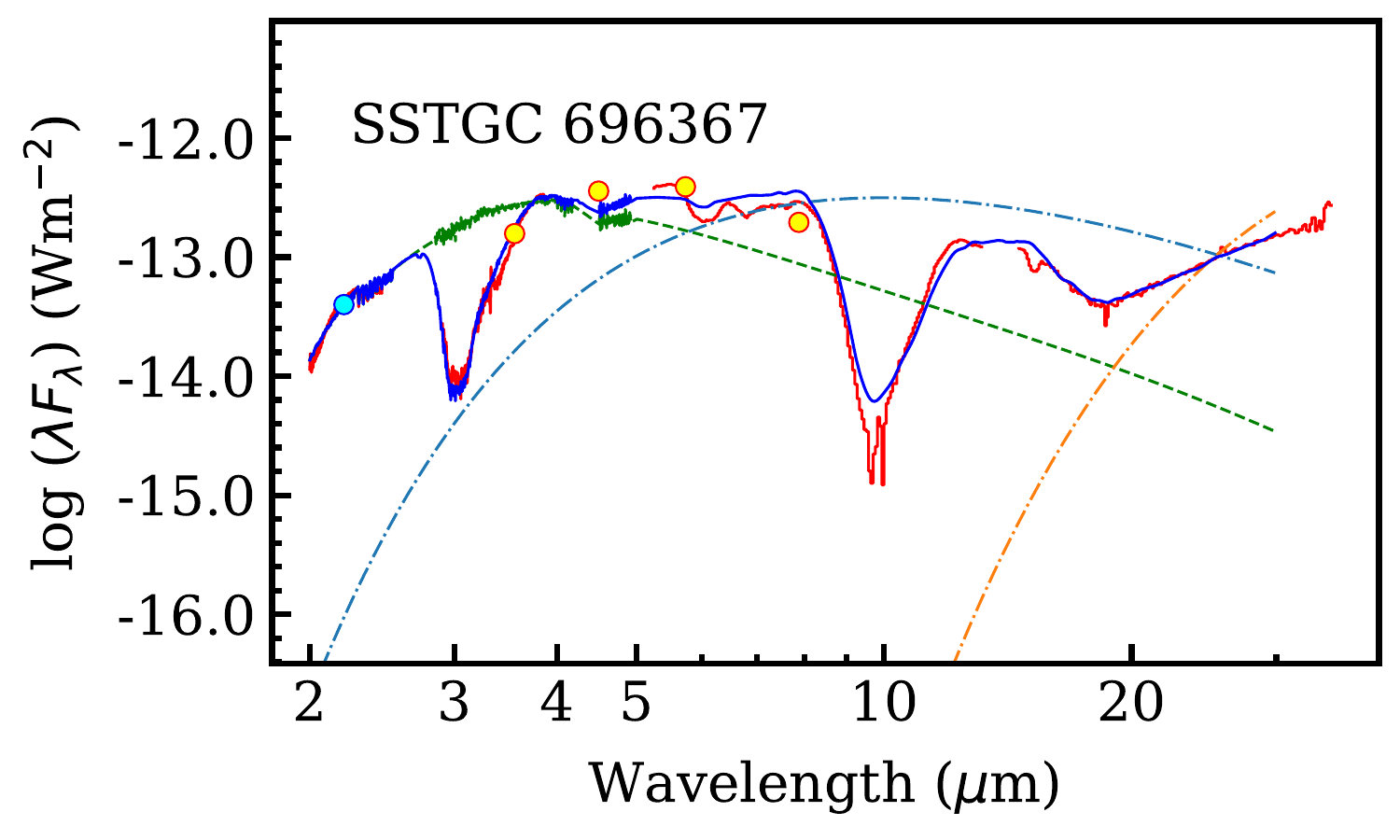}
\caption{Spectra of the sample sources (continued).}
\end{figure*}

\begin{figure*}[h]
\centering
\ContinuedFloat
\includegraphics[width=0.45\textwidth]{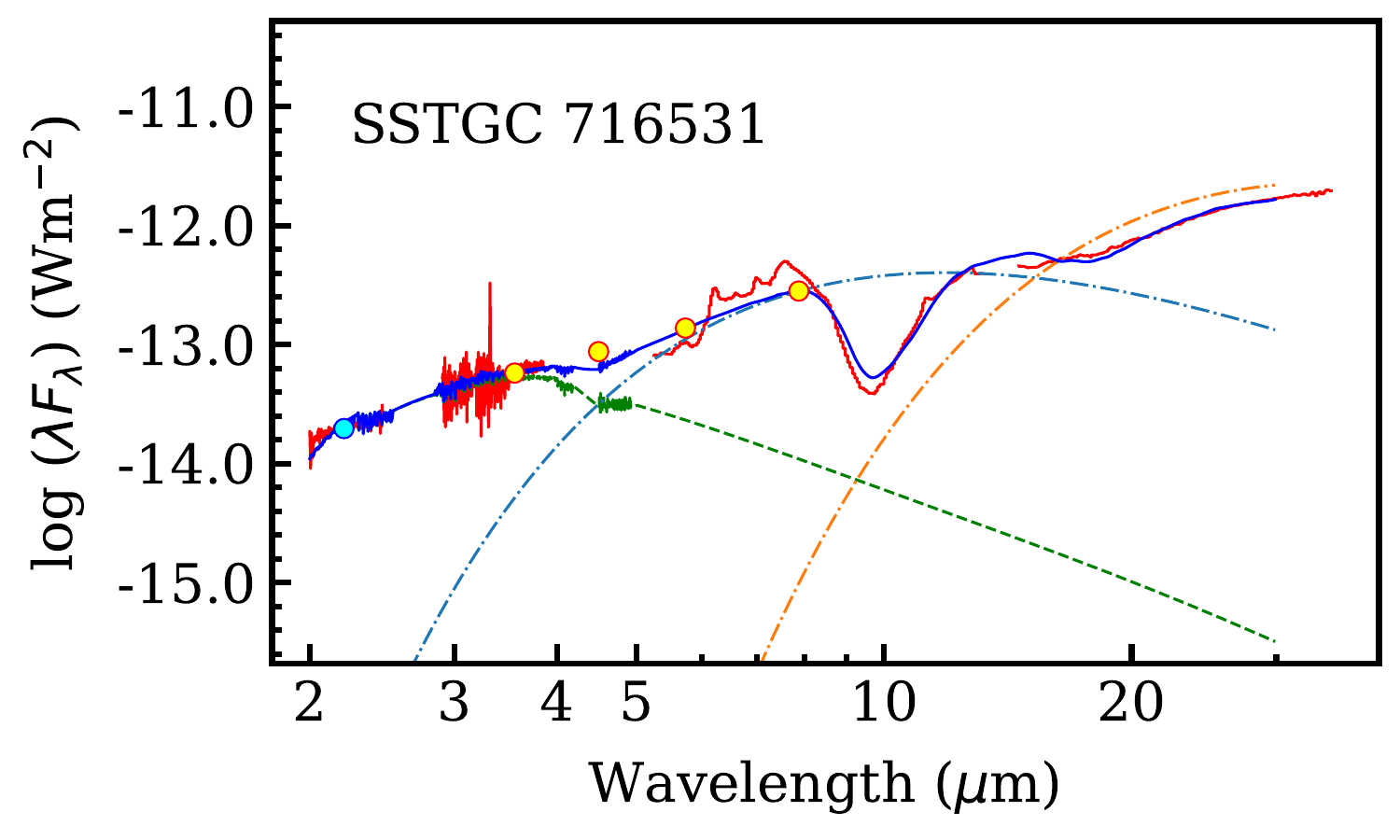}
\includegraphics[width=0.45\textwidth]{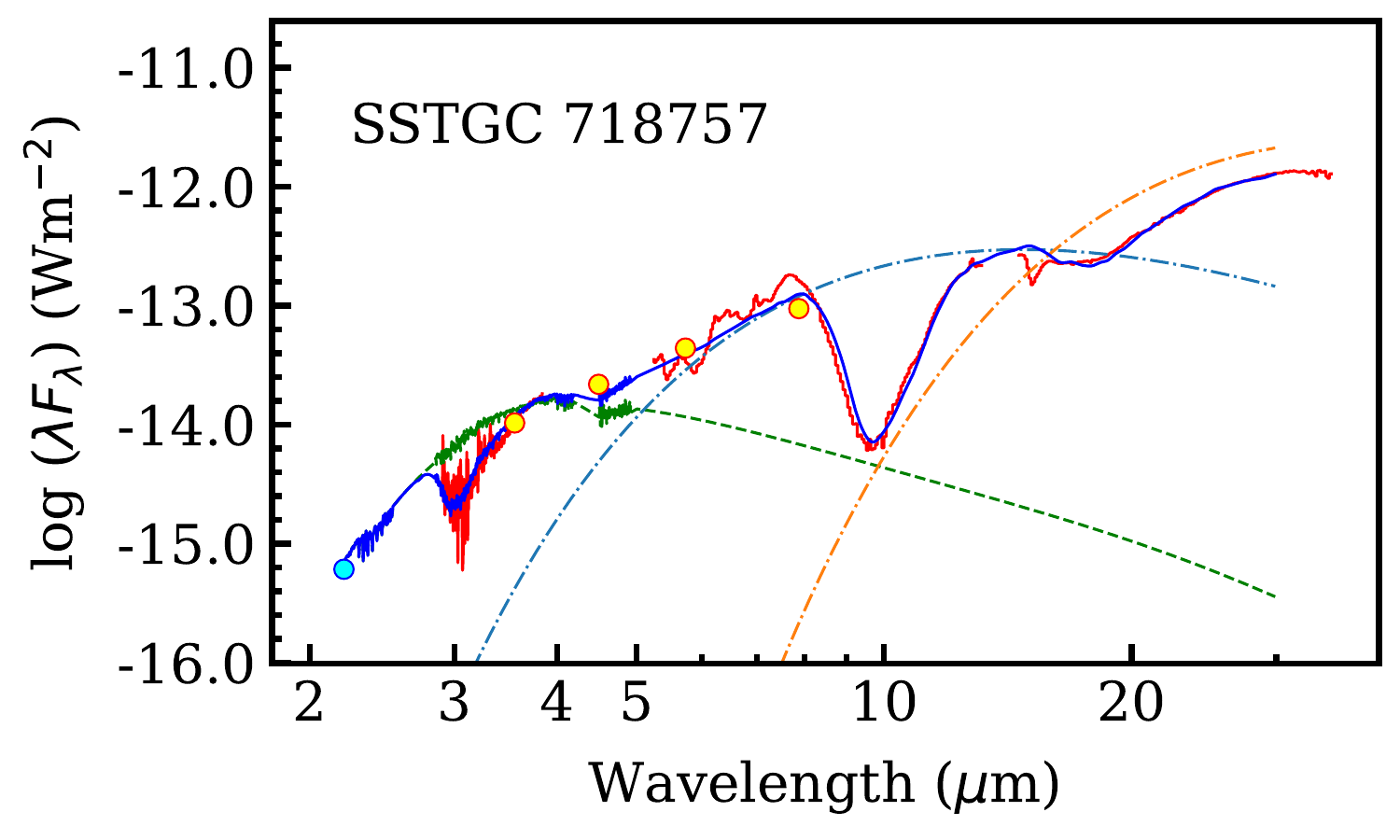}
\includegraphics[width=0.45\textwidth]{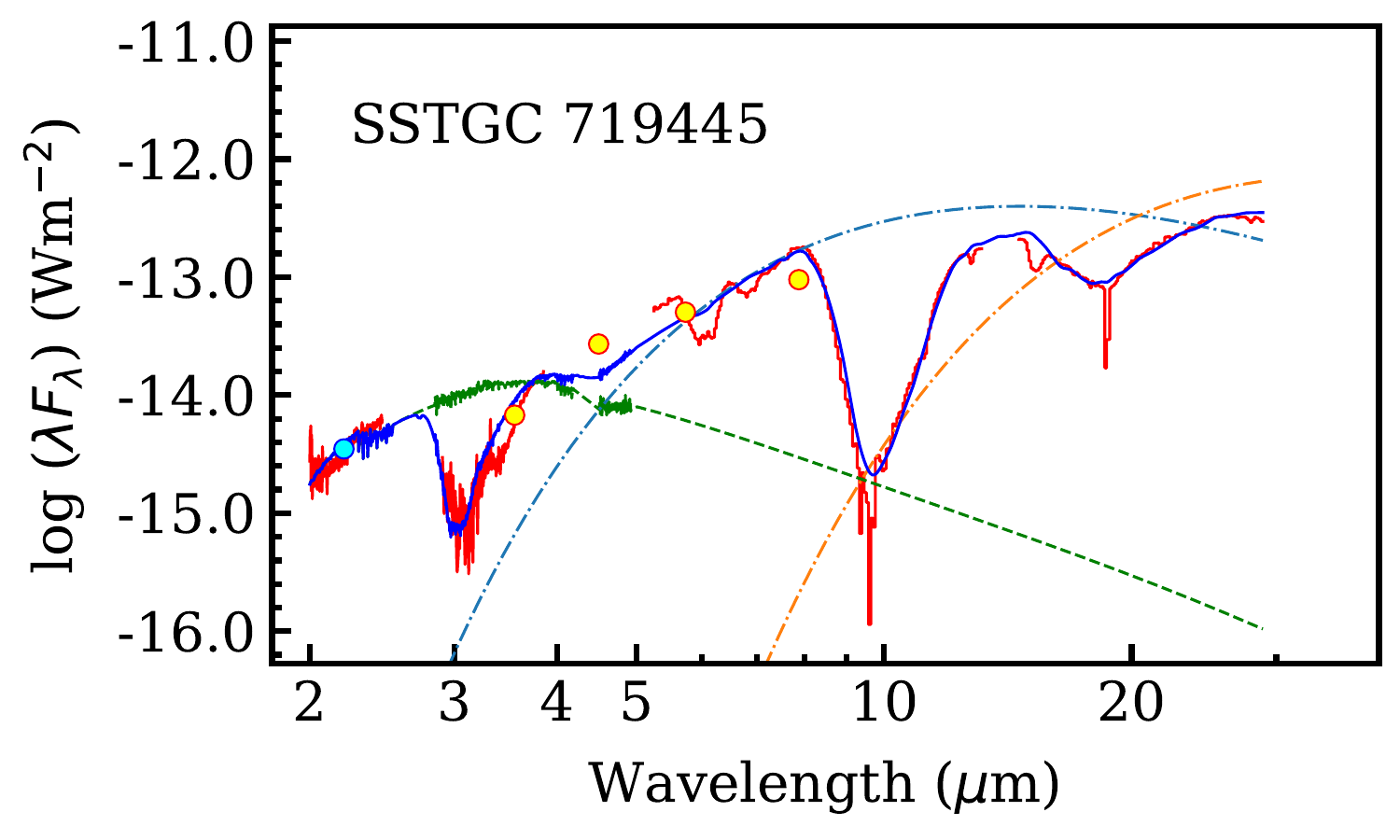}
\includegraphics[width=0.45\textwidth]{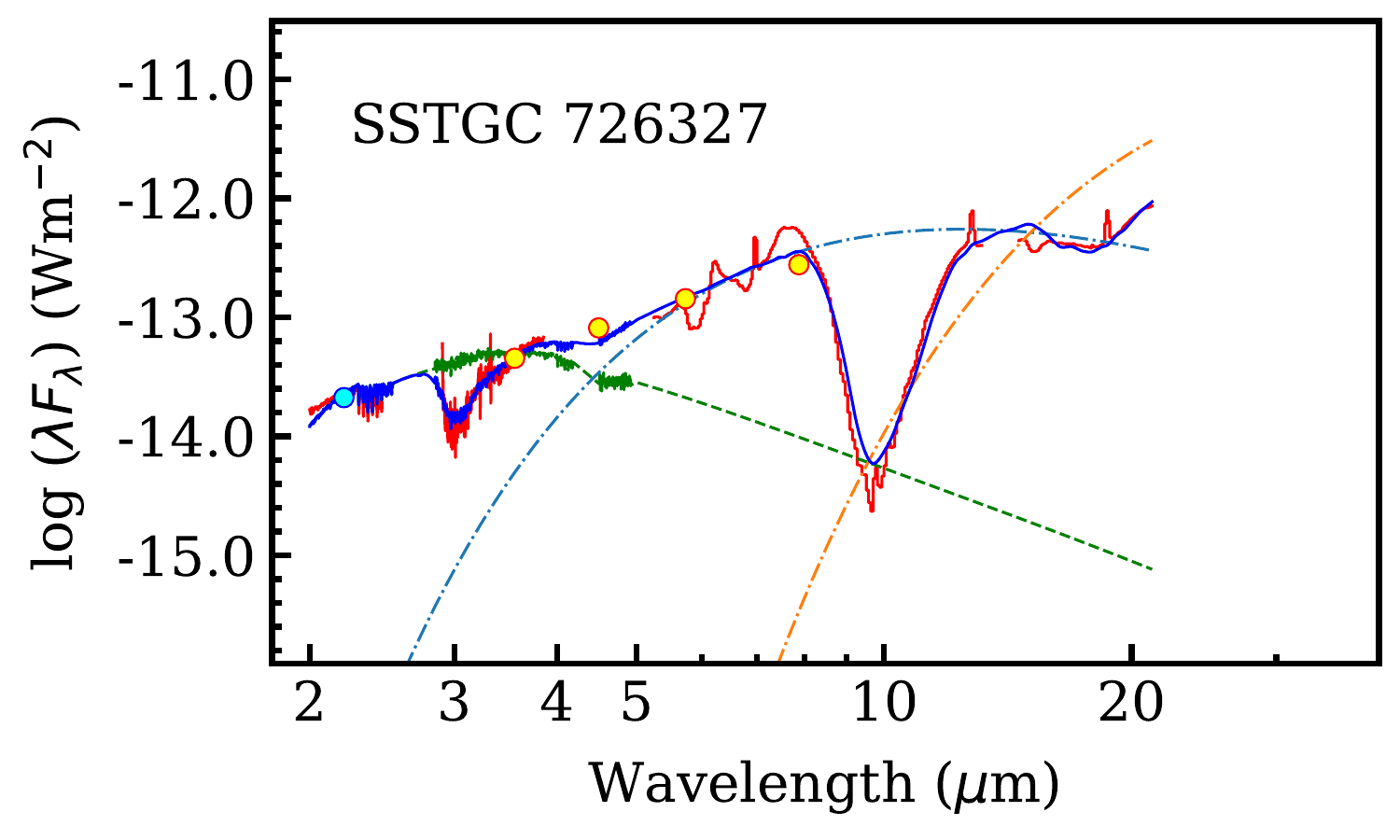}
\includegraphics[width=0.45\textwidth]{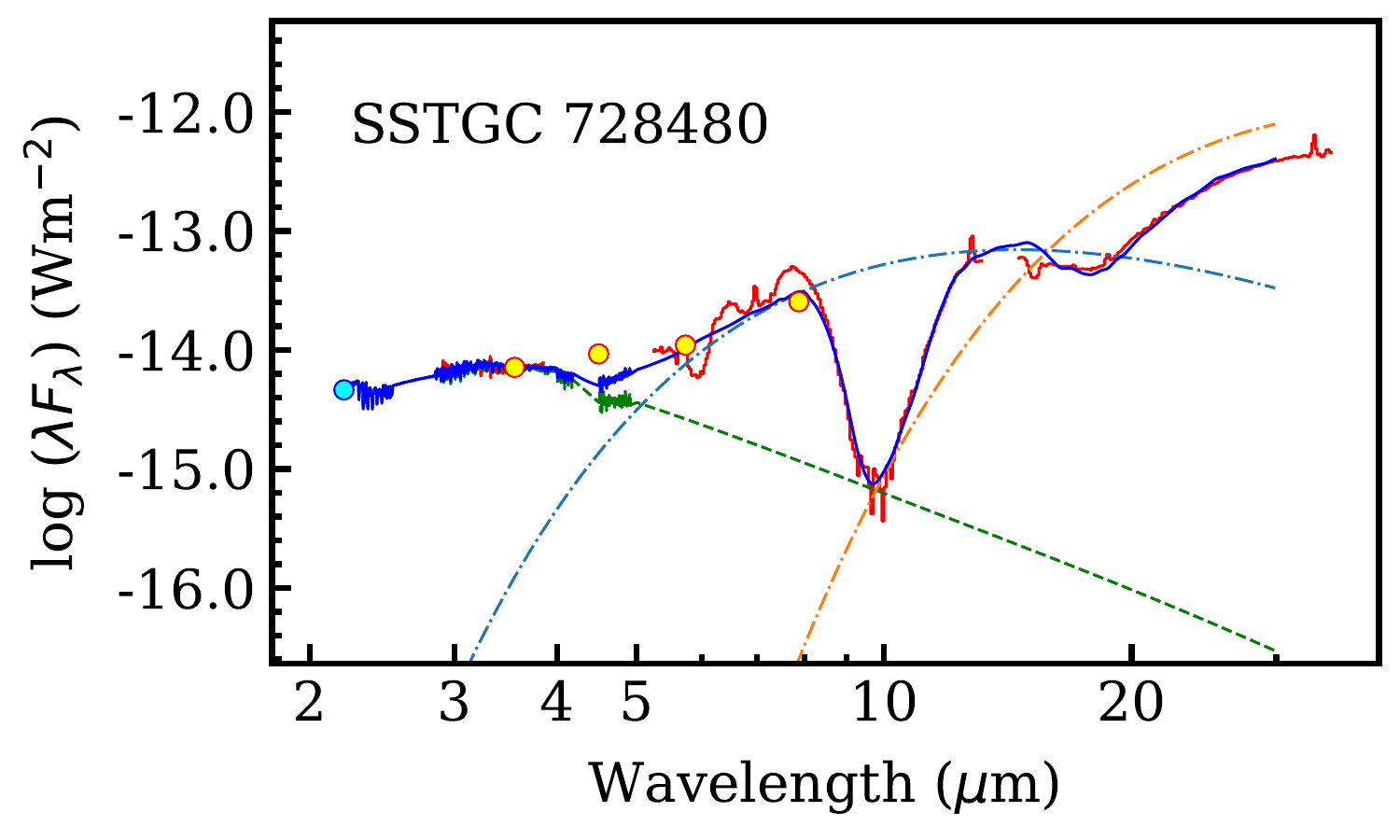}    
\includegraphics[width=0.45\textwidth]{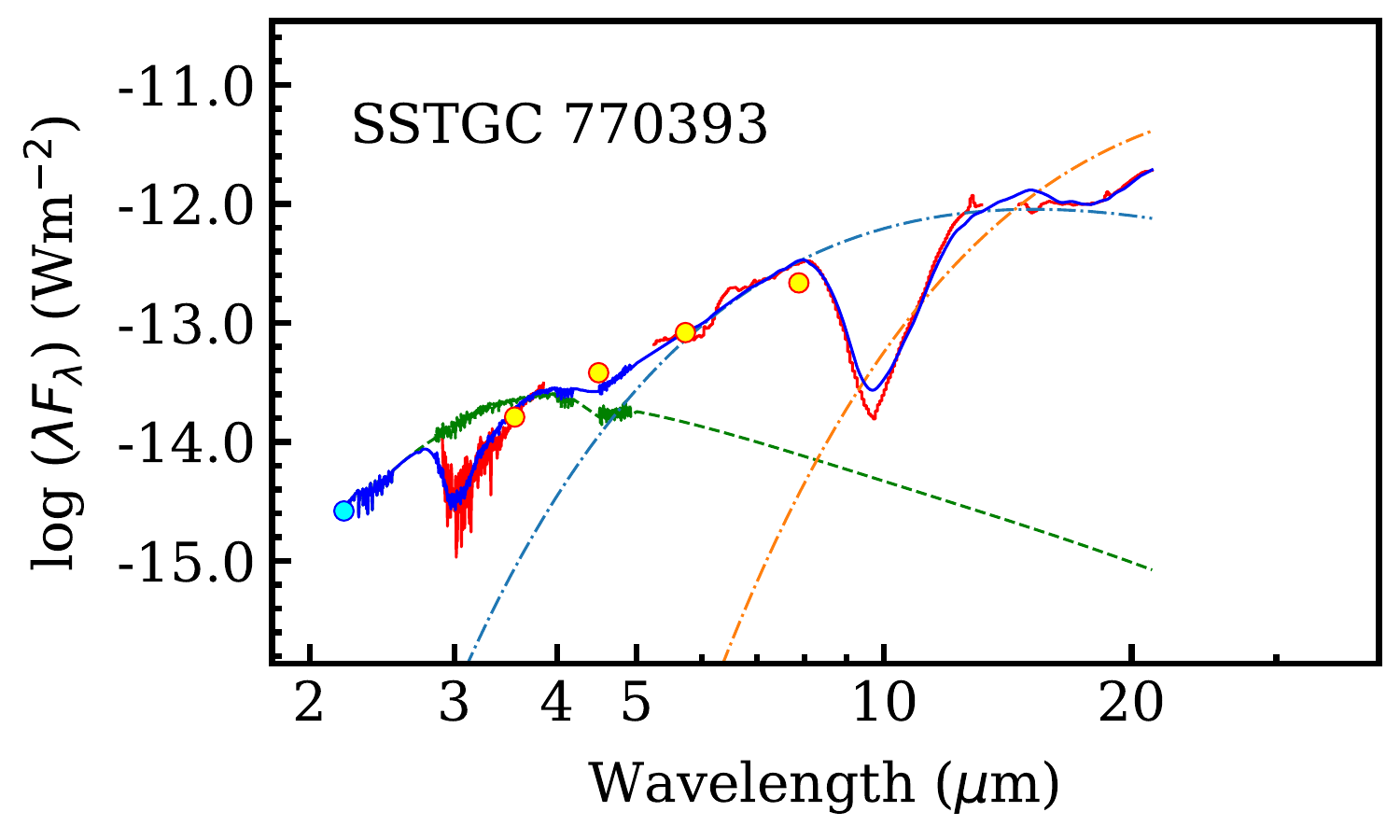}    
\includegraphics[width=0.45\textwidth]{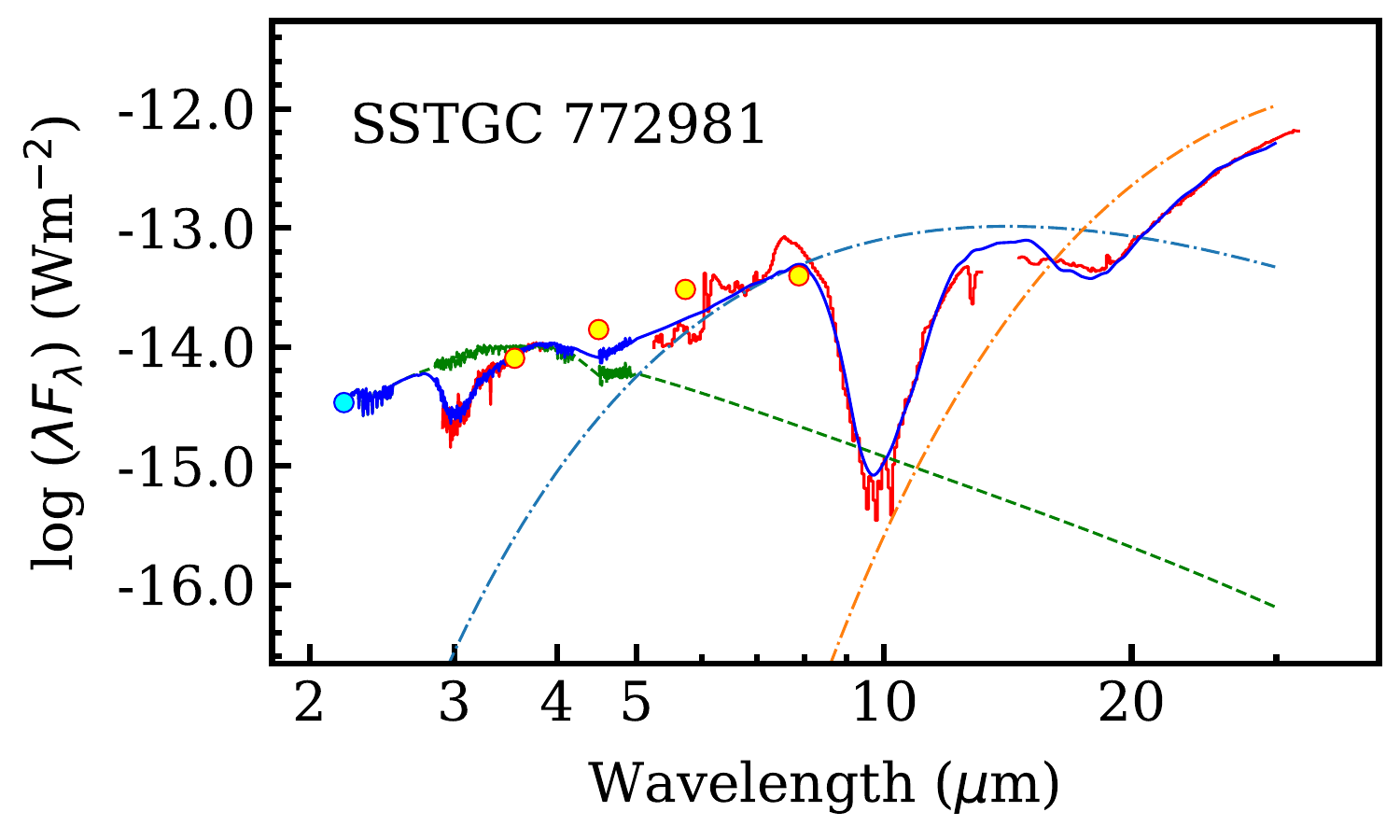}
\includegraphics[width=0.45\textwidth]{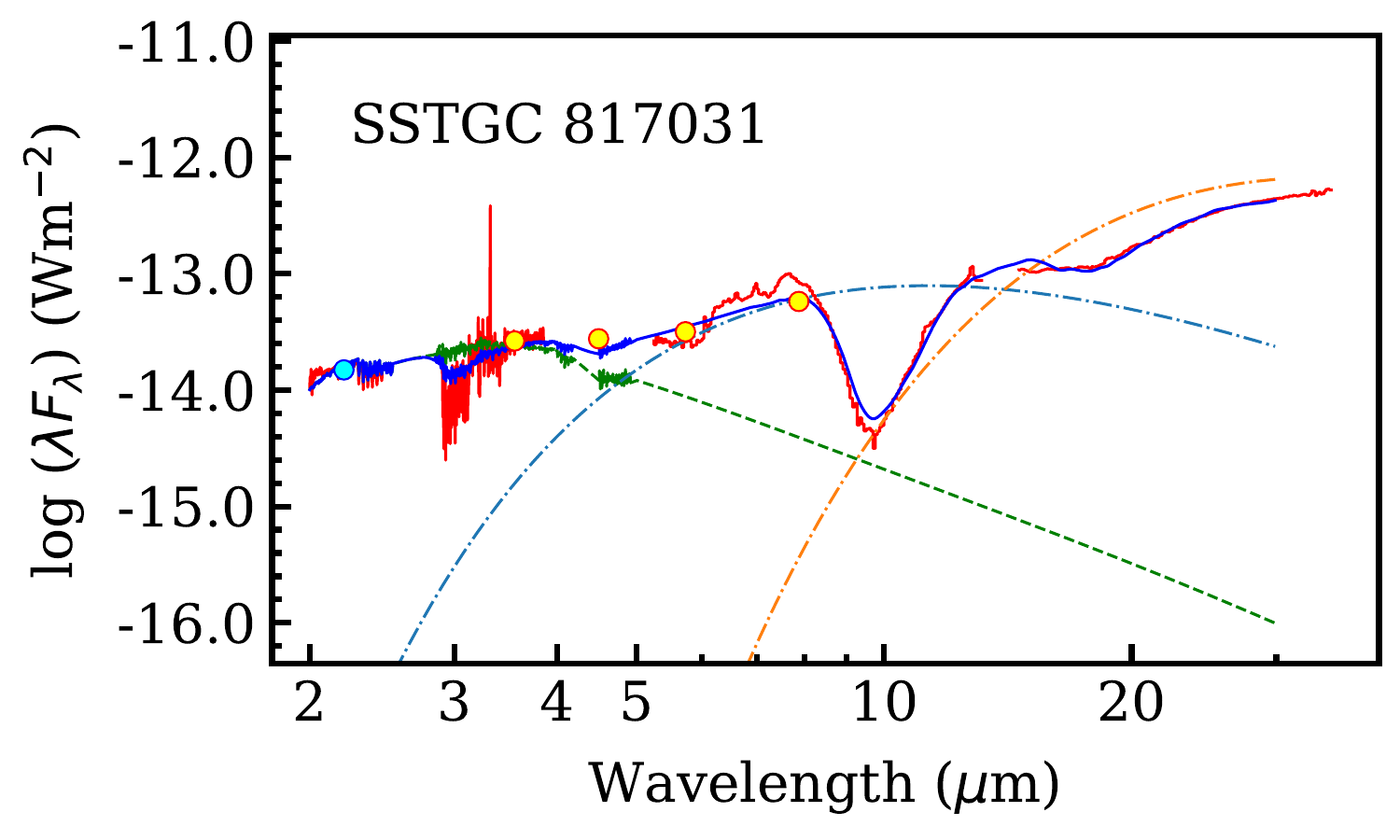}
\caption{Spectra of the sample sources (continued).}
\end{figure*}

\FloatBarrier

\section{Additional optical depth spectra centred at $3.53\ \mu$m}\label{sec:ch3oh2}

\begin{figure*}[h]
\centering
\includegraphics[width=0.25\textwidth]{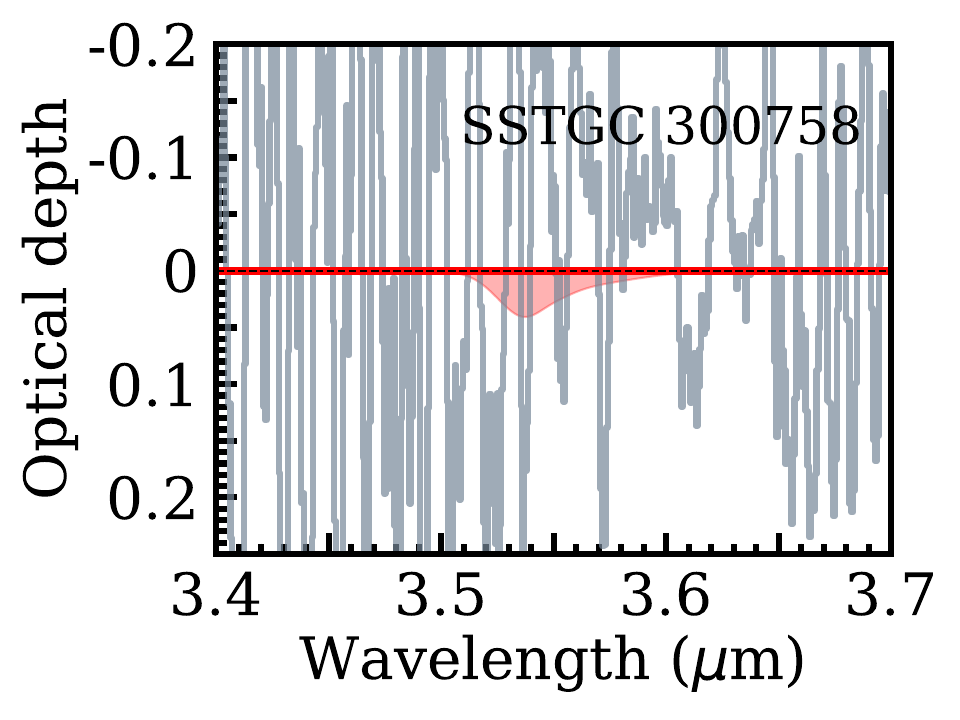}
\includegraphics[width=0.25\textwidth]{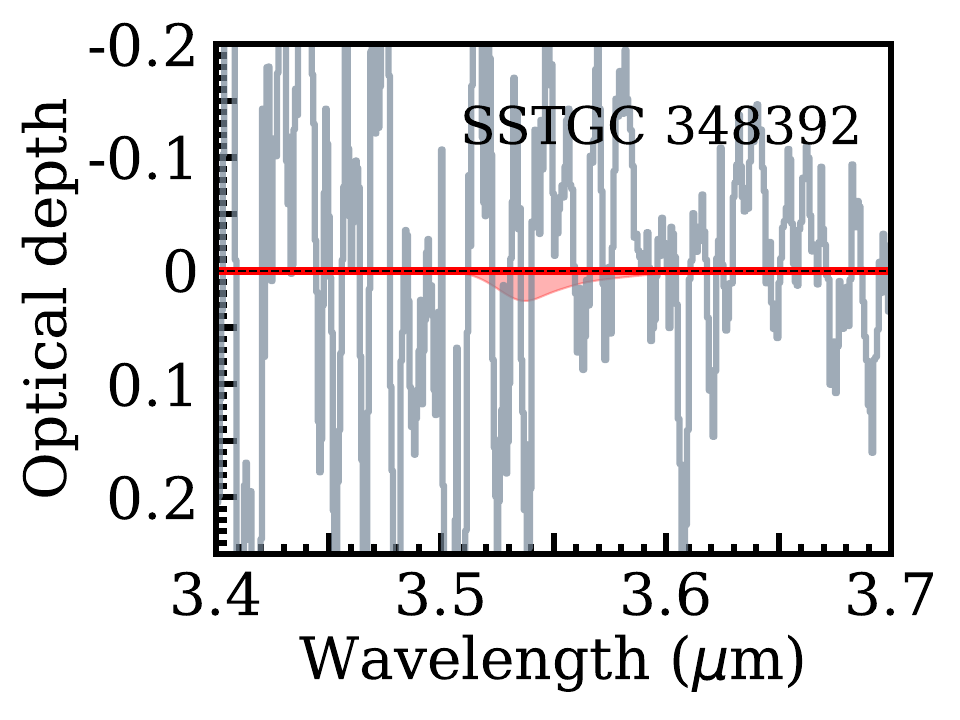}
\includegraphics[width=0.25\textwidth]{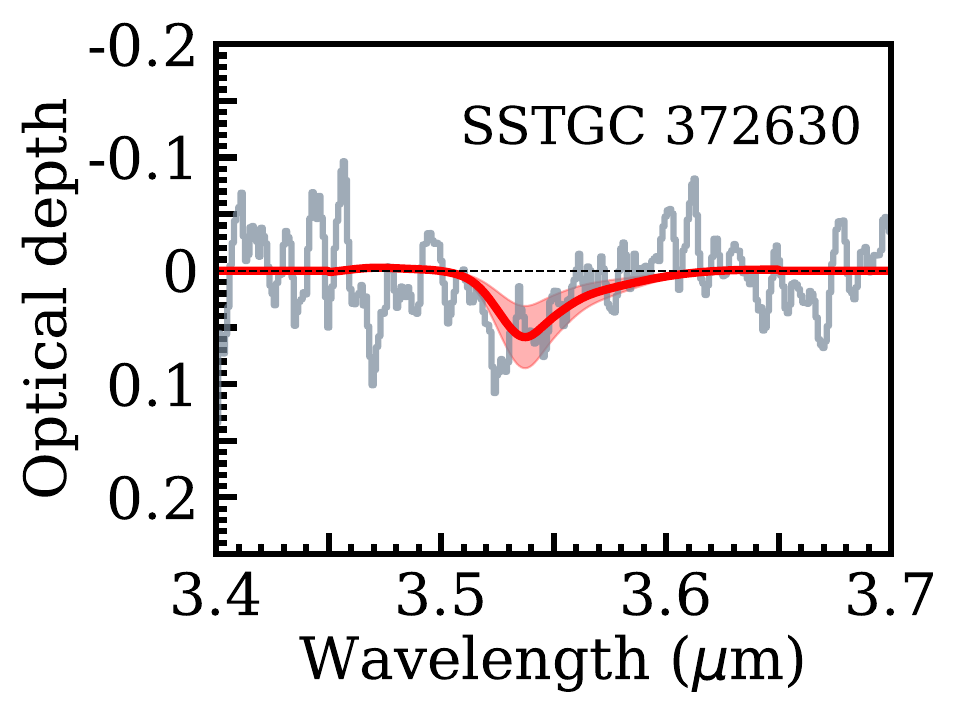}
\includegraphics[width=0.25\textwidth]{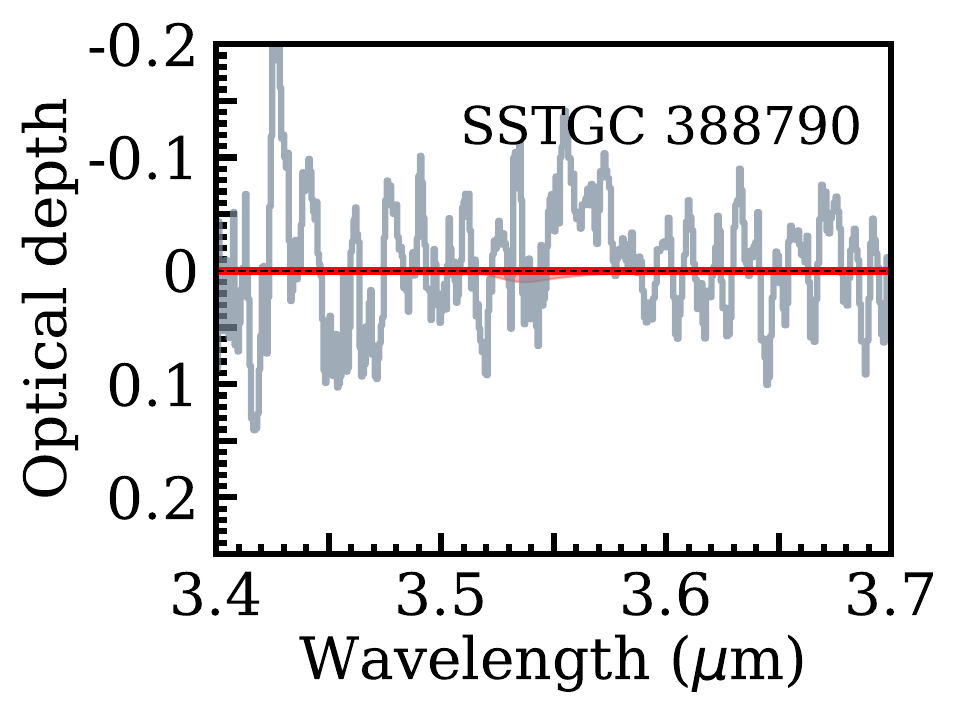}
\includegraphics[width=0.25\textwidth]{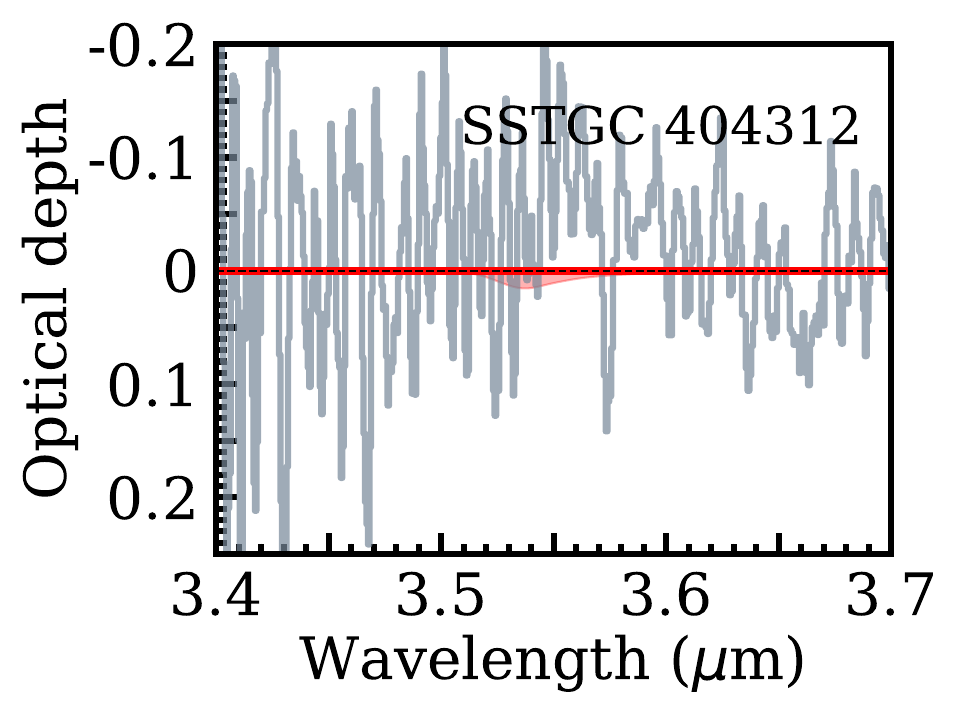}
\includegraphics[width=0.25\textwidth]{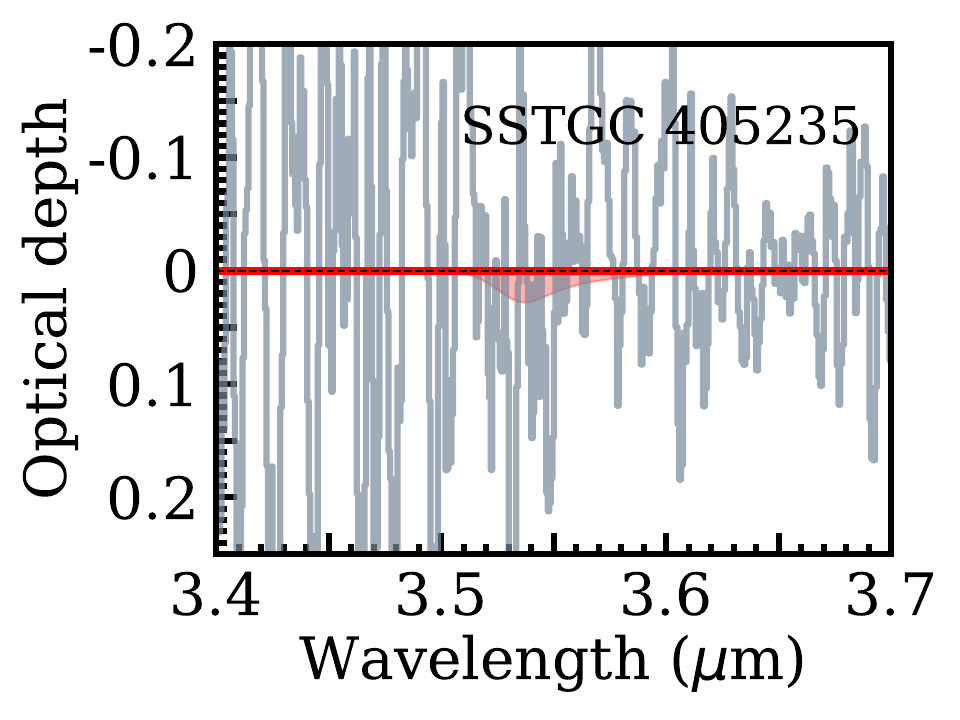}
\includegraphics[width=0.25\textwidth]{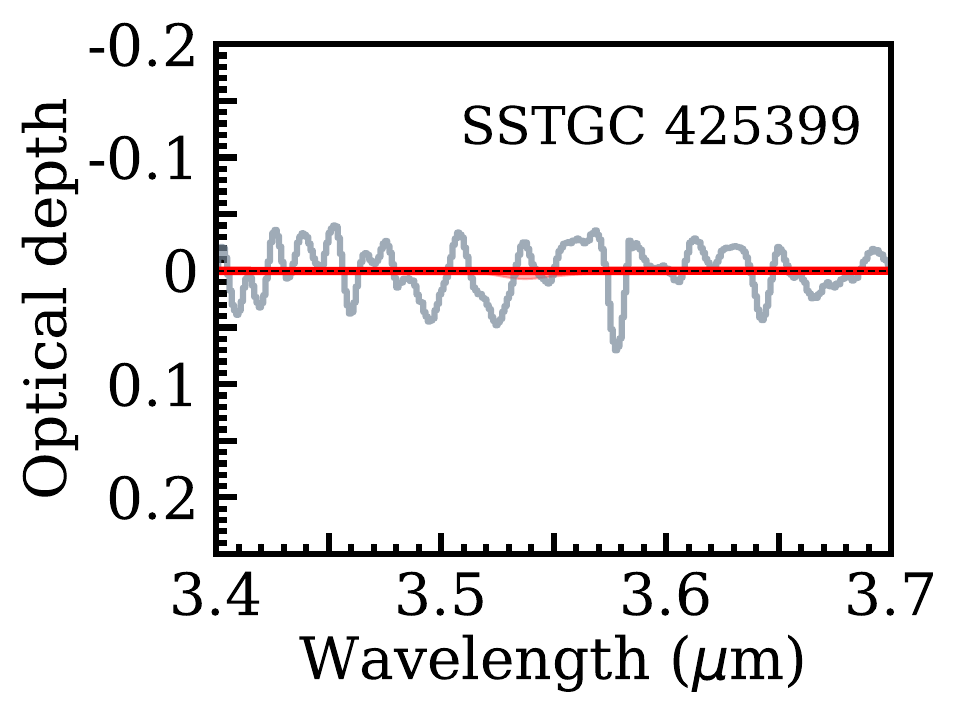}
\includegraphics[width=0.25\textwidth]{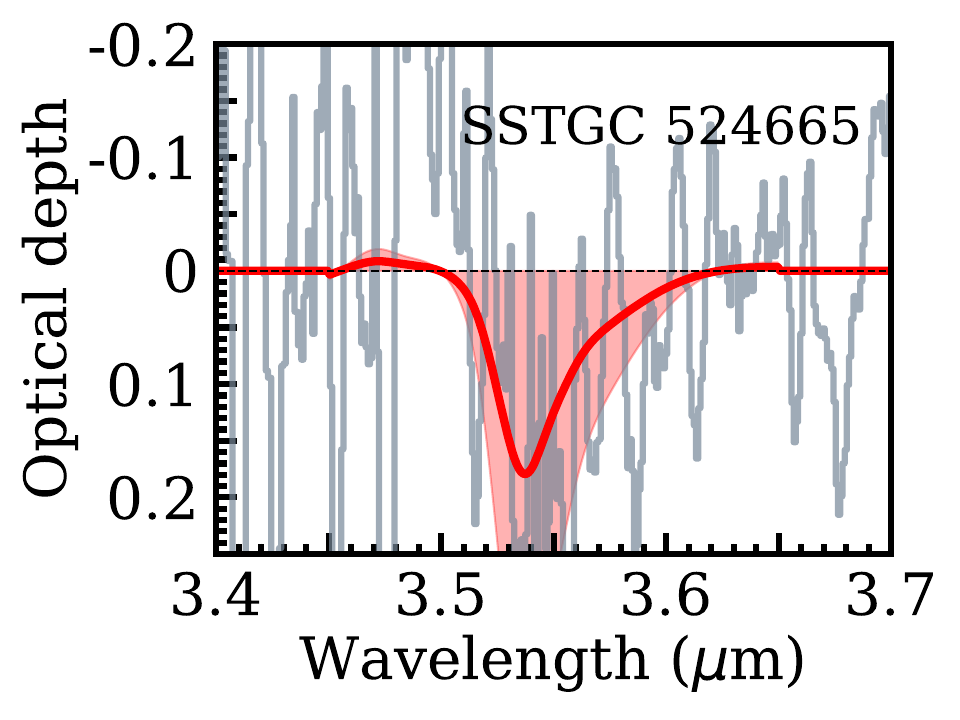}
\includegraphics[width=0.25\textwidth]{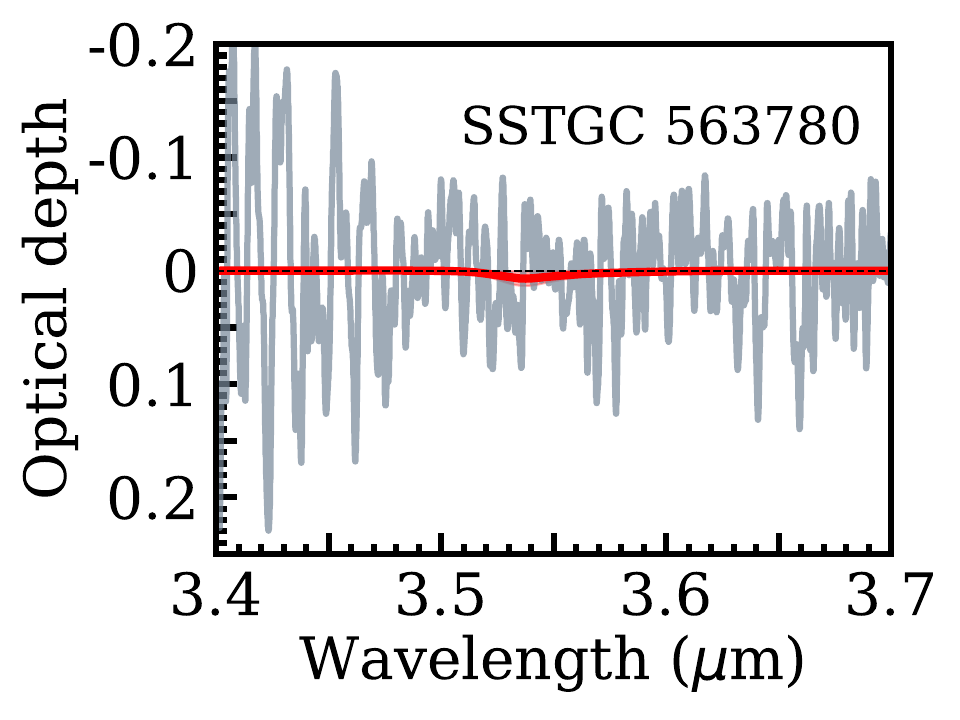} 
\includegraphics[width=0.25\textwidth]{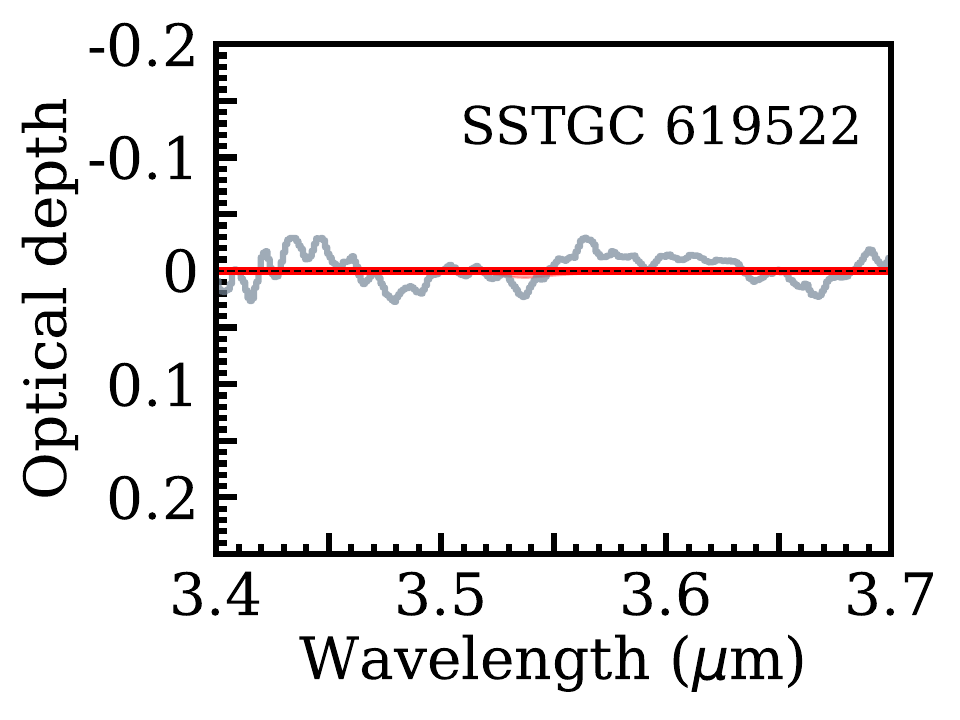}
\includegraphics[width=0.25\textwidth]{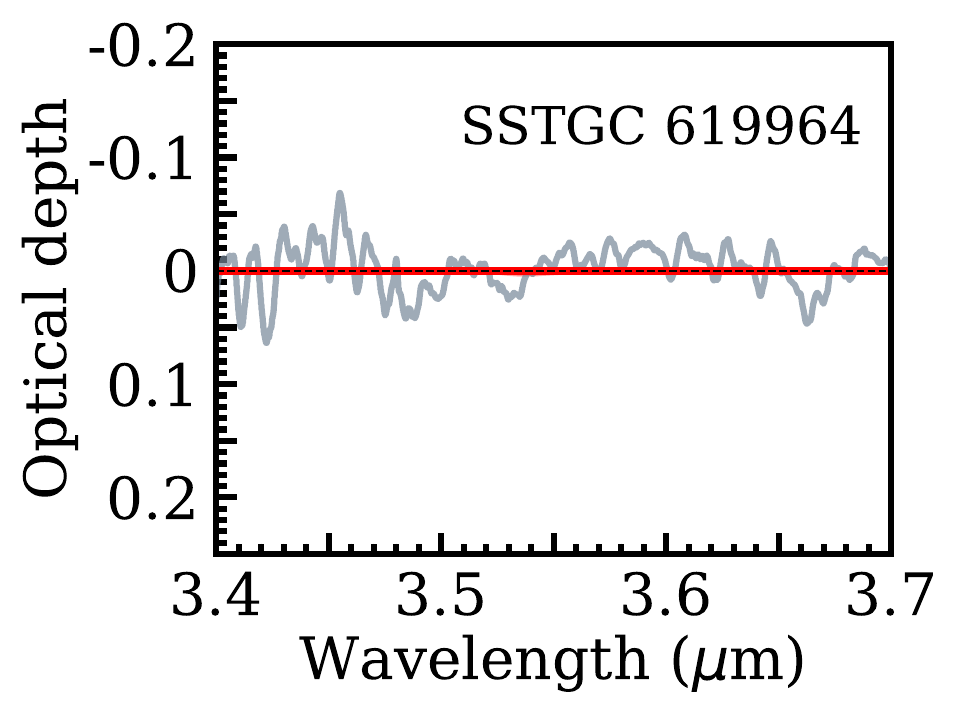}
\includegraphics[width=0.25\textwidth]{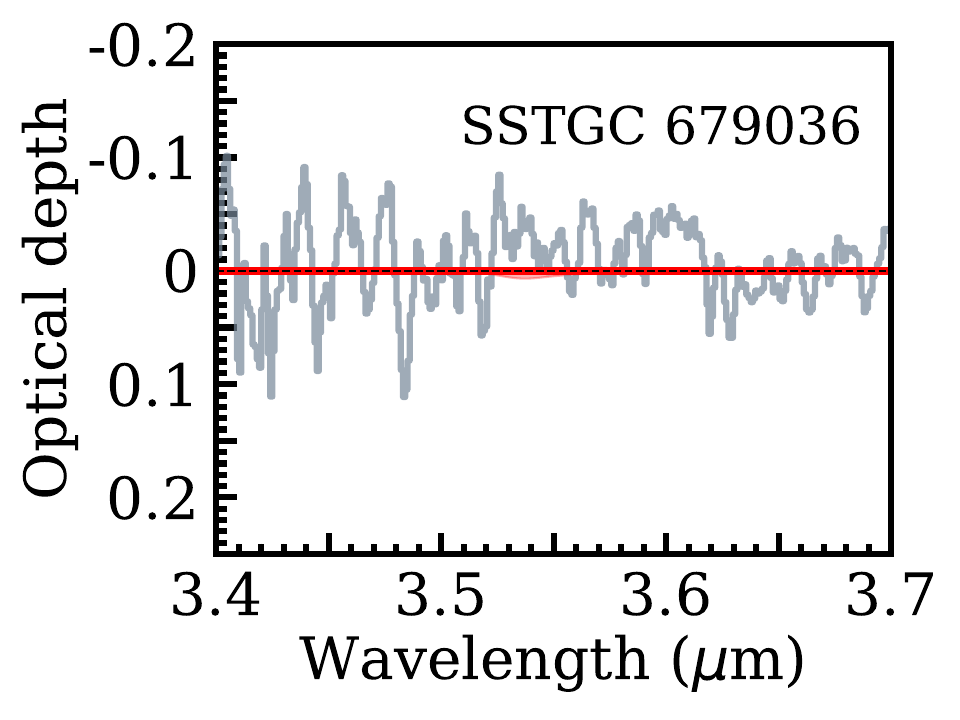}
\includegraphics[width=0.25\textwidth]{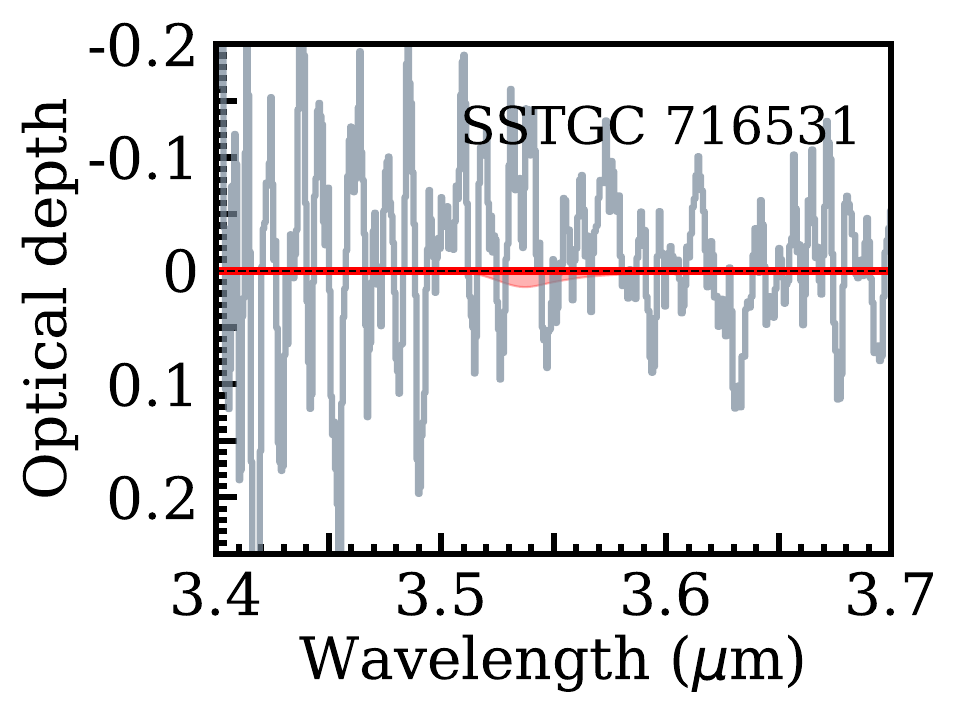}
\includegraphics[width=0.25\textwidth]{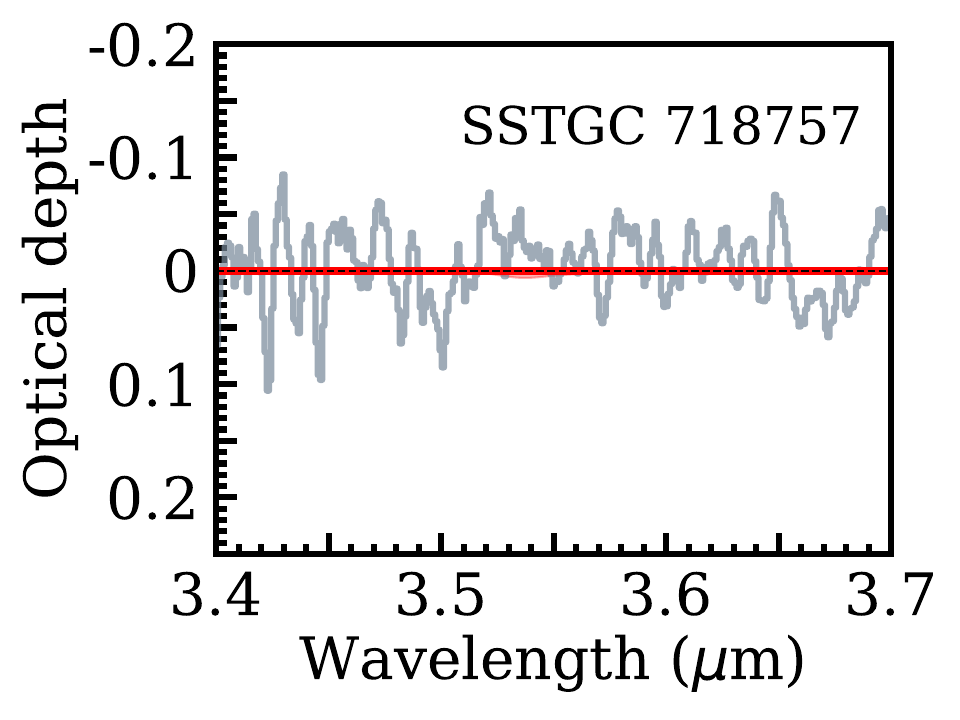}
\includegraphics[width=0.25\textwidth]{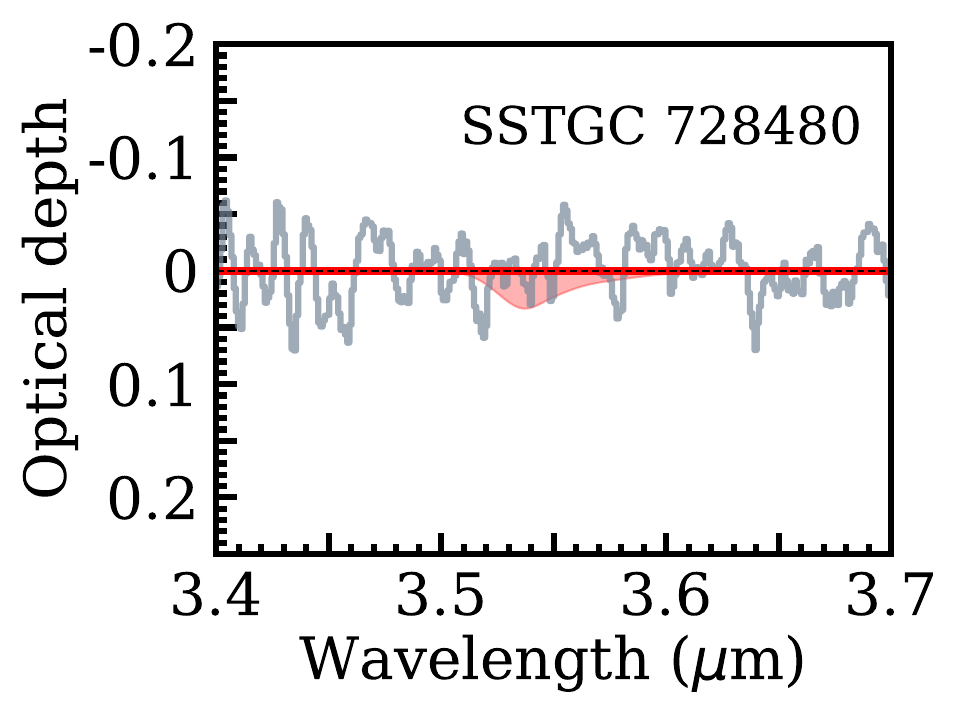}
\includegraphics[width=0.25\textwidth]{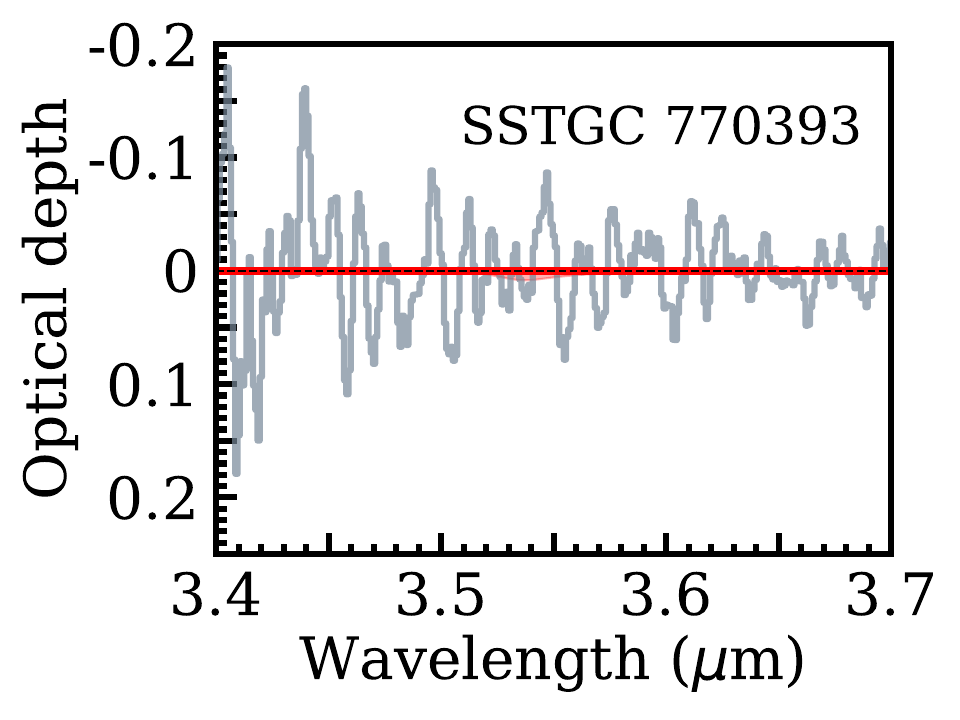}
\includegraphics[width=0.25\textwidth]{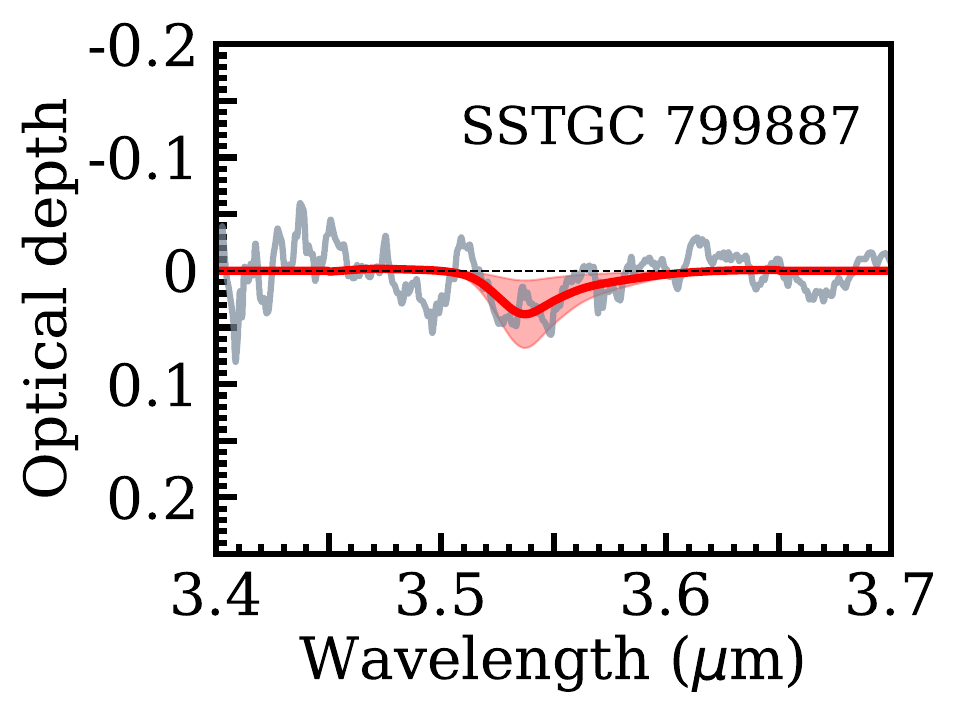}
\caption{Same as Fig.~\ref{fig:ch3oh}, but showing the optical-depth spectra of the remaining CMZ sources with non-significant $3.535\ \mu$m absorption.}
\label{fig:ch3oh2}
\end{figure*}

Figure~\ref{fig:ch3oh2} presents the remaining set of observed optical depth spectra centred on the $3.535\ \mu$m CH$_3$OH band. Each panel follows the same format as Fig.~\ref{fig:ch3oh}, which shows only the sources with significant detections. The observed spectra are shown in grey, and the red curves indicate the best-fitting models described in Sect.~\ref{sec:ch3oh}. These plots complement the fitting results listed in Table~\ref{tab:ch3oh}.

\FloatBarrier

\section{Methanol ice column densities from spectra around $9.7\ \mu$m}\label{sec:97}

In addition to the CH$_3$OH ice derived from the $3.535\ \mu$m absorption band, we used Spitzer/IRS spectra \citep{an:11} to measure the peak optical depth and column density of CH$_3$OH ice from the $9.74\ \mu$m band. Because the band depth is highly sensitive to the precise shape of the local continuum and the feature lies within the core of the stronger silicate absorption, we adopted a third-order polynomial to define the continuum across the band rather than relying on the global SED silicate profile (Sect.~\ref{sec:sed}). The results are listed in Table~\ref{tab:ch3oh97}, with uncertainties that include both statistical errors and systematics, such as those due to continuum placement. Owing to the very strong $10\ \mu$m silicate absorption towards our extremely red CMZ sources, the S/N at the band centre is low, and meaningful constraints could be derived for only a few objects.

\begin{figure*}[h]
\centering
\includegraphics[width=0.27\textwidth]{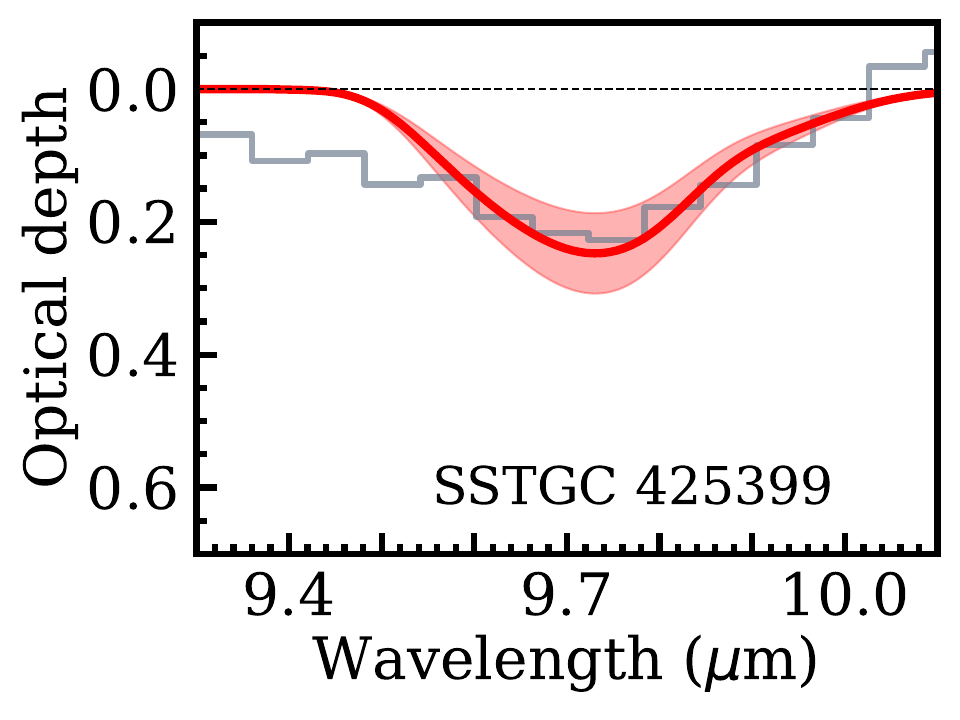}
\includegraphics[width=0.27\textwidth]{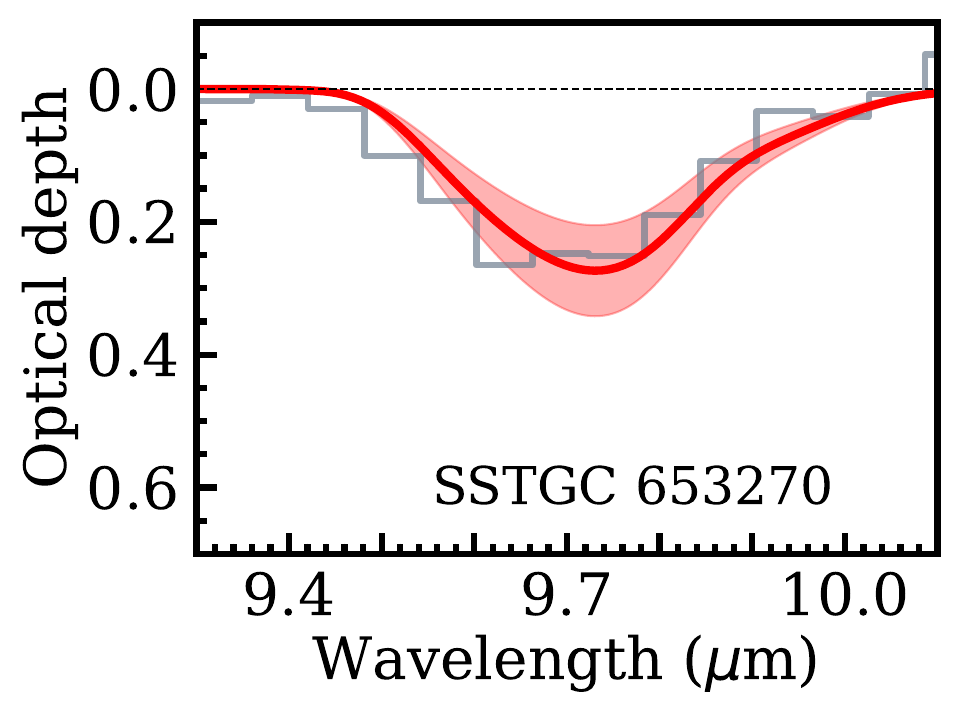}
\includegraphics[width=0.27\textwidth]{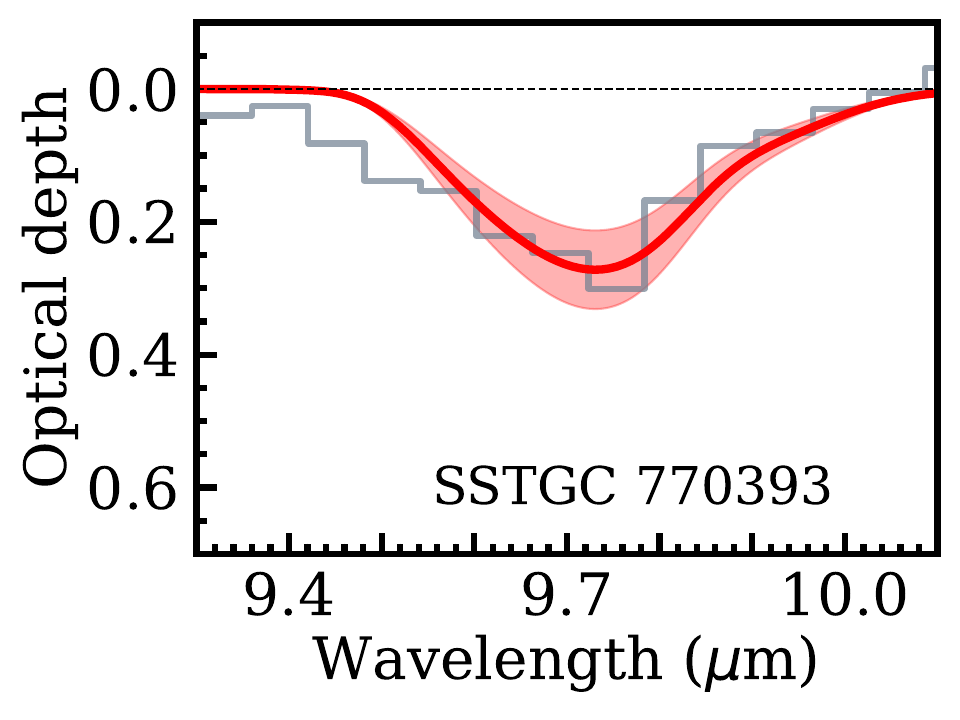}
\caption{Optical depth spectra of CMZ objects showing the $9.74\ \mu$m absorption feature of CH$_3$OH ice. Only the three sources detected at $>2.5\sigma$ are displayed.}
\label{fig:ch3oh97}
\end{figure*}

Figure~\ref{fig:ch3oh97} presents the optical-depth spectra of the three CMZ objects with significant ($>2.5\sigma$) detections of the $9.74\ \mu$m CH$_3$OH absorption. The $3.535\ \mu$m measurements (Table~\ref{tab:ch3oh}) show a mild detection for SSTGC~653270, but SSTGC~425399 and SSTGC~770393 exhibit non-detections. The discrepancy between the near-IR (Table\ref{tab:ch3oh}) and mid-IR (Table~\ref{tab:ch3oh97}) abundances may reflect the low S/N at the bottom of the silicate band affecting the $9.7\ \mu$m measurements, or could arise from dual sightlines, with the near- and mid-IR bands tracing different columns along the line of sight.

\begin{table}[h]
\centering
\caption{Peak optical depths and column densities of solid CH$_3$OH from $9.74\ \mu$m}
\label{tab:ch3oh97}
\begin{tabular}{ccc}
\hline\hline
Object ID & $\tau_{9.74}$ & $\nmethanol$ \\ 
(SSTGG) & & ($10^{17}\ \mathrm{cm}^{-2}$) \\
\hline
300758 & $<0.5$    & $<10.3$ \\
348392 & $<0.2$   & $<3.3$  \\
372630 & $<0.7$ & $<13.5$ \\
388790 & $<0.3$ & $<6.3$ \\
404312 & $<0.3$ & $<6.8$ \\
405235 & $<0.1$ & $<2.8$ \\
425399 & 0.3 $\pm$ 0.1 & 5.0 $\pm$ 1.5 \\
524665 & $<0.7$  & $<14.1$ \\
563780 & $<4.9$ & $<97.8$ \\
619522 & $<1.2$ & $<24.5$ \\
619964 & $<0.3$ & $<5.3$ \\
653270 & 0.3 $\pm$ 0.1 & 5.5 $\pm$ 1.6 \\
679036 & $<19.5$ & $<394.0$ \\
696367 & $<0.8$ & $<15.8$ \\
716531 & $<0.3$ & $<6.8$ \\
718757 & $<0.2$ & $<3.8$ \\
719445 & $<0.7$ & $<13.1$ \\
726327 & $<0.8$ & $<15.9$ \\
728480 & $<0.8$ & $<15.3$ \\
770393 & $0.3 \pm 0.1$ & 5.5 $\pm$ 1.5 \\
772981 & $<0.9$ & $<18.9$  \\
799887 & $<0.8$ & $<15.8$ \\
817031 & $<0.3$ & $<6.1$ \\
\hline
\end{tabular}
\end{table}

\FloatBarrier
\clearpage

\section{Maser observations of the full \citet{an:11} survey sample}\label{sec:maser}

We conducted a simultaneous survey of H$_2$O $6{23} - 5_{23}$ (22.23508 GHz) and class~I CH$_3$OH $7_0 - 6_1$ A$^+$ (44.06943 GHz) masers towards all 107 sources in June 2011 and April 2012. Observations were carried out with the Korean VLBI Network (KVN) Tamna telescope in single-dish mode \citep{kim:11,lee:11}. The main beam sizes were $130\arcsec$ at 22 GHz and $65\arcsec$ at 44 GHz, with corresponding aperture efficiencies of 0.70 and 0.69. Pointing and focus checks were performed every 2--3 h, with pointing accuracy better than $5\arcsec$. A digital spectrometer backend provided 4096 channels across a 64 MHz bandwidth. All spectra were obtained using position switching. Typical system temperatures were 100--200 K at 22 GHz and 150--350 K at 44 GHz. The on-source integration time was typically 10 min, yielding an rms noise level of $\sim$0.5 Jy at a velocity resolution of 0.2 km s$^{-1}$. Data were calibrated with the standard chopper-wheel method and placed on the T$_{\rm A}^*$ scale. The conversion factors from T$_{\rm A}^*$ to flux density were 11.1 Jy K$^{-1}$ at 22 GHz and 11.6 Jy K$^{-1}$ at 44 GHz.

We detected H$_2$O maser emission in 7 sources (7\%) -- SSTGC~304239, 354683, 358370, 360559, 511261, 524665, and 381931 -- and CH$_3$OH maser emission in 6 sources (6\%) — SSTGC~511261, 524665, 543691, 564417, 536969, and 381931. Three sources (SSTGC~381931, 511261, and 524665) exhibit both water and methanol maser emission. The detection rates are much lower than those reported in previous surveys of high-mass proto-stetllar objects and ultra-compact H II regions, e.g. 42\% for H$_2$O masers \citep{sridharan:02} and 31\% for 44 GHz CH$_3$OH masers \citep{fontani:10}. The detection rates of these two maser species tend to increase as the central objects evolve (\citet{kim:18}; \citet{kim:19}).

No H$_2$O or Class~I CH$_3$OH maser emission was detected towards any of the YSOs in our sample except SSTGC~524665, despite their initial classification as massive YSO candidates. H$_2$O masers were detected both in YSOs and evolved stars, while CH$_3$OH masers were predominantly detected towards high-mass YSOs, with Class~I masers tracing shocked gas in outflows \citep{menten:91, ellingsen:06}. The almost complete absence of maser emission in our sample therefore suggests that these objects may be less evolved than previously thought -- possibly in the deeply embedded Class 0/I stage -- or that their masses are overestimated, placing them in a regime where the necessary shock conditions for maser pumping are absent.

\begin{table}[h]
\centering
\caption{Maser Detections from KVN Observation}
\label{tab:kvn}
\begin{tabular}{lccc}
\hline
\hline
 & \multicolumn{2}{c}{Peak flux (Jy)} & \\
\cline{2-3}
SSTGC & H$_2$O maser & CH$_3$OH maser & This Study\\
\hline
304239 & 10.0 & &  \\
354683 & ~6.7 & & \\
358370 & 13.3 & & \\
360559 & ~7.8 & & \\
381931 & ~3.3 & ~4.6 & \\
511261 & 15.5 & ~4.6 & \\
524665 & 222.0 & 69.6 & \checkmark \\
536969 & & ~1.2 & \\
543691 & & ~2.3 & \\
564417 & & ~4.6 & \\
\hline
\end{tabular} 
\end{table}

\end{appendix}

\end{document}